\documentclass[12pt]{article}
\usepackage{epsfig,amsfonts,amssymb}
\usepackage{hyperref}
\usepackage{cite}
\input epsf.sty
\usepackage{cite}

\newcommand{\hab}{}

\def\ZZZ{{\hbox{ Z\kern-1.6mm Z}}}
\def\RRR{{\hbox{ R\kern-2.4mm R}}}
\def\CCC{{\hbox{ C\kern-2.0mm C}}}
\def\zzz{{\hbox{z\kern-1mm z}}}

\newcommand{\ten}{{(10)}}
\newcommand{\bet}{{( b )}}

\newcommand{\qq}{k}
\newcommand{\pp}{l}
\newcommand{\nn}{\nonumber \\}

\newcommand{\vt}{\vartheta}

\newcommand{\vtau} {\vec \tau}
\newcommand{\vj} {\vec J}
\newcommand{\vxi} {\vec \xi}
\newcommand{\vu} {\vec u}
\newcommand{\htau} {\vec \eta}
\newcommand{\vc}{\vec\chi}
\newcommand{\vpsi} {\vec \psi}

\newcommand{\qeq}{{\hbox{=\kern-2.3mm ? \kern.5mm }}}
\renewcommand{\qeq}{=}

\newcommand{\rrho}{r}
\newcommand{\bA}{{\bf A}}
\newcommand{\tx}{\wt x}
\newcommand{\bG}{{\bf G}}
\newcommand{\bF}{{\bar F}}
\newcommand{\bbb}{{\bar b}}
\newcommand{\gam}{\tau}
\newcommand{\eps}{\epsilon}
\newcommand{\vareps}{\varepsilon}
\newcommand{\ra}{\rangle}
\newcommand{\la}{\langle}
\newcommand{\T}{\chi_{T}(k)}
\newcommand{\Tm}{\chi_{T}(k')}
\newcommand{\Cn}{{\cal C}_n}
\newcommand{\vp}{\varphi}
\newcommand{\ve}{\varepsilon}
\newcommand{\tl}{\lambda}
\newcommand{\dt}{(\vec \nabla T)^2}
\newcommand{\hp}{{\wh\Phi}}
\newcommand{\hq}{{\wh Q_B}}
\newcommand{\he}{{\wh\eta_0}}
\newcommand{\ha}{{\wh{A}}}
\newcommand{\lllb}{\Bigl\langle\Bigl\langle}
\newcommand{\rrrb}{\Bigr\rangle\Bigr\rangle}
\newcommand{\tf}{\wt f}
\newcommand{\sss}{{\cal L}_{av}}
\newcommand{\bx}{\bar x}
\newcommand{\bw}{\bar w}
\newcommand{\ws}{{\wt\sigma}}
\newcommand{\wrh}{{\wt\rho}}
\newcommand{\wv}{{\wt v}}

\newcommand{\vv} {\bar v}
\newcommand{\uu} {\bar u}

\newcommand{\K}{{\rm K_1}}
\newcommand{\Kt}{{\rm \widetilde K_1}}

\newcommand{\bB}{\bar b'}
\newcommand{\Bu}{B_{\vec u}}
\newcommand{\VV}{{\cal V}}
\newcommand{\BB}{{\cal B}}
\newcommand{\DD}{{\cal D}}
\newcommand{\BBB}{{\bf B}}
\newcommand{\II}{{\cal I}}
\newcommand{\AAA}{{\cal A}}
\newcommand{\GG}{{\cal G}}
\newcommand{\KK}{{\cal K}}
\newcommand{\fff}{{\bf f}}
\newcommand{\ccc}{{\bf c}}
\newcommand{\FF}{{\cal F}}
\newcommand{\JJ}{{\cal J}}
\newcommand{\HH}{{\cal H}}
\newcommand{\MM}{{\cal M}}
\newcommand{\CC}{{\cal C}}
\newcommand{\bC}{{\bf C}}
\newcommand{\OO}{{\cal O}}
\newcommand{\QQ}{{\cal Q}}
\newcommand{\PP}{{\cal P}}
\newcommand{\EE}{{\cal E}}
\newcommand{\LL}{{\cal L}}
 \newcommand{\rrr}{\rangle\rangle}
\newcommand{\half}{{1\over 2}}
\newcommand{\wt}{\widetilde}
\newcommand{\wh}{\widehat}
\newcommand{\wc}{\wt}
\newcommand{\wb}{\bar}
\newcommand{\RR}{{\cal R}}
\newcommand{\NN}{{\cal N}}
\newcommand{\TT}{{\cal T}}
\newcommand{\bg}{\bar g}
\newcommand{\ba}{\bar a}
\newcommand{\bc}{\bar c}
\newcommand{\bd}{\bar d}
\newcommand{\bb}{\bar b}
\newcommand{\bT}{\bar \Theta}
\newcommand{\SSS}{{\cal S}}
\newcommand{\tlx}{\left(\tilde \lambda ; X^0(0) \right)}
\newcommand{\al}{\alpha}

\newcommand{\tk}{\tilde \kappa}

\newcommand{\ppp}{\prime\prime}

\newcommand{\omk}{\omega_n(\vec k)}
\newcommand{\onk}{\omega^{(N)}_{\vec k_\perp}}
\newcommand{\tI}{\wt\II}
\newcommand{\hI}{\wh\II}
\newcommand{\nI}{\II}

\newcommand{\cp}{\check\Phi}
\newcommand{\cps}{\Psi}
\newcommand{\crh}{\check\rho}
\newcommand{\cs}{\check\sigma}
\newcommand{\cv}{\check v}
\newcommand{\com}{\check\Omega}

\newcommand{\be}{\begin{equation}}
\newcommand{\ee}{\end{equation}}
\newcommand{\ben}{\begin{eqnarray}\displaystyle}
\newcommand{\een}{\end{eqnarray}}

\newcommand{\refb}[1]{(\ref{#1})}
\newcommand{\p}{\partial}
\newcommand{\sectiono}[1]{\section{#1}\setcounter{equation}{0}}
\newcommand{\subsectiono}[1]{\subsection{#1}\setcounter{equation}{0}}

\newcommand{\zet}{\zeta}

\newcommand{\gsim}{\stackrel{>}{\sim}}
\newcommand{\lsim}{\stackrel{<}{\sim}}

\newcommand{\Lamb}{\Lambda}

\def\one{{\hbox{ 1\kern-.8mm l}}}
\def\zero{{\hbox{ 0\kern-1.5mm 0}}}

\def\wa{{\wh a}}
\def\wb{{\wh b}}
\def\wc{{\wh c}}
\def\wc{\check}
\def\wdd{{\wh d}}

\newcommand{\bi}{{\bf i}}

\renewcommand{\theequation}{\thesection.\arabic{equation}}

\newcommand{\bea}[1]{\begin{eqnarray}\label{#1} }
\newcommand{\eea}{\end{eqnarray}}

\newcommand{\wJ}{\wt J}
\newcommand{\bN}{{\bf N}}

\newcommand{\aaa}{b}

\newcommand{\eqref}{\refb}

\newcommand{\un}{{\rm u}}

\newcommand{\XX}{{\cal X}}

\usepackage{bm}
\usepackage[table]{xcolor}
\def\rpnote#1{{\color{magenta} #1}}
\def\arnote#1{{\color{blue} #1}}
\def\asnote#1{{\color{red} #1}}

\def\figextra{

\ifx\JPicScale\undefined\def\JPicScale{1}\fi
\unitlength \JPicScale mm
\begin{picture}(100,80)(0,0)
\linethickness{0.3mm}
\put(71.01,39.25){\line(0,1){0.5}}
\multiput(70.99,40.25)(0.02,-0.5){1}{\line(0,-1){0.5}}
\multiput(70.94,40.75)(0.05,-0.5){1}{\line(0,-1){0.5}}
\multiput(70.87,41.24)(0.07,-0.5){1}{\line(0,-1){0.5}}
\multiput(70.77,41.73)(0.09,-0.49){1}{\line(0,-1){0.49}}
\multiput(70.65,42.22)(0.12,-0.49){1}{\line(0,-1){0.49}}
\multiput(70.51,42.7)(0.14,-0.48){1}{\line(0,-1){0.48}}
\multiput(70.35,43.17)(0.16,-0.47){1}{\line(0,-1){0.47}}
\multiput(70.16,43.64)(0.09,-0.23){2}{\line(0,-1){0.23}}
\multiput(69.96,44.09)(0.1,-0.23){2}{\line(0,-1){0.23}}
\multiput(69.73,44.54)(0.11,-0.22){2}{\line(0,-1){0.22}}
\multiput(69.48,44.97)(0.13,-0.22){2}{\line(0,-1){0.22}}
\multiput(69.21,45.39)(0.14,-0.21){2}{\line(0,-1){0.21}}
\multiput(68.92,45.8)(0.15,-0.2){2}{\line(0,-1){0.2}}
\multiput(68.61,46.19)(0.1,-0.13){3}{\line(0,-1){0.13}}
\multiput(68.28,46.57)(0.11,-0.13){3}{\line(0,-1){0.13}}
\multiput(67.93,46.93)(0.12,-0.12){3}{\line(0,-1){0.12}}
\multiput(67.57,47.28)(0.12,-0.12){3}{\line(1,0){0.12}}
\multiput(67.19,47.61)(0.13,-0.11){3}{\line(1,0){0.13}}
\multiput(66.8,47.92)(0.13,-0.1){3}{\line(1,0){0.13}}
\multiput(66.39,48.21)(0.2,-0.15){2}{\line(1,0){0.2}}
\multiput(65.97,48.48)(0.21,-0.14){2}{\line(1,0){0.21}}
\multiput(65.54,48.73)(0.22,-0.13){2}{\line(1,0){0.22}}
\multiput(65.09,48.96)(0.22,-0.11){2}{\line(1,0){0.22}}
\multiput(64.64,49.16)(0.23,-0.1){2}{\line(1,0){0.23}}
\multiput(64.17,49.35)(0.23,-0.09){2}{\line(1,0){0.23}}
\multiput(63.7,49.51)(0.47,-0.16){1}{\line(1,0){0.47}}
\multiput(63.22,49.65)(0.48,-0.14){1}{\line(1,0){0.48}}
\multiput(62.73,49.77)(0.49,-0.12){1}{\line(1,0){0.49}}
\multiput(62.24,49.87)(0.49,-0.09){1}{\line(1,0){0.49}}
\multiput(61.75,49.94)(0.5,-0.07){1}{\line(1,0){0.5}}
\multiput(61.25,49.99)(0.5,-0.05){1}{\line(1,0){0.5}}
\multiput(60.75,50.01)(0.5,-0.02){1}{\line(1,0){0.5}}
\put(60.25,50.01){\line(1,0){0.5}}
\multiput(59.75,49.99)(0.5,0.02){1}{\line(1,0){0.5}}
\multiput(59.25,49.94)(0.5,0.05){1}{\line(1,0){0.5}}
\multiput(58.76,49.87)(0.5,0.07){1}{\line(1,0){0.5}}
\multiput(58.27,49.77)(0.49,0.09){1}{\line(1,0){0.49}}
\multiput(57.78,49.65)(0.49,0.12){1}{\line(1,0){0.49}}
\multiput(57.3,49.51)(0.48,0.14){1}{\line(1,0){0.48}}
\multiput(56.83,49.35)(0.47,0.16){1}{\line(1,0){0.47}}
\multiput(56.36,49.16)(0.23,0.09){2}{\line(1,0){0.23}}
\multiput(55.91,48.96)(0.23,0.1){2}{\line(1,0){0.23}}
\multiput(55.46,48.73)(0.22,0.11){2}{\line(1,0){0.22}}
\multiput(55.03,48.48)(0.22,0.13){2}{\line(1,0){0.22}}
\multiput(54.61,48.21)(0.21,0.14){2}{\line(1,0){0.21}}
\multiput(54.2,47.92)(0.2,0.15){2}{\line(1,0){0.2}}
\multiput(53.81,47.61)(0.13,0.1){3}{\line(1,0){0.13}}
\multiput(53.43,47.28)(0.13,0.11){3}{\line(1,0){0.13}}
\multiput(53.07,46.93)(0.12,0.12){3}{\line(1,0){0.12}}
\multiput(52.72,46.57)(0.12,0.12){3}{\line(0,1){0.12}}
\multiput(52.39,46.19)(0.11,0.13){3}{\line(0,1){0.13}}
\multiput(52.08,45.8)(0.1,0.13){3}{\line(0,1){0.13}}
\multiput(51.79,45.39)(0.15,0.2){2}{\line(0,1){0.2}}
\multiput(51.52,44.97)(0.14,0.21){2}{\line(0,1){0.21}}
\multiput(51.27,44.54)(0.13,0.22){2}{\line(0,1){0.22}}
\multiput(51.04,44.09)(0.11,0.22){2}{\line(0,1){0.22}}
\multiput(50.84,43.64)(0.1,0.23){2}{\line(0,1){0.23}}
\multiput(50.65,43.17)(0.09,0.23){2}{\line(0,1){0.23}}
\multiput(50.49,42.7)(0.16,0.47){1}{\line(0,1){0.47}}
\multiput(50.35,42.22)(0.14,0.48){1}{\line(0,1){0.48}}
\multiput(50.23,41.73)(0.12,0.49){1}{\line(0,1){0.49}}
\multiput(50.13,41.24)(0.09,0.49){1}{\line(0,1){0.49}}
\multiput(50.06,40.75)(0.07,0.5){1}{\line(0,1){0.5}}
\multiput(50.01,40.25)(0.05,0.5){1}{\line(0,1){0.5}}
\multiput(49.99,39.75)(0.02,0.5){1}{\line(0,1){0.5}}
\put(49.99,39.25){\line(0,1){0.5}}
\multiput(49.99,39.25)(0.02,-0.5){1}{\line(0,-1){0.5}}
\multiput(50.01,38.75)(0.05,-0.5){1}{\line(0,-1){0.5}}
\multiput(50.06,38.25)(0.07,-0.5){1}{\line(0,-1){0.5}}
\multiput(50.13,37.76)(0.09,-0.49){1}{\line(0,-1){0.49}}
\multiput(50.23,37.27)(0.12,-0.49){1}{\line(0,-1){0.49}}
\multiput(50.35,36.78)(0.14,-0.48){1}{\line(0,-1){0.48}}
\multiput(50.49,36.3)(0.16,-0.47){1}{\line(0,-1){0.47}}
\multiput(50.65,35.83)(0.09,-0.23){2}{\line(0,-1){0.23}}
\multiput(50.84,35.36)(0.1,-0.23){2}{\line(0,-1){0.23}}
\multiput(51.04,34.91)(0.11,-0.22){2}{\line(0,-1){0.22}}
\multiput(51.27,34.46)(0.13,-0.22){2}{\line(0,-1){0.22}}
\multiput(51.52,34.03)(0.14,-0.21){2}{\line(0,-1){0.21}}
\multiput(51.79,33.61)(0.15,-0.2){2}{\line(0,-1){0.2}}
\multiput(52.08,33.2)(0.1,-0.13){3}{\line(0,-1){0.13}}
\multiput(52.39,32.81)(0.11,-0.13){3}{\line(0,-1){0.13}}
\multiput(52.72,32.43)(0.12,-0.12){3}{\line(0,-1){0.12}}
\multiput(53.07,32.07)(0.12,-0.12){3}{\line(1,0){0.12}}
\multiput(53.43,31.72)(0.13,-0.11){3}{\line(1,0){0.13}}
\multiput(53.81,31.39)(0.13,-0.1){3}{\line(1,0){0.13}}
\multiput(54.2,31.08)(0.2,-0.15){2}{\line(1,0){0.2}}
\multiput(54.61,30.79)(0.21,-0.14){2}{\line(1,0){0.21}}
\multiput(55.03,30.52)(0.22,-0.13){2}{\line(1,0){0.22}}
\multiput(55.46,30.27)(0.22,-0.11){2}{\line(1,0){0.22}}
\multiput(55.91,30.04)(0.23,-0.1){2}{\line(1,0){0.23}}
\multiput(56.36,29.84)(0.23,-0.09){2}{\line(1,0){0.23}}
\multiput(56.83,29.65)(0.47,-0.16){1}{\line(1,0){0.47}}
\multiput(57.3,29.49)(0.48,-0.14){1}{\line(1,0){0.48}}
\multiput(57.78,29.35)(0.49,-0.12){1}{\line(1,0){0.49}}
\multiput(58.27,29.23)(0.49,-0.09){1}{\line(1,0){0.49}}
\multiput(58.76,29.13)(0.5,-0.07){1}{\line(1,0){0.5}}
\multiput(59.25,29.06)(0.5,-0.05){1}{\line(1,0){0.5}}
\multiput(59.75,29.01)(0.5,-0.02){1}{\line(1,0){0.5}}
\put(60.25,28.99){\line(1,0){0.5}}
\multiput(60.75,28.99)(0.5,0.02){1}{\line(1,0){0.5}}
\multiput(61.25,29.01)(0.5,0.05){1}{\line(1,0){0.5}}
\multiput(61.75,29.06)(0.5,0.07){1}{\line(1,0){0.5}}
\multiput(62.24,29.13)(0.49,0.09){1}{\line(1,0){0.49}}
\multiput(62.73,29.23)(0.49,0.12){1}{\line(1,0){0.49}}
\multiput(63.22,29.35)(0.48,0.14){1}{\line(1,0){0.48}}
\multiput(63.7,29.49)(0.47,0.16){1}{\line(1,0){0.47}}
\multiput(64.17,29.65)(0.23,0.09){2}{\line(1,0){0.23}}
\multiput(64.64,29.84)(0.23,0.1){2}{\line(1,0){0.23}}
\multiput(65.09,30.04)(0.22,0.11){2}{\line(1,0){0.22}}
\multiput(65.54,30.27)(0.22,0.13){2}{\line(1,0){0.22}}
\multiput(65.97,30.52)(0.21,0.14){2}{\line(1,0){0.21}}
\multiput(66.39,30.79)(0.2,0.15){2}{\line(1,0){0.2}}
\multiput(66.8,31.08)(0.13,0.1){3}{\line(1,0){0.13}}
\multiput(67.19,31.39)(0.13,0.11){3}{\line(1,0){0.13}}
\multiput(67.57,31.72)(0.12,0.12){3}{\line(1,0){0.12}}
\multiput(67.93,32.07)(0.12,0.12){3}{\line(0,1){0.12}}
\multiput(68.28,32.43)(0.11,0.13){3}{\line(0,1){0.13}}
\multiput(68.61,32.81)(0.1,0.13){3}{\line(0,1){0.13}}
\multiput(68.92,33.2)(0.15,0.2){2}{\line(0,1){0.2}}
\multiput(69.21,33.61)(0.14,0.21){2}{\line(0,1){0.21}}
\multiput(69.48,34.03)(0.13,0.22){2}{\line(0,1){0.22}}
\multiput(69.73,34.46)(0.11,0.22){2}{\line(0,1){0.22}}
\multiput(69.96,34.91)(0.1,0.23){2}{\line(0,1){0.23}}
\multiput(70.16,35.36)(0.09,0.23){2}{\line(0,1){0.23}}
\multiput(70.35,35.83)(0.16,0.47){1}{\line(0,1){0.47}}
\multiput(70.51,36.3)(0.14,0.48){1}{\line(0,1){0.48}}
\multiput(70.65,36.78)(0.12,0.49){1}{\line(0,1){0.49}}
\multiput(70.77,37.27)(0.09,0.49){1}{\line(0,1){0.49}}
\multiput(70.87,37.76)(0.07,0.5){1}{\line(0,1){0.5}}
\multiput(70.94,38.25)(0.05,0.5){1}{\line(0,1){0.5}}
\multiput(70.99,38.75)(0.02,0.5){1}{\line(0,1){0.5}}

\linethickness{1mm}
\multiput(10,70)(0.3,-0.12){167}{\line(1,0){0.3}}
\linethickness{1mm}
\multiput(11,10)(0.27,0.12){167}{\line(1,0){0.27}}
\linethickness{0.3mm}
\put(38,2){\line(0,1){20}}
\linethickness{0.3mm}
\put(40,58){\line(0,1){20}}
\linethickness{0.3mm}
\put(18,13){\line(0,1){54}}
\linethickness{1mm}
\multiput(69,45)(0.14,0.12){208}{\line(1,0){0.14}}
\linethickness{1mm}
\multiput(68,32)(0.18,-0.12){167}{\line(1,0){0.18}}
\put(85,50){\makebox(0,0)[cc]{$\cdot$}}

\put(85,40){\makebox(0,0)[cc]{$\cdot$}}

\put(85,30){\makebox(0,0)[cc]{$\cdot$}}

\put(5,75){\makebox(0,0)[cc]{$p_j$}}

\put(10,5){\makebox(0,0)[cc]{$p_i$}}

\put(14,40){\makebox(0,0)[cc]{$\ell$}}

\put(30,68){\makebox(0,0)[cc]{$p_j-\ell$}}

\put(45,75){\makebox(0,0)[cc]{$k_1$}}

\put(55,60){\makebox(0,0)[cc]{$p_j-\ell+k_1$}}

\put(35,5){\makebox(0,0)[cc]{$k_2$}}

\put(28,13){\makebox(0,0)[cc]{$p_i+\ell$}}

\put(50,20){\makebox(0,0)[cc]{$p_i+\ell+k_2$}}

\put(60,40){\makebox(0,0)[cc]{$\Gamma$}}

\end{picture}

}

\def\figone{

\def\JPicScale{0.8}
\ifx\JPicScale\undefined\def\JPicScale{1}\fi
\unitlength \JPicScale mm
\begin{picture}(95,65.72)(0,0)
\linethickness{0.3mm}
\put(70.72,50.11){\line(0,1){0.5}}
\multiput(70.7,51.11)(0.02,-0.5){1}{\line(0,-1){0.5}}
\multiput(70.67,51.61)(0.03,-0.5){1}{\line(0,-1){0.5}}
\multiput(70.62,52.11)(0.05,-0.5){1}{\line(0,-1){0.5}}
\multiput(70.55,52.61)(0.07,-0.5){1}{\line(0,-1){0.5}}
\multiput(70.47,53.11)(0.08,-0.5){1}{\line(0,-1){0.5}}
\multiput(70.37,53.6)(0.1,-0.49){1}{\line(0,-1){0.49}}
\multiput(70.26,54.09)(0.11,-0.49){1}{\line(0,-1){0.49}}
\multiput(70.13,54.58)(0.13,-0.49){1}{\line(0,-1){0.49}}
\multiput(69.98,55.06)(0.15,-0.48){1}{\line(0,-1){0.48}}
\multiput(69.82,55.53)(0.16,-0.48){1}{\line(0,-1){0.48}}
\multiput(69.64,56)(0.18,-0.47){1}{\line(0,-1){0.47}}
\multiput(69.45,56.47)(0.1,-0.23){2}{\line(0,-1){0.23}}
\multiput(69.24,56.92)(0.1,-0.23){2}{\line(0,-1){0.23}}
\multiput(69.02,57.38)(0.11,-0.23){2}{\line(0,-1){0.23}}
\multiput(68.78,57.82)(0.12,-0.22){2}{\line(0,-1){0.22}}
\multiput(68.53,58.25)(0.13,-0.22){2}{\line(0,-1){0.22}}
\multiput(68.27,58.68)(0.13,-0.21){2}{\line(0,-1){0.21}}
\multiput(67.99,59.1)(0.14,-0.21){2}{\line(0,-1){0.21}}
\multiput(67.7,59.51)(0.15,-0.2){2}{\line(0,-1){0.2}}
\multiput(67.39,59.91)(0.1,-0.13){3}{\line(0,-1){0.13}}
\multiput(67.07,60.3)(0.11,-0.13){3}{\line(0,-1){0.13}}
\multiput(66.74,60.67)(0.11,-0.13){3}{\line(0,-1){0.13}}
\multiput(66.4,61.04)(0.11,-0.12){3}{\line(0,-1){0.12}}
\multiput(66.04,61.4)(0.12,-0.12){3}{\line(1,0){0.12}}
\multiput(65.67,61.74)(0.12,-0.11){3}{\line(1,0){0.12}}
\multiput(65.3,62.07)(0.13,-0.11){3}{\line(1,0){0.13}}
\multiput(64.91,62.39)(0.13,-0.11){3}{\line(1,0){0.13}}
\multiput(64.51,62.7)(0.13,-0.1){3}{\line(1,0){0.13}}
\multiput(64.1,62.99)(0.2,-0.15){2}{\line(1,0){0.2}}
\multiput(63.68,63.27)(0.21,-0.14){2}{\line(1,0){0.21}}
\multiput(63.25,63.53)(0.21,-0.13){2}{\line(1,0){0.21}}
\multiput(62.82,63.78)(0.22,-0.13){2}{\line(1,0){0.22}}
\multiput(62.38,64.02)(0.22,-0.12){2}{\line(1,0){0.22}}
\multiput(61.92,64.24)(0.23,-0.11){2}{\line(1,0){0.23}}
\multiput(61.47,64.45)(0.23,-0.1){2}{\line(1,0){0.23}}
\multiput(61,64.64)(0.23,-0.1){2}{\line(1,0){0.23}}
\multiput(60.53,64.82)(0.47,-0.18){1}{\line(1,0){0.47}}
\multiput(60.06,64.98)(0.48,-0.16){1}{\line(1,0){0.48}}
\multiput(59.58,65.13)(0.48,-0.15){1}{\line(1,0){0.48}}
\multiput(59.09,65.26)(0.49,-0.13){1}{\line(1,0){0.49}}
\multiput(58.6,65.37)(0.49,-0.11){1}{\line(1,0){0.49}}
\multiput(58.11,65.47)(0.49,-0.1){1}{\line(1,0){0.49}}
\multiput(57.61,65.55)(0.5,-0.08){1}{\line(1,0){0.5}}
\multiput(57.11,65.62)(0.5,-0.07){1}{\line(1,0){0.5}}
\multiput(56.61,65.67)(0.5,-0.05){1}{\line(1,0){0.5}}
\multiput(56.11,65.7)(0.5,-0.03){1}{\line(1,0){0.5}}
\multiput(55.61,65.72)(0.5,-0.02){1}{\line(1,0){0.5}}
\put(55.11,65.72){\line(1,0){0.5}}
\multiput(54.6,65.7)(0.5,0.02){1}{\line(1,0){0.5}}
\multiput(54.1,65.67)(0.5,0.03){1}{\line(1,0){0.5}}
\multiput(53.6,65.62)(0.5,0.05){1}{\line(1,0){0.5}}
\multiput(53.1,65.55)(0.5,0.07){1}{\line(1,0){0.5}}
\multiput(52.61,65.47)(0.5,0.08){1}{\line(1,0){0.5}}
\multiput(52.11,65.37)(0.49,0.1){1}{\line(1,0){0.49}}
\multiput(51.62,65.26)(0.49,0.11){1}{\line(1,0){0.49}}
\multiput(51.14,65.13)(0.49,0.13){1}{\line(1,0){0.49}}
\multiput(50.66,64.98)(0.48,0.15){1}{\line(1,0){0.48}}
\multiput(50.18,64.82)(0.48,0.16){1}{\line(1,0){0.48}}
\multiput(49.71,64.64)(0.47,0.18){1}{\line(1,0){0.47}}
\multiput(49.25,64.45)(0.23,0.1){2}{\line(1,0){0.23}}
\multiput(48.79,64.24)(0.23,0.1){2}{\line(1,0){0.23}}
\multiput(48.34,64.02)(0.23,0.11){2}{\line(1,0){0.23}}
\multiput(47.9,63.78)(0.22,0.12){2}{\line(1,0){0.22}}
\multiput(47.46,63.53)(0.22,0.13){2}{\line(1,0){0.22}}
\multiput(47.03,63.27)(0.21,0.13){2}{\line(1,0){0.21}}
\multiput(46.62,62.99)(0.21,0.14){2}{\line(1,0){0.21}}
\multiput(46.21,62.7)(0.2,0.15){2}{\line(1,0){0.2}}
\multiput(45.81,62.39)(0.13,0.1){3}{\line(1,0){0.13}}
\multiput(45.42,62.07)(0.13,0.11){3}{\line(1,0){0.13}}
\multiput(45.04,61.74)(0.13,0.11){3}{\line(1,0){0.13}}
\multiput(44.67,61.4)(0.12,0.11){3}{\line(1,0){0.12}}
\multiput(44.32,61.04)(0.12,0.12){3}{\line(1,0){0.12}}
\multiput(43.98,60.67)(0.11,0.12){3}{\line(0,1){0.12}}
\multiput(43.64,60.3)(0.11,0.13){3}{\line(0,1){0.13}}
\multiput(43.32,59.91)(0.11,0.13){3}{\line(0,1){0.13}}
\multiput(43.02,59.51)(0.1,0.13){3}{\line(0,1){0.13}}
\multiput(42.73,59.1)(0.15,0.2){2}{\line(0,1){0.2}}
\multiput(42.45,58.68)(0.14,0.21){2}{\line(0,1){0.21}}
\multiput(42.18,58.25)(0.13,0.21){2}{\line(0,1){0.21}}
\multiput(41.93,57.82)(0.13,0.22){2}{\line(0,1){0.22}}
\multiput(41.69,57.38)(0.12,0.22){2}{\line(0,1){0.22}}
\multiput(41.47,56.92)(0.11,0.23){2}{\line(0,1){0.23}}
\multiput(41.26,56.47)(0.1,0.23){2}{\line(0,1){0.23}}
\multiput(41.07,56)(0.1,0.23){2}{\line(0,1){0.23}}
\multiput(40.89,55.53)(0.18,0.47){1}{\line(0,1){0.47}}
\multiput(40.73,55.06)(0.16,0.48){1}{\line(0,1){0.48}}
\multiput(40.59,54.58)(0.15,0.48){1}{\line(0,1){0.48}}
\multiput(40.46,54.09)(0.13,0.49){1}{\line(0,1){0.49}}
\multiput(40.34,53.6)(0.11,0.49){1}{\line(0,1){0.49}}
\multiput(40.24,53.11)(0.1,0.49){1}{\line(0,1){0.49}}
\multiput(40.16,52.61)(0.08,0.5){1}{\line(0,1){0.5}}
\multiput(40.1,52.11)(0.07,0.5){1}{\line(0,1){0.5}}
\multiput(40.05,51.61)(0.05,0.5){1}{\line(0,1){0.5}}
\multiput(40.01,51.11)(0.03,0.5){1}{\line(0,1){0.5}}
\multiput(40,50.61)(0.02,0.5){1}{\line(0,1){0.5}}
\put(40,50.11){\line(0,1){0.5}}
\multiput(40,50.11)(0.02,-0.5){1}{\line(0,-1){0.5}}
\multiput(40.01,49.6)(0.03,-0.5){1}{\line(0,-1){0.5}}
\multiput(40.05,49.1)(0.05,-0.5){1}{\line(0,-1){0.5}}
\multiput(40.1,48.6)(0.07,-0.5){1}{\line(0,-1){0.5}}
\multiput(40.16,48.1)(0.08,-0.5){1}{\line(0,-1){0.5}}
\multiput(40.24,47.61)(0.1,-0.49){1}{\line(0,-1){0.49}}
\multiput(40.34,47.11)(0.11,-0.49){1}{\line(0,-1){0.49}}
\multiput(40.46,46.62)(0.13,-0.49){1}{\line(0,-1){0.49}}
\multiput(40.59,46.14)(0.15,-0.48){1}{\line(0,-1){0.48}}
\multiput(40.73,45.66)(0.16,-0.48){1}{\line(0,-1){0.48}}
\multiput(40.89,45.18)(0.18,-0.47){1}{\line(0,-1){0.47}}
\multiput(41.07,44.71)(0.1,-0.23){2}{\line(0,-1){0.23}}
\multiput(41.26,44.25)(0.1,-0.23){2}{\line(0,-1){0.23}}
\multiput(41.47,43.79)(0.11,-0.23){2}{\line(0,-1){0.23}}
\multiput(41.69,43.34)(0.12,-0.22){2}{\line(0,-1){0.22}}
\multiput(41.93,42.9)(0.13,-0.22){2}{\line(0,-1){0.22}}
\multiput(42.18,42.46)(0.13,-0.21){2}{\line(0,-1){0.21}}
\multiput(42.45,42.03)(0.14,-0.21){2}{\line(0,-1){0.21}}
\multiput(42.73,41.62)(0.15,-0.2){2}{\line(0,-1){0.2}}
\multiput(43.02,41.21)(0.1,-0.13){3}{\line(0,-1){0.13}}
\multiput(43.32,40.81)(0.11,-0.13){3}{\line(0,-1){0.13}}
\multiput(43.64,40.42)(0.11,-0.13){3}{\line(0,-1){0.13}}
\multiput(43.98,40.04)(0.11,-0.12){3}{\line(0,-1){0.12}}
\multiput(44.32,39.67)(0.12,-0.12){3}{\line(1,0){0.12}}
\multiput(44.67,39.32)(0.12,-0.11){3}{\line(1,0){0.12}}
\multiput(45.04,38.98)(0.13,-0.11){3}{\line(1,0){0.13}}
\multiput(45.42,38.64)(0.13,-0.11){3}{\line(1,0){0.13}}
\multiput(45.81,38.32)(0.13,-0.1){3}{\line(1,0){0.13}}
\multiput(46.21,38.02)(0.2,-0.15){2}{\line(1,0){0.2}}
\multiput(46.62,37.73)(0.21,-0.14){2}{\line(1,0){0.21}}
\multiput(47.03,37.45)(0.21,-0.13){2}{\line(1,0){0.21}}
\multiput(47.46,37.18)(0.22,-0.13){2}{\line(1,0){0.22}}
\multiput(47.9,36.93)(0.22,-0.12){2}{\line(1,0){0.22}}
\multiput(48.34,36.69)(0.23,-0.11){2}{\line(1,0){0.23}}
\multiput(48.79,36.47)(0.23,-0.1){2}{\line(1,0){0.23}}
\multiput(49.25,36.26)(0.23,-0.1){2}{\line(1,0){0.23}}
\multiput(49.71,36.07)(0.47,-0.18){1}{\line(1,0){0.47}}
\multiput(50.18,35.89)(0.48,-0.16){1}{\line(1,0){0.48}}
\multiput(50.66,35.73)(0.48,-0.15){1}{\line(1,0){0.48}}
\multiput(51.14,35.59)(0.49,-0.13){1}{\line(1,0){0.49}}
\multiput(51.62,35.46)(0.49,-0.11){1}{\line(1,0){0.49}}
\multiput(52.11,35.34)(0.49,-0.1){1}{\line(1,0){0.49}}
\multiput(52.61,35.24)(0.5,-0.08){1}{\line(1,0){0.5}}
\multiput(53.1,35.16)(0.5,-0.07){1}{\line(1,0){0.5}}
\multiput(53.6,35.1)(0.5,-0.05){1}{\line(1,0){0.5}}
\multiput(54.1,35.05)(0.5,-0.03){1}{\line(1,0){0.5}}
\multiput(54.6,35.01)(0.5,-0.02){1}{\line(1,0){0.5}}
\put(55.11,35){\line(1,0){0.5}}
\multiput(55.61,35)(0.5,0.02){1}{\line(1,0){0.5}}
\multiput(56.11,35.01)(0.5,0.03){1}{\line(1,0){0.5}}
\multiput(56.61,35.05)(0.5,0.05){1}{\line(1,0){0.5}}
\multiput(57.11,35.1)(0.5,0.07){1}{\line(1,0){0.5}}
\multiput(57.61,35.16)(0.5,0.08){1}{\line(1,0){0.5}}
\multiput(58.11,35.24)(0.49,0.1){1}{\line(1,0){0.49}}
\multiput(58.6,35.34)(0.49,0.11){1}{\line(1,0){0.49}}
\multiput(59.09,35.46)(0.49,0.13){1}{\line(1,0){0.49}}
\multiput(59.58,35.59)(0.48,0.15){1}{\line(1,0){0.48}}
\multiput(60.06,35.73)(0.48,0.16){1}{\line(1,0){0.48}}
\multiput(60.53,35.89)(0.47,0.18){1}{\line(1,0){0.47}}
\multiput(61,36.07)(0.23,0.1){2}{\line(1,0){0.23}}
\multiput(61.47,36.26)(0.23,0.1){2}{\line(1,0){0.23}}
\multiput(61.92,36.47)(0.23,0.11){2}{\line(1,0){0.23}}
\multiput(62.38,36.69)(0.22,0.12){2}{\line(1,0){0.22}}
\multiput(62.82,36.93)(0.22,0.13){2}{\line(1,0){0.22}}
\multiput(63.25,37.18)(0.21,0.13){2}{\line(1,0){0.21}}
\multiput(63.68,37.45)(0.21,0.14){2}{\line(1,0){0.21}}
\multiput(64.1,37.73)(0.2,0.15){2}{\line(1,0){0.2}}
\multiput(64.51,38.02)(0.13,0.1){3}{\line(1,0){0.13}}
\multiput(64.91,38.32)(0.13,0.11){3}{\line(1,0){0.13}}
\multiput(65.3,38.64)(0.13,0.11){3}{\line(1,0){0.13}}
\multiput(65.67,38.98)(0.12,0.11){3}{\line(1,0){0.12}}
\multiput(66.04,39.32)(0.12,0.12){3}{\line(1,0){0.12}}
\multiput(66.4,39.67)(0.11,0.12){3}{\line(0,1){0.12}}
\multiput(66.74,40.04)(0.11,0.13){3}{\line(0,1){0.13}}
\multiput(67.07,40.42)(0.11,0.13){3}{\line(0,1){0.13}}
\multiput(67.39,40.81)(0.1,0.13){3}{\line(0,1){0.13}}
\multiput(67.7,41.21)(0.15,0.2){2}{\line(0,1){0.2}}
\multiput(67.99,41.62)(0.14,0.21){2}{\line(0,1){0.21}}
\multiput(68.27,42.03)(0.13,0.21){2}{\line(0,1){0.21}}
\multiput(68.53,42.46)(0.13,0.22){2}{\line(0,1){0.22}}
\multiput(68.78,42.9)(0.12,0.22){2}{\line(0,1){0.22}}
\multiput(69.02,43.34)(0.11,0.23){2}{\line(0,1){0.23}}
\multiput(69.24,43.79)(0.1,0.23){2}{\line(0,1){0.23}}
\multiput(69.45,44.25)(0.1,0.23){2}{\line(0,1){0.23}}
\multiput(69.64,44.71)(0.18,0.47){1}{\line(0,1){0.47}}
\multiput(69.82,45.18)(0.16,0.48){1}{\line(0,1){0.48}}
\multiput(69.98,45.66)(0.15,0.48){1}{\line(0,1){0.48}}
\multiput(70.13,46.14)(0.13,0.49){1}{\line(0,1){0.49}}
\multiput(70.26,46.62)(0.11,0.49){1}{\line(0,1){0.49}}
\multiput(70.37,47.11)(0.1,0.49){1}{\line(0,1){0.49}}
\multiput(70.47,47.61)(0.08,0.5){1}{\line(0,1){0.5}}
\multiput(70.55,48.1)(0.07,0.5){1}{\line(0,1){0.5}}
\multiput(70.62,48.6)(0.05,0.5){1}{\line(0,1){0.5}}
\multiput(70.67,49.1)(0.03,0.5){1}{\line(0,1){0.5}}
\multiput(70.7,49.6)(0.02,0.5){1}{\line(0,1){0.5}}

\linethickness{1mm}
\put(20,50){\line(1,0){20}}
\linethickness{1mm}
\put(71,50){\line(1,0){25}}
\linethickness{0.3mm}
\put(55,15){\line(0,1){20}}
\put(55,50){\makebox(0,0)[cc]{$\Gamma^{(3)}$}}

\put(25,55){\makebox(0,0)[cc]{$p$}}

\put(80,55){\makebox(0,0)[cc]{$-p-k$}}

\put(60,25){\makebox(0,0)[cc]{$k$}}

\put(25,45){\makebox(0,0)[cc]{$\alpha$}}

\put(80,45){\makebox(0,0)[cc]{$\beta$}}

\put(50,25){\makebox(0,0)[cc]{$\ve$}}

\end{picture}

}

\def\figtwo{

\def\JPicScale{0.9}
\ifx\JPicScale\undefined\def\JPicScale{1}\fi
\unitlength \JPicScale mm

}

\def\figthree{

\def\JPicScale{0.8}
\ifx\JPicScale\undefined\def\JPicScale{1}\fi
\unitlength \JPicScale mm
\begin{picture}(110,60)(0,0)
\linethickness{0.3mm}
\put(50,39.75){\line(0,1){0.51}}
\multiput(49.97,40.76)(0.03,-0.51){1}{\line(0,-1){0.51}}
\multiput(49.92,41.26)(0.05,-0.5){1}{\line(0,-1){0.5}}
\multiput(49.84,41.76)(0.08,-0.5){1}{\line(0,-1){0.5}}
\multiput(49.74,42.26)(0.1,-0.5){1}{\line(0,-1){0.5}}
\multiput(49.61,42.75)(0.13,-0.49){1}{\line(0,-1){0.49}}
\multiput(49.46,43.23)(0.15,-0.48){1}{\line(0,-1){0.48}}
\multiput(49.29,43.71)(0.18,-0.48){1}{\line(0,-1){0.48}}
\multiput(49.09,44.18)(0.1,-0.23){2}{\line(0,-1){0.23}}
\multiput(48.86,44.63)(0.11,-0.23){2}{\line(0,-1){0.23}}
\multiput(48.62,45.07)(0.12,-0.22){2}{\line(0,-1){0.22}}
\multiput(48.35,45.5)(0.13,-0.21){2}{\line(0,-1){0.21}}
\multiput(48.06,45.92)(0.14,-0.21){2}{\line(0,-1){0.21}}
\multiput(47.75,46.32)(0.1,-0.13){3}{\line(0,-1){0.13}}
\multiput(47.42,46.7)(0.11,-0.13){3}{\line(0,-1){0.13}}
\multiput(47.07,47.07)(0.12,-0.12){3}{\line(0,-1){0.12}}
\multiput(46.7,47.42)(0.12,-0.12){3}{\line(1,0){0.12}}
\multiput(46.32,47.75)(0.13,-0.11){3}{\line(1,0){0.13}}
\multiput(45.92,48.06)(0.13,-0.1){3}{\line(1,0){0.13}}
\multiput(45.5,48.35)(0.21,-0.14){2}{\line(1,0){0.21}}
\multiput(45.07,48.62)(0.21,-0.13){2}{\line(1,0){0.21}}
\multiput(44.63,48.86)(0.22,-0.12){2}{\line(1,0){0.22}}
\multiput(44.18,49.09)(0.23,-0.11){2}{\line(1,0){0.23}}
\multiput(43.71,49.29)(0.23,-0.1){2}{\line(1,0){0.23}}
\multiput(43.23,49.46)(0.48,-0.18){1}{\line(1,0){0.48}}
\multiput(42.75,49.61)(0.48,-0.15){1}{\line(1,0){0.48}}
\multiput(42.26,49.74)(0.49,-0.13){1}{\line(1,0){0.49}}
\multiput(41.76,49.84)(0.5,-0.1){1}{\line(1,0){0.5}}
\multiput(41.26,49.92)(0.5,-0.08){1}{\line(1,0){0.5}}
\multiput(40.76,49.97)(0.5,-0.05){1}{\line(1,0){0.5}}
\multiput(40.25,50)(0.51,-0.03){1}{\line(1,0){0.51}}
\put(39.75,50){\line(1,0){0.51}}
\multiput(39.24,49.97)(0.51,0.03){1}{\line(1,0){0.51}}
\multiput(38.74,49.92)(0.5,0.05){1}{\line(1,0){0.5}}
\multiput(38.24,49.84)(0.5,0.08){1}{\line(1,0){0.5}}
\multiput(37.74,49.74)(0.5,0.1){1}{\line(1,0){0.5}}
\multiput(37.25,49.61)(0.49,0.13){1}{\line(1,0){0.49}}
\multiput(36.77,49.46)(0.48,0.15){1}{\line(1,0){0.48}}
\multiput(36.29,49.29)(0.48,0.18){1}{\line(1,0){0.48}}
\multiput(35.82,49.09)(0.23,0.1){2}{\line(1,0){0.23}}
\multiput(35.37,48.86)(0.23,0.11){2}{\line(1,0){0.23}}
\multiput(34.93,48.62)(0.22,0.12){2}{\line(1,0){0.22}}
\multiput(34.5,48.35)(0.21,0.13){2}{\line(1,0){0.21}}
\multiput(34.08,48.06)(0.21,0.14){2}{\line(1,0){0.21}}
\multiput(33.68,47.75)(0.13,0.1){3}{\line(1,0){0.13}}
\multiput(33.3,47.42)(0.13,0.11){3}{\line(1,0){0.13}}
\multiput(32.93,47.07)(0.12,0.12){3}{\line(1,0){0.12}}
\multiput(32.58,46.7)(0.12,0.12){3}{\line(0,1){0.12}}
\multiput(32.25,46.32)(0.11,0.13){3}{\line(0,1){0.13}}
\multiput(31.94,45.92)(0.1,0.13){3}{\line(0,1){0.13}}
\multiput(31.65,45.5)(0.14,0.21){2}{\line(0,1){0.21}}
\multiput(31.38,45.07)(0.13,0.21){2}{\line(0,1){0.21}}
\multiput(31.14,44.63)(0.12,0.22){2}{\line(0,1){0.22}}
\multiput(30.91,44.18)(0.11,0.23){2}{\line(0,1){0.23}}
\multiput(30.71,43.71)(0.1,0.23){2}{\line(0,1){0.23}}
\multiput(30.54,43.23)(0.18,0.48){1}{\line(0,1){0.48}}
\multiput(30.39,42.75)(0.15,0.48){1}{\line(0,1){0.48}}
\multiput(30.26,42.26)(0.13,0.49){1}{\line(0,1){0.49}}
\multiput(30.16,41.76)(0.1,0.5){1}{\line(0,1){0.5}}
\multiput(30.08,41.26)(0.08,0.5){1}{\line(0,1){0.5}}
\multiput(30.03,40.76)(0.05,0.5){1}{\line(0,1){0.5}}
\multiput(30,40.25)(0.03,0.51){1}{\line(0,1){0.51}}
\put(30,39.75){\line(0,1){0.51}}
\multiput(30,39.75)(0.03,-0.51){1}{\line(0,-1){0.51}}
\multiput(30.03,39.24)(0.05,-0.5){1}{\line(0,-1){0.5}}
\multiput(30.08,38.74)(0.08,-0.5){1}{\line(0,-1){0.5}}
\multiput(30.16,38.24)(0.1,-0.5){1}{\line(0,-1){0.5}}
\multiput(30.26,37.74)(0.13,-0.49){1}{\line(0,-1){0.49}}
\multiput(30.39,37.25)(0.15,-0.48){1}{\line(0,-1){0.48}}
\multiput(30.54,36.77)(0.18,-0.48){1}{\line(0,-1){0.48}}
\multiput(30.71,36.29)(0.1,-0.23){2}{\line(0,-1){0.23}}
\multiput(30.91,35.82)(0.11,-0.23){2}{\line(0,-1){0.23}}
\multiput(31.14,35.37)(0.12,-0.22){2}{\line(0,-1){0.22}}
\multiput(31.38,34.93)(0.13,-0.21){2}{\line(0,-1){0.21}}
\multiput(31.65,34.5)(0.14,-0.21){2}{\line(0,-1){0.21}}
\multiput(31.94,34.08)(0.1,-0.13){3}{\line(0,-1){0.13}}
\multiput(32.25,33.68)(0.11,-0.13){3}{\line(0,-1){0.13}}
\multiput(32.58,33.3)(0.12,-0.12){3}{\line(0,-1){0.12}}
\multiput(32.93,32.93)(0.12,-0.12){3}{\line(1,0){0.12}}
\multiput(33.3,32.58)(0.13,-0.11){3}{\line(1,0){0.13}}
\multiput(33.68,32.25)(0.13,-0.1){3}{\line(1,0){0.13}}
\multiput(34.08,31.94)(0.21,-0.14){2}{\line(1,0){0.21}}
\multiput(34.5,31.65)(0.21,-0.13){2}{\line(1,0){0.21}}
\multiput(34.93,31.38)(0.22,-0.12){2}{\line(1,0){0.22}}
\multiput(35.37,31.14)(0.23,-0.11){2}{\line(1,0){0.23}}
\multiput(35.82,30.91)(0.23,-0.1){2}{\line(1,0){0.23}}
\multiput(36.29,30.71)(0.48,-0.18){1}{\line(1,0){0.48}}
\multiput(36.77,30.54)(0.48,-0.15){1}{\line(1,0){0.48}}
\multiput(37.25,30.39)(0.49,-0.13){1}{\line(1,0){0.49}}
\multiput(37.74,30.26)(0.5,-0.1){1}{\line(1,0){0.5}}
\multiput(38.24,30.16)(0.5,-0.08){1}{\line(1,0){0.5}}
\multiput(38.74,30.08)(0.5,-0.05){1}{\line(1,0){0.5}}
\multiput(39.24,30.03)(0.51,-0.03){1}{\line(1,0){0.51}}
\put(39.75,30){\line(1,0){0.51}}
\multiput(40.25,30)(0.51,0.03){1}{\line(1,0){0.51}}
\multiput(40.76,30.03)(0.5,0.05){1}{\line(1,0){0.5}}
\multiput(41.26,30.08)(0.5,0.08){1}{\line(1,0){0.5}}
\multiput(41.76,30.16)(0.5,0.1){1}{\line(1,0){0.5}}
\multiput(42.26,30.26)(0.49,0.13){1}{\line(1,0){0.49}}
\multiput(42.75,30.39)(0.48,0.15){1}{\line(1,0){0.48}}
\multiput(43.23,30.54)(0.48,0.18){1}{\line(1,0){0.48}}
\multiput(43.71,30.71)(0.23,0.1){2}{\line(1,0){0.23}}
\multiput(44.18,30.91)(0.23,0.11){2}{\line(1,0){0.23}}
\multiput(44.63,31.14)(0.22,0.12){2}{\line(1,0){0.22}}
\multiput(45.07,31.38)(0.21,0.13){2}{\line(1,0){0.21}}
\multiput(45.5,31.65)(0.21,0.14){2}{\line(1,0){0.21}}
\multiput(45.92,31.94)(0.13,0.1){3}{\line(1,0){0.13}}
\multiput(46.32,32.25)(0.13,0.11){3}{\line(1,0){0.13}}
\multiput(46.7,32.58)(0.12,0.12){3}{\line(1,0){0.12}}
\multiput(47.07,32.93)(0.12,0.12){3}{\line(0,1){0.12}}
\multiput(47.42,33.3)(0.11,0.13){3}{\line(0,1){0.13}}
\multiput(47.75,33.68)(0.1,0.13){3}{\line(0,1){0.13}}
\multiput(48.06,34.08)(0.14,0.21){2}{\line(0,1){0.21}}
\multiput(48.35,34.5)(0.13,0.21){2}{\line(0,1){0.21}}
\multiput(48.62,34.93)(0.12,0.22){2}{\line(0,1){0.22}}
\multiput(48.86,35.37)(0.11,0.23){2}{\line(0,1){0.23}}
\multiput(49.09,35.82)(0.1,0.23){2}{\line(0,1){0.23}}
\multiput(49.29,36.29)(0.18,0.48){1}{\line(0,1){0.48}}
\multiput(49.46,36.77)(0.15,0.48){1}{\line(0,1){0.48}}
\multiput(49.61,37.25)(0.13,0.49){1}{\line(0,1){0.49}}
\multiput(49.74,37.74)(0.1,0.5){1}{\line(0,1){0.5}}
\multiput(49.84,38.24)(0.08,0.5){1}{\line(0,1){0.5}}
\multiput(49.92,38.74)(0.05,0.5){1}{\line(0,1){0.5}}
\multiput(49.97,39.24)(0.03,0.51){1}{\line(0,1){0.51}}

\linethickness{0.3mm}
\put(100,39.75){\line(0,1){0.51}}
\multiput(99.97,40.76)(0.03,-0.51){1}{\line(0,-1){0.51}}
\multiput(99.92,41.26)(0.05,-0.5){1}{\line(0,-1){0.5}}
\multiput(99.84,41.76)(0.08,-0.5){1}{\line(0,-1){0.5}}
\multiput(99.74,42.26)(0.1,-0.5){1}{\line(0,-1){0.5}}
\multiput(99.61,42.75)(0.13,-0.49){1}{\line(0,-1){0.49}}
\multiput(99.46,43.23)(0.15,-0.48){1}{\line(0,-1){0.48}}
\multiput(99.29,43.71)(0.18,-0.48){1}{\line(0,-1){0.48}}
\multiput(99.09,44.18)(0.1,-0.23){2}{\line(0,-1){0.23}}
\multiput(98.86,44.63)(0.11,-0.23){2}{\line(0,-1){0.23}}
\multiput(98.62,45.07)(0.12,-0.22){2}{\line(0,-1){0.22}}
\multiput(98.35,45.5)(0.13,-0.21){2}{\line(0,-1){0.21}}
\multiput(98.06,45.92)(0.14,-0.21){2}{\line(0,-1){0.21}}
\multiput(97.75,46.32)(0.1,-0.13){3}{\line(0,-1){0.13}}
\multiput(97.42,46.7)(0.11,-0.13){3}{\line(0,-1){0.13}}
\multiput(97.07,47.07)(0.12,-0.12){3}{\line(0,-1){0.12}}
\multiput(96.7,47.42)(0.12,-0.12){3}{\line(1,0){0.12}}
\multiput(96.32,47.75)(0.13,-0.11){3}{\line(1,0){0.13}}
\multiput(95.92,48.06)(0.13,-0.1){3}{\line(1,0){0.13}}
\multiput(95.5,48.35)(0.21,-0.14){2}{\line(1,0){0.21}}
\multiput(95.07,48.62)(0.21,-0.13){2}{\line(1,0){0.21}}
\multiput(94.63,48.86)(0.22,-0.12){2}{\line(1,0){0.22}}
\multiput(94.18,49.09)(0.23,-0.11){2}{\line(1,0){0.23}}
\multiput(93.71,49.29)(0.23,-0.1){2}{\line(1,0){0.23}}
\multiput(93.23,49.46)(0.48,-0.18){1}{\line(1,0){0.48}}
\multiput(92.75,49.61)(0.48,-0.15){1}{\line(1,0){0.48}}
\multiput(92.26,49.74)(0.49,-0.13){1}{\line(1,0){0.49}}
\multiput(91.76,49.84)(0.5,-0.1){1}{\line(1,0){0.5}}
\multiput(91.26,49.92)(0.5,-0.08){1}{\line(1,0){0.5}}
\multiput(90.76,49.97)(0.5,-0.05){1}{\line(1,0){0.5}}
\multiput(90.25,50)(0.51,-0.03){1}{\line(1,0){0.51}}
\put(89.75,50){\line(1,0){0.51}}
\multiput(89.24,49.97)(0.51,0.03){1}{\line(1,0){0.51}}
\multiput(88.74,49.92)(0.5,0.05){1}{\line(1,0){0.5}}
\multiput(88.24,49.84)(0.5,0.08){1}{\line(1,0){0.5}}
\multiput(87.74,49.74)(0.5,0.1){1}{\line(1,0){0.5}}
\multiput(87.25,49.61)(0.49,0.13){1}{\line(1,0){0.49}}
\multiput(86.77,49.46)(0.48,0.15){1}{\line(1,0){0.48}}
\multiput(86.29,49.29)(0.48,0.18){1}{\line(1,0){0.48}}
\multiput(85.82,49.09)(0.23,0.1){2}{\line(1,0){0.23}}
\multiput(85.37,48.86)(0.23,0.11){2}{\line(1,0){0.23}}
\multiput(84.93,48.62)(0.22,0.12){2}{\line(1,0){0.22}}
\multiput(84.5,48.35)(0.21,0.13){2}{\line(1,0){0.21}}
\multiput(84.08,48.06)(0.21,0.14){2}{\line(1,0){0.21}}
\multiput(83.68,47.75)(0.13,0.1){3}{\line(1,0){0.13}}
\multiput(83.3,47.42)(0.13,0.11){3}{\line(1,0){0.13}}
\multiput(82.93,47.07)(0.12,0.12){3}{\line(1,0){0.12}}
\multiput(82.58,46.7)(0.12,0.12){3}{\line(0,1){0.12}}
\multiput(82.25,46.32)(0.11,0.13){3}{\line(0,1){0.13}}
\multiput(81.94,45.92)(0.1,0.13){3}{\line(0,1){0.13}}
\multiput(81.65,45.5)(0.14,0.21){2}{\line(0,1){0.21}}
\multiput(81.38,45.07)(0.13,0.21){2}{\line(0,1){0.21}}
\multiput(81.14,44.63)(0.12,0.22){2}{\line(0,1){0.22}}
\multiput(80.91,44.18)(0.11,0.23){2}{\line(0,1){0.23}}
\multiput(80.71,43.71)(0.1,0.23){2}{\line(0,1){0.23}}
\multiput(80.54,43.23)(0.18,0.48){1}{\line(0,1){0.48}}
\multiput(80.39,42.75)(0.15,0.48){1}{\line(0,1){0.48}}
\multiput(80.26,42.26)(0.13,0.49){1}{\line(0,1){0.49}}
\multiput(80.16,41.76)(0.1,0.5){1}{\line(0,1){0.5}}
\multiput(80.08,41.26)(0.08,0.5){1}{\line(0,1){0.5}}
\multiput(80.03,40.76)(0.05,0.5){1}{\line(0,1){0.5}}
\multiput(80,40.25)(0.03,0.51){1}{\line(0,1){0.51}}
\put(80,39.75){\line(0,1){0.51}}
\multiput(80,39.75)(0.03,-0.51){1}{\line(0,-1){0.51}}
\multiput(80.03,39.24)(0.05,-0.5){1}{\line(0,-1){0.5}}
\multiput(80.08,38.74)(0.08,-0.5){1}{\line(0,-1){0.5}}
\multiput(80.16,38.24)(0.1,-0.5){1}{\line(0,-1){0.5}}
\multiput(80.26,37.74)(0.13,-0.49){1}{\line(0,-1){0.49}}
\multiput(80.39,37.25)(0.15,-0.48){1}{\line(0,-1){0.48}}
\multiput(80.54,36.77)(0.18,-0.48){1}{\line(0,-1){0.48}}
\multiput(80.71,36.29)(0.1,-0.23){2}{\line(0,-1){0.23}}
\multiput(80.91,35.82)(0.11,-0.23){2}{\line(0,-1){0.23}}
\multiput(81.14,35.37)(0.12,-0.22){2}{\line(0,-1){0.22}}
\multiput(81.38,34.93)(0.13,-0.21){2}{\line(0,-1){0.21}}
\multiput(81.65,34.5)(0.14,-0.21){2}{\line(0,-1){0.21}}
\multiput(81.94,34.08)(0.1,-0.13){3}{\line(0,-1){0.13}}
\multiput(82.25,33.68)(0.11,-0.13){3}{\line(0,-1){0.13}}
\multiput(82.58,33.3)(0.12,-0.12){3}{\line(0,-1){0.12}}
\multiput(82.93,32.93)(0.12,-0.12){3}{\line(1,0){0.12}}
\multiput(83.3,32.58)(0.13,-0.11){3}{\line(1,0){0.13}}
\multiput(83.68,32.25)(0.13,-0.1){3}{\line(1,0){0.13}}
\multiput(84.08,31.94)(0.21,-0.14){2}{\line(1,0){0.21}}
\multiput(84.5,31.65)(0.21,-0.13){2}{\line(1,0){0.21}}
\multiput(84.93,31.38)(0.22,-0.12){2}{\line(1,0){0.22}}
\multiput(85.37,31.14)(0.23,-0.11){2}{\line(1,0){0.23}}
\multiput(85.82,30.91)(0.23,-0.1){2}{\line(1,0){0.23}}
\multiput(86.29,30.71)(0.48,-0.18){1}{\line(1,0){0.48}}
\multiput(86.77,30.54)(0.48,-0.15){1}{\line(1,0){0.48}}
\multiput(87.25,30.39)(0.49,-0.13){1}{\line(1,0){0.49}}
\multiput(87.74,30.26)(0.5,-0.1){1}{\line(1,0){0.5}}
\multiput(88.24,30.16)(0.5,-0.08){1}{\line(1,0){0.5}}
\multiput(88.74,30.08)(0.5,-0.05){1}{\line(1,0){0.5}}
\multiput(89.24,30.03)(0.51,-0.03){1}{\line(1,0){0.51}}
\put(89.75,30){\line(1,0){0.51}}
\multiput(90.25,30)(0.51,0.03){1}{\line(1,0){0.51}}
\multiput(90.76,30.03)(0.5,0.05){1}{\line(1,0){0.5}}
\multiput(91.26,30.08)(0.5,0.08){1}{\line(1,0){0.5}}
\multiput(91.76,30.16)(0.5,0.1){1}{\line(1,0){0.5}}
\multiput(92.26,30.26)(0.49,0.13){1}{\line(1,0){0.49}}
\multiput(92.75,30.39)(0.48,0.15){1}{\line(1,0){0.48}}
\multiput(93.23,30.54)(0.48,0.18){1}{\line(1,0){0.48}}
\multiput(93.71,30.71)(0.23,0.1){2}{\line(1,0){0.23}}
\multiput(94.18,30.91)(0.23,0.11){2}{\line(1,0){0.23}}
\multiput(94.63,31.14)(0.22,0.12){2}{\line(1,0){0.22}}
\multiput(95.07,31.38)(0.21,0.13){2}{\line(1,0){0.21}}
\multiput(95.5,31.65)(0.21,0.14){2}{\line(1,0){0.21}}
\multiput(95.92,31.94)(0.13,0.1){3}{\line(1,0){0.13}}
\multiput(96.32,32.25)(0.13,0.11){3}{\line(1,0){0.13}}
\multiput(96.7,32.58)(0.12,0.12){3}{\line(1,0){0.12}}
\multiput(97.07,32.93)(0.12,0.12){3}{\line(0,1){0.12}}
\multiput(97.42,33.3)(0.11,0.13){3}{\line(0,1){0.13}}
\multiput(97.75,33.68)(0.1,0.13){3}{\line(0,1){0.13}}
\multiput(98.06,34.08)(0.14,0.21){2}{\line(0,1){0.21}}
\multiput(98.35,34.5)(0.13,0.21){2}{\line(0,1){0.21}}
\multiput(98.62,34.93)(0.12,0.22){2}{\line(0,1){0.22}}
\multiput(98.86,35.37)(0.11,0.23){2}{\line(0,1){0.23}}
\multiput(99.09,35.82)(0.1,0.23){2}{\line(0,1){0.23}}
\multiput(99.29,36.29)(0.18,0.48){1}{\line(0,1){0.48}}
\multiput(99.46,36.77)(0.15,0.48){1}{\line(0,1){0.48}}
\multiput(99.61,37.25)(0.13,0.49){1}{\line(0,1){0.49}}
\multiput(99.74,37.74)(0.1,0.5){1}{\line(0,1){0.5}}
\multiput(99.84,38.24)(0.08,0.5){1}{\line(0,1){0.5}}
\multiput(99.92,38.74)(0.05,0.5){1}{\line(0,1){0.5}}
\multiput(99.97,39.24)(0.03,0.51){1}{\line(0,1){0.51}}

\linethickness{1mm}
\put(10,40){\line(1,0){20}}
\linethickness{1mm}
\put(50,40){\line(1,0){30}}
\linethickness{0.3mm}
\put(40,10){\line(0,1){20}}
\linethickness{1mm}
\multiput(90,50)(0.24,0.12){83}{\line(1,0){0.24}}
\linethickness{1mm}
\multiput(90,30)(0.24,-0.12){83}{\line(1,0){0.24}}
\put(105,50){\makebox(0,0)[cc]{$\cdot$}}

\put(105,30){\makebox(0,0)[cc]{$\cdot$}}

\put(110,40){\makebox(0,0)[cc]{$\cdot$}}

\put(20,45){\makebox(0,0)[cc]{$p_i$}}

\put(40,40){\makebox(0,0)[cc]{$\Gamma^{(3)}$}}

\put(90,40){\makebox(0,0)[cc]{$\wt\Gamma$}}

\put(20,35){\makebox(0,0)[cc]{$\eps_i$}}

\put(45,20){\makebox(0,0)[cc]{$k_1$}}

\put(35,20){\makebox(0,0)[cc]{$\ve_1$}}

\put(65,45){\makebox(0,0)[cc]{$-p_i-k_1$}}

\linethickness{0.3mm}
\put(85,10){\line(0,1){22}}
\put(90,20){\makebox(0,0)[cc]{$k_2$}}

\put(80,20){\makebox(0,0)[cc]{$\ve_2$}}

\end{picture}

}

\def\figfour{

\def\JPicScale{0.8}
\ifx\JPicScale\undefined\def\JPicScale{1}\fi
\unitlength \JPicScale mm
\begin{picture}(80,60)(0,0)
\linethickness{0.3mm}
\put(60,39.75){\line(0,1){0.51}}
\multiput(59.97,40.76)(0.03,-0.51){1}{\line(0,-1){0.51}}
\multiput(59.92,41.26)(0.05,-0.5){1}{\line(0,-1){0.5}}
\multiput(59.84,41.76)(0.08,-0.5){1}{\line(0,-1){0.5}}
\multiput(59.74,42.26)(0.1,-0.5){1}{\line(0,-1){0.5}}
\multiput(59.61,42.75)(0.13,-0.49){1}{\line(0,-1){0.49}}
\multiput(59.46,43.23)(0.15,-0.48){1}{\line(0,-1){0.48}}
\multiput(59.29,43.71)(0.18,-0.48){1}{\line(0,-1){0.48}}
\multiput(59.09,44.18)(0.1,-0.23){2}{\line(0,-1){0.23}}
\multiput(58.86,44.63)(0.11,-0.23){2}{\line(0,-1){0.23}}
\multiput(58.62,45.07)(0.12,-0.22){2}{\line(0,-1){0.22}}
\multiput(58.35,45.5)(0.13,-0.21){2}{\line(0,-1){0.21}}
\multiput(58.06,45.92)(0.14,-0.21){2}{\line(0,-1){0.21}}
\multiput(57.75,46.32)(0.1,-0.13){3}{\line(0,-1){0.13}}
\multiput(57.42,46.7)(0.11,-0.13){3}{\line(0,-1){0.13}}
\multiput(57.07,47.07)(0.12,-0.12){3}{\line(0,-1){0.12}}
\multiput(56.7,47.42)(0.12,-0.12){3}{\line(1,0){0.12}}
\multiput(56.32,47.75)(0.13,-0.11){3}{\line(1,0){0.13}}
\multiput(55.92,48.06)(0.13,-0.1){3}{\line(1,0){0.13}}
\multiput(55.5,48.35)(0.21,-0.14){2}{\line(1,0){0.21}}
\multiput(55.07,48.62)(0.21,-0.13){2}{\line(1,0){0.21}}
\multiput(54.63,48.86)(0.22,-0.12){2}{\line(1,0){0.22}}
\multiput(54.18,49.09)(0.23,-0.11){2}{\line(1,0){0.23}}
\multiput(53.71,49.29)(0.23,-0.1){2}{\line(1,0){0.23}}
\multiput(53.23,49.46)(0.48,-0.18){1}{\line(1,0){0.48}}
\multiput(52.75,49.61)(0.48,-0.15){1}{\line(1,0){0.48}}
\multiput(52.26,49.74)(0.49,-0.13){1}{\line(1,0){0.49}}
\multiput(51.76,49.84)(0.5,-0.1){1}{\line(1,0){0.5}}
\multiput(51.26,49.92)(0.5,-0.08){1}{\line(1,0){0.5}}
\multiput(50.76,49.97)(0.5,-0.05){1}{\line(1,0){0.5}}
\multiput(50.25,50)(0.51,-0.03){1}{\line(1,0){0.51}}
\put(49.75,50){\line(1,0){0.51}}
\multiput(49.24,49.97)(0.51,0.03){1}{\line(1,0){0.51}}
\multiput(48.74,49.92)(0.5,0.05){1}{\line(1,0){0.5}}
\multiput(48.24,49.84)(0.5,0.08){1}{\line(1,0){0.5}}
\multiput(47.74,49.74)(0.5,0.1){1}{\line(1,0){0.5}}
\multiput(47.25,49.61)(0.49,0.13){1}{\line(1,0){0.49}}
\multiput(46.77,49.46)(0.48,0.15){1}{\line(1,0){0.48}}
\multiput(46.29,49.29)(0.48,0.18){1}{\line(1,0){0.48}}
\multiput(45.82,49.09)(0.23,0.1){2}{\line(1,0){0.23}}
\multiput(45.37,48.86)(0.23,0.11){2}{\line(1,0){0.23}}
\multiput(44.93,48.62)(0.22,0.12){2}{\line(1,0){0.22}}
\multiput(44.5,48.35)(0.21,0.13){2}{\line(1,0){0.21}}
\multiput(44.08,48.06)(0.21,0.14){2}{\line(1,0){0.21}}
\multiput(43.68,47.75)(0.13,0.1){3}{\line(1,0){0.13}}
\multiput(43.3,47.42)(0.13,0.11){3}{\line(1,0){0.13}}
\multiput(42.93,47.07)(0.12,0.12){3}{\line(1,0){0.12}}
\multiput(42.58,46.7)(0.12,0.12){3}{\line(0,1){0.12}}
\multiput(42.25,46.32)(0.11,0.13){3}{\line(0,1){0.13}}
\multiput(41.94,45.92)(0.1,0.13){3}{\line(0,1){0.13}}
\multiput(41.65,45.5)(0.14,0.21){2}{\line(0,1){0.21}}
\multiput(41.38,45.07)(0.13,0.21){2}{\line(0,1){0.21}}
\multiput(41.14,44.63)(0.12,0.22){2}{\line(0,1){0.22}}
\multiput(40.91,44.18)(0.11,0.23){2}{\line(0,1){0.23}}
\multiput(40.71,43.71)(0.1,0.23){2}{\line(0,1){0.23}}
\multiput(40.54,43.23)(0.18,0.48){1}{\line(0,1){0.48}}
\multiput(40.39,42.75)(0.15,0.48){1}{\line(0,1){0.48}}
\multiput(40.26,42.26)(0.13,0.49){1}{\line(0,1){0.49}}
\multiput(40.16,41.76)(0.1,0.5){1}{\line(0,1){0.5}}
\multiput(40.08,41.26)(0.08,0.5){1}{\line(0,1){0.5}}
\multiput(40.03,40.76)(0.05,0.5){1}{\line(0,1){0.5}}
\multiput(40,40.25)(0.03,0.51){1}{\line(0,1){0.51}}
\put(40,39.75){\line(0,1){0.51}}
\multiput(40,39.75)(0.03,-0.51){1}{\line(0,-1){0.51}}
\multiput(40.03,39.24)(0.05,-0.5){1}{\line(0,-1){0.5}}
\multiput(40.08,38.74)(0.08,-0.5){1}{\line(0,-1){0.5}}
\multiput(40.16,38.24)(0.1,-0.5){1}{\line(0,-1){0.5}}
\multiput(40.26,37.74)(0.13,-0.49){1}{\line(0,-1){0.49}}
\multiput(40.39,37.25)(0.15,-0.48){1}{\line(0,-1){0.48}}
\multiput(40.54,36.77)(0.18,-0.48){1}{\line(0,-1){0.48}}
\multiput(40.71,36.29)(0.1,-0.23){2}{\line(0,-1){0.23}}
\multiput(40.91,35.82)(0.11,-0.23){2}{\line(0,-1){0.23}}
\multiput(41.14,35.37)(0.12,-0.22){2}{\line(0,-1){0.22}}
\multiput(41.38,34.93)(0.13,-0.21){2}{\line(0,-1){0.21}}
\multiput(41.65,34.5)(0.14,-0.21){2}{\line(0,-1){0.21}}
\multiput(41.94,34.08)(0.1,-0.13){3}{\line(0,-1){0.13}}
\multiput(42.25,33.68)(0.11,-0.13){3}{\line(0,-1){0.13}}
\multiput(42.58,33.3)(0.12,-0.12){3}{\line(0,-1){0.12}}
\multiput(42.93,32.93)(0.12,-0.12){3}{\line(1,0){0.12}}
\multiput(43.3,32.58)(0.13,-0.11){3}{\line(1,0){0.13}}
\multiput(43.68,32.25)(0.13,-0.1){3}{\line(1,0){0.13}}
\multiput(44.08,31.94)(0.21,-0.14){2}{\line(1,0){0.21}}
\multiput(44.5,31.65)(0.21,-0.13){2}{\line(1,0){0.21}}
\multiput(44.93,31.38)(0.22,-0.12){2}{\line(1,0){0.22}}
\multiput(45.37,31.14)(0.23,-0.11){2}{\line(1,0){0.23}}
\multiput(45.82,30.91)(0.23,-0.1){2}{\line(1,0){0.23}}
\multiput(46.29,30.71)(0.48,-0.18){1}{\line(1,0){0.48}}
\multiput(46.77,30.54)(0.48,-0.15){1}{\line(1,0){0.48}}
\multiput(47.25,30.39)(0.49,-0.13){1}{\line(1,0){0.49}}
\multiput(47.74,30.26)(0.5,-0.1){1}{\line(1,0){0.5}}
\multiput(48.24,30.16)(0.5,-0.08){1}{\line(1,0){0.5}}
\multiput(48.74,30.08)(0.5,-0.05){1}{\line(1,0){0.5}}
\multiput(49.24,30.03)(0.51,-0.03){1}{\line(1,0){0.51}}
\put(49.75,30){\line(1,0){0.51}}
\multiput(50.25,30)(0.51,0.03){1}{\line(1,0){0.51}}
\multiput(50.76,30.03)(0.5,0.05){1}{\line(1,0){0.5}}
\multiput(51.26,30.08)(0.5,0.08){1}{\line(1,0){0.5}}
\multiput(51.76,30.16)(0.5,0.1){1}{\line(1,0){0.5}}
\multiput(52.26,30.26)(0.49,0.13){1}{\line(1,0){0.49}}
\multiput(52.75,30.39)(0.48,0.15){1}{\line(1,0){0.48}}
\multiput(53.23,30.54)(0.48,0.18){1}{\line(1,0){0.48}}
\multiput(53.71,30.71)(0.23,0.1){2}{\line(1,0){0.23}}
\multiput(54.18,30.91)(0.23,0.11){2}{\line(1,0){0.23}}
\multiput(54.63,31.14)(0.22,0.12){2}{\line(1,0){0.22}}
\multiput(55.07,31.38)(0.21,0.13){2}{\line(1,0){0.21}}
\multiput(55.5,31.65)(0.21,0.14){2}{\line(1,0){0.21}}
\multiput(55.92,31.94)(0.13,0.1){3}{\line(1,0){0.13}}
\multiput(56.32,32.25)(0.13,0.11){3}{\line(1,0){0.13}}
\multiput(56.7,32.58)(0.12,0.12){3}{\line(1,0){0.12}}
\multiput(57.07,32.93)(0.12,0.12){3}{\line(0,1){0.12}}
\multiput(57.42,33.3)(0.11,0.13){3}{\line(0,1){0.13}}
\multiput(57.75,33.68)(0.1,0.13){3}{\line(0,1){0.13}}
\multiput(58.06,34.08)(0.14,0.21){2}{\line(0,1){0.21}}
\multiput(58.35,34.5)(0.13,0.21){2}{\line(0,1){0.21}}
\multiput(58.62,34.93)(0.12,0.22){2}{\line(0,1){0.22}}
\multiput(58.86,35.37)(0.11,0.23){2}{\line(0,1){0.23}}
\multiput(59.09,35.82)(0.1,0.23){2}{\line(0,1){0.23}}
\multiput(59.29,36.29)(0.18,0.48){1}{\line(0,1){0.48}}
\multiput(59.46,36.77)(0.15,0.48){1}{\line(0,1){0.48}}
\multiput(59.61,37.25)(0.13,0.49){1}{\line(0,1){0.49}}
\multiput(59.74,37.74)(0.1,0.5){1}{\line(0,1){0.5}}
\multiput(59.84,38.24)(0.08,0.5){1}{\line(0,1){0.5}}
\multiput(59.92,38.74)(0.05,0.5){1}{\line(0,1){0.5}}
\multiput(59.97,39.24)(0.03,0.51){1}{\line(0,1){0.51}}

\linethickness{0.3mm}
\put(20,40){\line(1,0){20}}
\linethickness{1mm}
\multiput(50,50)(0.36,0.12){83}{\line(1,0){0.36}}
\linethickness{1mm}
\multiput(50,30)(0.36,-0.12){83}{\line(1,0){0.36}}
\put(25,45){\makebox(0,0)[cc]{$k$}}

\put(25,35){\makebox(0,0)[cc]{$\ve$}}

\put(50,40){\makebox(0,0)[cc]{$\wt\Gamma$}}

\put(65,50){\makebox(0,0)[cc]{$\cdot$}}

\put(65,40){\makebox(0,0)[cc]{$\cdot$}}

\put(65,30){\makebox(0,0)[cc]{$\cdot$}}

\put(70,60){\makebox(0,0)[cc]{$\alpha_1$}}

\put(70,52){\makebox(0,0)[cc]{$q_1$}}

\put(70,27){\makebox(0,0)[cc]{$\alpha_N$}}

\put(70,19){\makebox(0,0)[cc]{$q_N$}}

\end{picture}

}

\def\figfive{

\def\JPicScale{0.8}
\ifx\JPicScale\undefined\def\JPicScale{1}\fi
\unitlength \JPicScale mm
\begin{picture}(120,60)(0,0)
\linethickness{0.3mm}
\put(50,39.75){\line(0,1){0.51}}
\multiput(49.97,40.76)(0.03,-0.51){1}{\line(0,-1){0.51}}
\multiput(49.92,41.26)(0.05,-0.5){1}{\line(0,-1){0.5}}
\multiput(49.84,41.76)(0.08,-0.5){1}{\line(0,-1){0.5}}
\multiput(49.74,42.26)(0.1,-0.5){1}{\line(0,-1){0.5}}
\multiput(49.61,42.75)(0.13,-0.49){1}{\line(0,-1){0.49}}
\multiput(49.46,43.23)(0.15,-0.48){1}{\line(0,-1){0.48}}
\multiput(49.29,43.71)(0.18,-0.48){1}{\line(0,-1){0.48}}
\multiput(49.09,44.18)(0.1,-0.23){2}{\line(0,-1){0.23}}
\multiput(48.86,44.63)(0.11,-0.23){2}{\line(0,-1){0.23}}
\multiput(48.62,45.07)(0.12,-0.22){2}{\line(0,-1){0.22}}
\multiput(48.35,45.5)(0.13,-0.21){2}{\line(0,-1){0.21}}
\multiput(48.06,45.92)(0.14,-0.21){2}{\line(0,-1){0.21}}
\multiput(47.75,46.32)(0.1,-0.13){3}{\line(0,-1){0.13}}
\multiput(47.42,46.7)(0.11,-0.13){3}{\line(0,-1){0.13}}
\multiput(47.07,47.07)(0.12,-0.12){3}{\line(0,-1){0.12}}
\multiput(46.7,47.42)(0.12,-0.12){3}{\line(1,0){0.12}}
\multiput(46.32,47.75)(0.13,-0.11){3}{\line(1,0){0.13}}
\multiput(45.92,48.06)(0.13,-0.1){3}{\line(1,0){0.13}}
\multiput(45.5,48.35)(0.21,-0.14){2}{\line(1,0){0.21}}
\multiput(45.07,48.62)(0.21,-0.13){2}{\line(1,0){0.21}}
\multiput(44.63,48.86)(0.22,-0.12){2}{\line(1,0){0.22}}
\multiput(44.18,49.09)(0.23,-0.11){2}{\line(1,0){0.23}}
\multiput(43.71,49.29)(0.23,-0.1){2}{\line(1,0){0.23}}
\multiput(43.23,49.46)(0.48,-0.18){1}{\line(1,0){0.48}}
\multiput(42.75,49.61)(0.48,-0.15){1}{\line(1,0){0.48}}
\multiput(42.26,49.74)(0.49,-0.13){1}{\line(1,0){0.49}}
\multiput(41.76,49.84)(0.5,-0.1){1}{\line(1,0){0.5}}
\multiput(41.26,49.92)(0.5,-0.08){1}{\line(1,0){0.5}}
\multiput(40.76,49.97)(0.5,-0.05){1}{\line(1,0){0.5}}
\multiput(40.25,50)(0.51,-0.03){1}{\line(1,0){0.51}}
\put(39.75,50){\line(1,0){0.51}}
\multiput(39.24,49.97)(0.51,0.03){1}{\line(1,0){0.51}}
\multiput(38.74,49.92)(0.5,0.05){1}{\line(1,0){0.5}}
\multiput(38.24,49.84)(0.5,0.08){1}{\line(1,0){0.5}}
\multiput(37.74,49.74)(0.5,0.1){1}{\line(1,0){0.5}}
\multiput(37.25,49.61)(0.49,0.13){1}{\line(1,0){0.49}}
\multiput(36.77,49.46)(0.48,0.15){1}{\line(1,0){0.48}}
\multiput(36.29,49.29)(0.48,0.18){1}{\line(1,0){0.48}}
\multiput(35.82,49.09)(0.23,0.1){2}{\line(1,0){0.23}}
\multiput(35.37,48.86)(0.23,0.11){2}{\line(1,0){0.23}}
\multiput(34.93,48.62)(0.22,0.12){2}{\line(1,0){0.22}}
\multiput(34.5,48.35)(0.21,0.13){2}{\line(1,0){0.21}}
\multiput(34.08,48.06)(0.21,0.14){2}{\line(1,0){0.21}}
\multiput(33.68,47.75)(0.13,0.1){3}{\line(1,0){0.13}}
\multiput(33.3,47.42)(0.13,0.11){3}{\line(1,0){0.13}}
\multiput(32.93,47.07)(0.12,0.12){3}{\line(1,0){0.12}}
\multiput(32.58,46.7)(0.12,0.12){3}{\line(0,1){0.12}}
\multiput(32.25,46.32)(0.11,0.13){3}{\line(0,1){0.13}}
\multiput(31.94,45.92)(0.1,0.13){3}{\line(0,1){0.13}}
\multiput(31.65,45.5)(0.14,0.21){2}{\line(0,1){0.21}}
\multiput(31.38,45.07)(0.13,0.21){2}{\line(0,1){0.21}}
\multiput(31.14,44.63)(0.12,0.22){2}{\line(0,1){0.22}}
\multiput(30.91,44.18)(0.11,0.23){2}{\line(0,1){0.23}}
\multiput(30.71,43.71)(0.1,0.23){2}{\line(0,1){0.23}}
\multiput(30.54,43.23)(0.18,0.48){1}{\line(0,1){0.48}}
\multiput(30.39,42.75)(0.15,0.48){1}{\line(0,1){0.48}}
\multiput(30.26,42.26)(0.13,0.49){1}{\line(0,1){0.49}}
\multiput(30.16,41.76)(0.1,0.5){1}{\line(0,1){0.5}}
\multiput(30.08,41.26)(0.08,0.5){1}{\line(0,1){0.5}}
\multiput(30.03,40.76)(0.05,0.5){1}{\line(0,1){0.5}}
\multiput(30,40.25)(0.03,0.51){1}{\line(0,1){0.51}}
\put(30,39.75){\line(0,1){0.51}}
\multiput(30,39.75)(0.03,-0.51){1}{\line(0,-1){0.51}}
\multiput(30.03,39.24)(0.05,-0.5){1}{\line(0,-1){0.5}}
\multiput(30.08,38.74)(0.08,-0.5){1}{\line(0,-1){0.5}}
\multiput(30.16,38.24)(0.1,-0.5){1}{\line(0,-1){0.5}}
\multiput(30.26,37.74)(0.13,-0.49){1}{\line(0,-1){0.49}}
\multiput(30.39,37.25)(0.15,-0.48){1}{\line(0,-1){0.48}}
\multiput(30.54,36.77)(0.18,-0.48){1}{\line(0,-1){0.48}}
\multiput(30.71,36.29)(0.1,-0.23){2}{\line(0,-1){0.23}}
\multiput(30.91,35.82)(0.11,-0.23){2}{\line(0,-1){0.23}}
\multiput(31.14,35.37)(0.12,-0.22){2}{\line(0,-1){0.22}}
\multiput(31.38,34.93)(0.13,-0.21){2}{\line(0,-1){0.21}}
\multiput(31.65,34.5)(0.14,-0.21){2}{\line(0,-1){0.21}}
\multiput(31.94,34.08)(0.1,-0.13){3}{\line(0,-1){0.13}}
\multiput(32.25,33.68)(0.11,-0.13){3}{\line(0,-1){0.13}}
\multiput(32.58,33.3)(0.12,-0.12){3}{\line(0,-1){0.12}}
\multiput(32.93,32.93)(0.12,-0.12){3}{\line(1,0){0.12}}
\multiput(33.3,32.58)(0.13,-0.11){3}{\line(1,0){0.13}}
\multiput(33.68,32.25)(0.13,-0.1){3}{\line(1,0){0.13}}
\multiput(34.08,31.94)(0.21,-0.14){2}{\line(1,0){0.21}}
\multiput(34.5,31.65)(0.21,-0.13){2}{\line(1,0){0.21}}
\multiput(34.93,31.38)(0.22,-0.12){2}{\line(1,0){0.22}}
\multiput(35.37,31.14)(0.23,-0.11){2}{\line(1,0){0.23}}
\multiput(35.82,30.91)(0.23,-0.1){2}{\line(1,0){0.23}}
\multiput(36.29,30.71)(0.48,-0.18){1}{\line(1,0){0.48}}
\multiput(36.77,30.54)(0.48,-0.15){1}{\line(1,0){0.48}}
\multiput(37.25,30.39)(0.49,-0.13){1}{\line(1,0){0.49}}
\multiput(37.74,30.26)(0.5,-0.1){1}{\line(1,0){0.5}}
\multiput(38.24,30.16)(0.5,-0.08){1}{\line(1,0){0.5}}
\multiput(38.74,30.08)(0.5,-0.05){1}{\line(1,0){0.5}}
\multiput(39.24,30.03)(0.51,-0.03){1}{\line(1,0){0.51}}
\put(39.75,30){\line(1,0){0.51}}
\multiput(40.25,30)(0.51,0.03){1}{\line(1,0){0.51}}
\multiput(40.76,30.03)(0.5,0.05){1}{\line(1,0){0.5}}
\multiput(41.26,30.08)(0.5,0.08){1}{\line(1,0){0.5}}
\multiput(41.76,30.16)(0.5,0.1){1}{\line(1,0){0.5}}
\multiput(42.26,30.26)(0.49,0.13){1}{\line(1,0){0.49}}
\multiput(42.75,30.39)(0.48,0.15){1}{\line(1,0){0.48}}
\multiput(43.23,30.54)(0.48,0.18){1}{\line(1,0){0.48}}
\multiput(43.71,30.71)(0.23,0.1){2}{\line(1,0){0.23}}
\multiput(44.18,30.91)(0.23,0.11){2}{\line(1,0){0.23}}
\multiput(44.63,31.14)(0.22,0.12){2}{\line(1,0){0.22}}
\multiput(45.07,31.38)(0.21,0.13){2}{\line(1,0){0.21}}
\multiput(45.5,31.65)(0.21,0.14){2}{\line(1,0){0.21}}
\multiput(45.92,31.94)(0.13,0.1){3}{\line(1,0){0.13}}
\multiput(46.32,32.25)(0.13,0.11){3}{\line(1,0){0.13}}
\multiput(46.7,32.58)(0.12,0.12){3}{\line(1,0){0.12}}
\multiput(47.07,32.93)(0.12,0.12){3}{\line(0,1){0.12}}
\multiput(47.42,33.3)(0.11,0.13){3}{\line(0,1){0.13}}
\multiput(47.75,33.68)(0.1,0.13){3}{\line(0,1){0.13}}
\multiput(48.06,34.08)(0.14,0.21){2}{\line(0,1){0.21}}
\multiput(48.35,34.5)(0.13,0.21){2}{\line(0,1){0.21}}
\multiput(48.62,34.93)(0.12,0.22){2}{\line(0,1){0.22}}
\multiput(48.86,35.37)(0.11,0.23){2}{\line(0,1){0.23}}
\multiput(49.09,35.82)(0.1,0.23){2}{\line(0,1){0.23}}
\multiput(49.29,36.29)(0.18,0.48){1}{\line(0,1){0.48}}
\multiput(49.46,36.77)(0.15,0.48){1}{\line(0,1){0.48}}
\multiput(49.61,37.25)(0.13,0.49){1}{\line(0,1){0.49}}
\multiput(49.74,37.74)(0.1,0.5){1}{\line(0,1){0.5}}
\multiput(49.84,38.24)(0.08,0.5){1}{\line(0,1){0.5}}
\multiput(49.92,38.74)(0.05,0.5){1}{\line(0,1){0.5}}
\multiput(49.97,39.24)(0.03,0.51){1}{\line(0,1){0.51}}

\linethickness{0.3mm}
\put(100,39.75){\line(0,1){0.51}}
\multiput(99.97,40.76)(0.03,-0.51){1}{\line(0,-1){0.51}}
\multiput(99.92,41.26)(0.05,-0.5){1}{\line(0,-1){0.5}}
\multiput(99.84,41.76)(0.08,-0.5){1}{\line(0,-1){0.5}}
\multiput(99.74,42.26)(0.1,-0.5){1}{\line(0,-1){0.5}}
\multiput(99.61,42.75)(0.13,-0.49){1}{\line(0,-1){0.49}}
\multiput(99.46,43.23)(0.15,-0.48){1}{\line(0,-1){0.48}}
\multiput(99.29,43.71)(0.18,-0.48){1}{\line(0,-1){0.48}}
\multiput(99.09,44.18)(0.1,-0.23){2}{\line(0,-1){0.23}}
\multiput(98.86,44.63)(0.11,-0.23){2}{\line(0,-1){0.23}}
\multiput(98.62,45.07)(0.12,-0.22){2}{\line(0,-1){0.22}}
\multiput(98.35,45.5)(0.13,-0.21){2}{\line(0,-1){0.21}}
\multiput(98.06,45.92)(0.14,-0.21){2}{\line(0,-1){0.21}}
\multiput(97.75,46.32)(0.1,-0.13){3}{\line(0,-1){0.13}}
\multiput(97.42,46.7)(0.11,-0.13){3}{\line(0,-1){0.13}}
\multiput(97.07,47.07)(0.12,-0.12){3}{\line(0,-1){0.12}}
\multiput(96.7,47.42)(0.12,-0.12){3}{\line(1,0){0.12}}
\multiput(96.32,47.75)(0.13,-0.11){3}{\line(1,0){0.13}}
\multiput(95.92,48.06)(0.13,-0.1){3}{\line(1,0){0.13}}
\multiput(95.5,48.35)(0.21,-0.14){2}{\line(1,0){0.21}}
\multiput(95.07,48.62)(0.21,-0.13){2}{\line(1,0){0.21}}
\multiput(94.63,48.86)(0.22,-0.12){2}{\line(1,0){0.22}}
\multiput(94.18,49.09)(0.23,-0.11){2}{\line(1,0){0.23}}
\multiput(93.71,49.29)(0.23,-0.1){2}{\line(1,0){0.23}}
\multiput(93.23,49.46)(0.48,-0.18){1}{\line(1,0){0.48}}
\multiput(92.75,49.61)(0.48,-0.15){1}{\line(1,0){0.48}}
\multiput(92.26,49.74)(0.49,-0.13){1}{\line(1,0){0.49}}
\multiput(91.76,49.84)(0.5,-0.1){1}{\line(1,0){0.5}}
\multiput(91.26,49.92)(0.5,-0.08){1}{\line(1,0){0.5}}
\multiput(90.76,49.97)(0.5,-0.05){1}{\line(1,0){0.5}}
\multiput(90.25,50)(0.51,-0.03){1}{\line(1,0){0.51}}
\put(89.75,50){\line(1,0){0.51}}
\multiput(89.24,49.97)(0.51,0.03){1}{\line(1,0){0.51}}
\multiput(88.74,49.92)(0.5,0.05){1}{\line(1,0){0.5}}
\multiput(88.24,49.84)(0.5,0.08){1}{\line(1,0){0.5}}
\multiput(87.74,49.74)(0.5,0.1){1}{\line(1,0){0.5}}
\multiput(87.25,49.61)(0.49,0.13){1}{\line(1,0){0.49}}
\multiput(86.77,49.46)(0.48,0.15){1}{\line(1,0){0.48}}
\multiput(86.29,49.29)(0.48,0.18){1}{\line(1,0){0.48}}
\multiput(85.82,49.09)(0.23,0.1){2}{\line(1,0){0.23}}
\multiput(85.37,48.86)(0.23,0.11){2}{\line(1,0){0.23}}
\multiput(84.93,48.62)(0.22,0.12){2}{\line(1,0){0.22}}
\multiput(84.5,48.35)(0.21,0.13){2}{\line(1,0){0.21}}
\multiput(84.08,48.06)(0.21,0.14){2}{\line(1,0){0.21}}
\multiput(83.68,47.75)(0.13,0.1){3}{\line(1,0){0.13}}
\multiput(83.3,47.42)(0.13,0.11){3}{\line(1,0){0.13}}
\multiput(82.93,47.07)(0.12,0.12){3}{\line(1,0){0.12}}
\multiput(82.58,46.7)(0.12,0.12){3}{\line(0,1){0.12}}
\multiput(82.25,46.32)(0.11,0.13){3}{\line(0,1){0.13}}
\multiput(81.94,45.92)(0.1,0.13){3}{\line(0,1){0.13}}
\multiput(81.65,45.5)(0.14,0.21){2}{\line(0,1){0.21}}
\multiput(81.38,45.07)(0.13,0.21){2}{\line(0,1){0.21}}
\multiput(81.14,44.63)(0.12,0.22){2}{\line(0,1){0.22}}
\multiput(80.91,44.18)(0.11,0.23){2}{\line(0,1){0.23}}
\multiput(80.71,43.71)(0.1,0.23){2}{\line(0,1){0.23}}
\multiput(80.54,43.23)(0.18,0.48){1}{\line(0,1){0.48}}
\multiput(80.39,42.75)(0.15,0.48){1}{\line(0,1){0.48}}
\multiput(80.26,42.26)(0.13,0.49){1}{\line(0,1){0.49}}
\multiput(80.16,41.76)(0.1,0.5){1}{\line(0,1){0.5}}
\multiput(80.08,41.26)(0.08,0.5){1}{\line(0,1){0.5}}
\multiput(80.03,40.76)(0.05,0.5){1}{\line(0,1){0.5}}
\multiput(80,40.25)(0.03,0.51){1}{\line(0,1){0.51}}
\put(80,39.75){\line(0,1){0.51}}
\multiput(80,39.75)(0.03,-0.51){1}{\line(0,-1){0.51}}
\multiput(80.03,39.24)(0.05,-0.5){1}{\line(0,-1){0.5}}
\multiput(80.08,38.74)(0.08,-0.5){1}{\line(0,-1){0.5}}
\multiput(80.16,38.24)(0.1,-0.5){1}{\line(0,-1){0.5}}
\multiput(80.26,37.74)(0.13,-0.49){1}{\line(0,-1){0.49}}
\multiput(80.39,37.25)(0.15,-0.48){1}{\line(0,-1){0.48}}
\multiput(80.54,36.77)(0.18,-0.48){1}{\line(0,-1){0.48}}
\multiput(80.71,36.29)(0.1,-0.23){2}{\line(0,-1){0.23}}
\multiput(80.91,35.82)(0.11,-0.23){2}{\line(0,-1){0.23}}
\multiput(81.14,35.37)(0.12,-0.22){2}{\line(0,-1){0.22}}
\multiput(81.38,34.93)(0.13,-0.21){2}{\line(0,-1){0.21}}
\multiput(81.65,34.5)(0.14,-0.21){2}{\line(0,-1){0.21}}
\multiput(81.94,34.08)(0.1,-0.13){3}{\line(0,-1){0.13}}
\multiput(82.25,33.68)(0.11,-0.13){3}{\line(0,-1){0.13}}
\multiput(82.58,33.3)(0.12,-0.12){3}{\line(0,-1){0.12}}
\multiput(82.93,32.93)(0.12,-0.12){3}{\line(1,0){0.12}}
\multiput(83.3,32.58)(0.13,-0.11){3}{\line(1,0){0.13}}
\multiput(83.68,32.25)(0.13,-0.1){3}{\line(1,0){0.13}}
\multiput(84.08,31.94)(0.21,-0.14){2}{\line(1,0){0.21}}
\multiput(84.5,31.65)(0.21,-0.13){2}{\line(1,0){0.21}}
\multiput(84.93,31.38)(0.22,-0.12){2}{\line(1,0){0.22}}
\multiput(85.37,31.14)(0.23,-0.11){2}{\line(1,0){0.23}}
\multiput(85.82,30.91)(0.23,-0.1){2}{\line(1,0){0.23}}
\multiput(86.29,30.71)(0.48,-0.18){1}{\line(1,0){0.48}}
\multiput(86.77,30.54)(0.48,-0.15){1}{\line(1,0){0.48}}
\multiput(87.25,30.39)(0.49,-0.13){1}{\line(1,0){0.49}}
\multiput(87.74,30.26)(0.5,-0.1){1}{\line(1,0){0.5}}
\multiput(88.24,30.16)(0.5,-0.08){1}{\line(1,0){0.5}}
\multiput(88.74,30.08)(0.5,-0.05){1}{\line(1,0){0.5}}
\multiput(89.24,30.03)(0.51,-0.03){1}{\line(1,0){0.51}}
\put(89.75,30){\line(1,0){0.51}}
\multiput(90.25,30)(0.51,0.03){1}{\line(1,0){0.51}}
\multiput(90.76,30.03)(0.5,0.05){1}{\line(1,0){0.5}}
\multiput(91.26,30.08)(0.5,0.08){1}{\line(1,0){0.5}}
\multiput(91.76,30.16)(0.5,0.1){1}{\line(1,0){0.5}}
\multiput(92.26,30.26)(0.49,0.13){1}{\line(1,0){0.49}}
\multiput(92.75,30.39)(0.48,0.15){1}{\line(1,0){0.48}}
\multiput(93.23,30.54)(0.48,0.18){1}{\line(1,0){0.48}}
\multiput(93.71,30.71)(0.23,0.1){2}{\line(1,0){0.23}}
\multiput(94.18,30.91)(0.23,0.11){2}{\line(1,0){0.23}}
\multiput(94.63,31.14)(0.22,0.12){2}{\line(1,0){0.22}}
\multiput(95.07,31.38)(0.21,0.13){2}{\line(1,0){0.21}}
\multiput(95.5,31.65)(0.21,0.14){2}{\line(1,0){0.21}}
\multiput(95.92,31.94)(0.13,0.1){3}{\line(1,0){0.13}}
\multiput(96.32,32.25)(0.13,0.11){3}{\line(1,0){0.13}}
\multiput(96.7,32.58)(0.12,0.12){3}{\line(1,0){0.12}}
\multiput(97.07,32.93)(0.12,0.12){3}{\line(0,1){0.12}}
\multiput(97.42,33.3)(0.11,0.13){3}{\line(0,1){0.13}}
\multiput(97.75,33.68)(0.1,0.13){3}{\line(0,1){0.13}}
\multiput(98.06,34.08)(0.14,0.21){2}{\line(0,1){0.21}}
\multiput(98.35,34.5)(0.13,0.21){2}{\line(0,1){0.21}}
\multiput(98.62,34.93)(0.12,0.22){2}{\line(0,1){0.22}}
\multiput(98.86,35.37)(0.11,0.23){2}{\line(0,1){0.23}}
\multiput(99.09,35.82)(0.1,0.23){2}{\line(0,1){0.23}}
\multiput(99.29,36.29)(0.18,0.48){1}{\line(0,1){0.48}}
\multiput(99.46,36.77)(0.15,0.48){1}{\line(0,1){0.48}}
\multiput(99.61,37.25)(0.13,0.49){1}{\line(0,1){0.49}}
\multiput(99.74,37.74)(0.1,0.5){1}{\line(0,1){0.5}}
\multiput(99.84,38.24)(0.08,0.5){1}{\line(0,1){0.5}}
\multiput(99.92,38.74)(0.05,0.5){1}{\line(0,1){0.5}}
\multiput(99.97,39.24)(0.03,0.51){1}{\line(0,1){0.51}}

\linethickness{1mm}
\put(10,40){\line(1,0){20}}
\linethickness{1mm}
\put(50,40){\line(1,0){30}}
\linethickness{1mm}
\multiput(90,50)(0.36,0.12){83}{\line(1,0){0.36}}
\linethickness{1mm}
\multiput(90,30)(0.36,-0.12){83}{\line(1,0){0.36}}
\linethickness{0.3mm}
\put(31,10){\line(0,1){25}}
\linethickness{0.3mm}
\put(49,10){\line(0,1){25}}
\put(15,45){\makebox(0,0)[cc]{$p_i$}}

\put(15,35){\makebox(0,0)[cc]{$\eps_i$}}

\put(65,45){\makebox(0,0)[cc]{$-p_i-k_1-k_2$}}

\put(90,40){\makebox(0,0)[cc]{$\Gamma$}}

\put(110,40){\makebox(0,0)[cc]{$\cdot$}}

\put(108,50){\makebox(0,0)[cc]{$\cdot$}}

\put(108,30){\makebox(0,0)[cc]{$\cdot$}}

\put(40,40){\makebox(0,0)[cc]{$\Gamma^{(4)}$}}

\put(27,20){\makebox(0,0)[cc]{$k_1$}}

\put(35,20){\makebox(0,0)[cc]{$\ve_1$}}

\put(45,20){\makebox(0,0)[cc]{$k_2$}}

\put(54,20){\makebox(0,0)[cc]{$\ve_2$}}

\end{picture}

}

\def\figeight{

\def\JPicScale{0.8}
\ifx\JPicScale\undefined\def\JPicScale{1}\fi
\unitlength \JPicScale mm
\begin{picture}(120,60)(0,0)
\linethickness{0.3mm}
\put(50,39.75){\line(0,1){0.51}}
\multiput(49.97,40.76)(0.03,-0.51){1}{\line(0,-1){0.51}}
\multiput(49.92,41.26)(0.05,-0.5){1}{\line(0,-1){0.5}}
\multiput(49.84,41.76)(0.08,-0.5){1}{\line(0,-1){0.5}}
\multiput(49.74,42.26)(0.1,-0.5){1}{\line(0,-1){0.5}}
\multiput(49.61,42.75)(0.13,-0.49){1}{\line(0,-1){0.49}}
\multiput(49.46,43.23)(0.15,-0.48){1}{\line(0,-1){0.48}}
\multiput(49.29,43.71)(0.18,-0.48){1}{\line(0,-1){0.48}}
\multiput(49.09,44.18)(0.1,-0.23){2}{\line(0,-1){0.23}}
\multiput(48.86,44.63)(0.11,-0.23){2}{\line(0,-1){0.23}}
\multiput(48.62,45.07)(0.12,-0.22){2}{\line(0,-1){0.22}}
\multiput(48.35,45.5)(0.13,-0.21){2}{\line(0,-1){0.21}}
\multiput(48.06,45.92)(0.14,-0.21){2}{\line(0,-1){0.21}}
\multiput(47.75,46.32)(0.1,-0.13){3}{\line(0,-1){0.13}}
\multiput(47.42,46.7)(0.11,-0.13){3}{\line(0,-1){0.13}}
\multiput(47.07,47.07)(0.12,-0.12){3}{\line(0,-1){0.12}}
\multiput(46.7,47.42)(0.12,-0.12){3}{\line(1,0){0.12}}
\multiput(46.32,47.75)(0.13,-0.11){3}{\line(1,0){0.13}}
\multiput(45.92,48.06)(0.13,-0.1){3}{\line(1,0){0.13}}
\multiput(45.5,48.35)(0.21,-0.14){2}{\line(1,0){0.21}}
\multiput(45.07,48.62)(0.21,-0.13){2}{\line(1,0){0.21}}
\multiput(44.63,48.86)(0.22,-0.12){2}{\line(1,0){0.22}}
\multiput(44.18,49.09)(0.23,-0.11){2}{\line(1,0){0.23}}
\multiput(43.71,49.29)(0.23,-0.1){2}{\line(1,0){0.23}}
\multiput(43.23,49.46)(0.48,-0.18){1}{\line(1,0){0.48}}
\multiput(42.75,49.61)(0.48,-0.15){1}{\line(1,0){0.48}}
\multiput(42.26,49.74)(0.49,-0.13){1}{\line(1,0){0.49}}
\multiput(41.76,49.84)(0.5,-0.1){1}{\line(1,0){0.5}}
\multiput(41.26,49.92)(0.5,-0.08){1}{\line(1,0){0.5}}
\multiput(40.76,49.97)(0.5,-0.05){1}{\line(1,0){0.5}}
\multiput(40.25,50)(0.51,-0.03){1}{\line(1,0){0.51}}
\put(39.75,50){\line(1,0){0.51}}
\multiput(39.24,49.97)(0.51,0.03){1}{\line(1,0){0.51}}
\multiput(38.74,49.92)(0.5,0.05){1}{\line(1,0){0.5}}
\multiput(38.24,49.84)(0.5,0.08){1}{\line(1,0){0.5}}
\multiput(37.74,49.74)(0.5,0.1){1}{\line(1,0){0.5}}
\multiput(37.25,49.61)(0.49,0.13){1}{\line(1,0){0.49}}
\multiput(36.77,49.46)(0.48,0.15){1}{\line(1,0){0.48}}
\multiput(36.29,49.29)(0.48,0.18){1}{\line(1,0){0.48}}
\multiput(35.82,49.09)(0.23,0.1){2}{\line(1,0){0.23}}
\multiput(35.37,48.86)(0.23,0.11){2}{\line(1,0){0.23}}
\multiput(34.93,48.62)(0.22,0.12){2}{\line(1,0){0.22}}
\multiput(34.5,48.35)(0.21,0.13){2}{\line(1,0){0.21}}
\multiput(34.08,48.06)(0.21,0.14){2}{\line(1,0){0.21}}
\multiput(33.68,47.75)(0.13,0.1){3}{\line(1,0){0.13}}
\multiput(33.3,47.42)(0.13,0.11){3}{\line(1,0){0.13}}
\multiput(32.93,47.07)(0.12,0.12){3}{\line(1,0){0.12}}
\multiput(32.58,46.7)(0.12,0.12){3}{\line(0,1){0.12}}
\multiput(32.25,46.32)(0.11,0.13){3}{\line(0,1){0.13}}
\multiput(31.94,45.92)(0.1,0.13){3}{\line(0,1){0.13}}
\multiput(31.65,45.5)(0.14,0.21){2}{\line(0,1){0.21}}
\multiput(31.38,45.07)(0.13,0.21){2}{\line(0,1){0.21}}
\multiput(31.14,44.63)(0.12,0.22){2}{\line(0,1){0.22}}
\multiput(30.91,44.18)(0.11,0.23){2}{\line(0,1){0.23}}
\multiput(30.71,43.71)(0.1,0.23){2}{\line(0,1){0.23}}
\multiput(30.54,43.23)(0.18,0.48){1}{\line(0,1){0.48}}
\multiput(30.39,42.75)(0.15,0.48){1}{\line(0,1){0.48}}
\multiput(30.26,42.26)(0.13,0.49){1}{\line(0,1){0.49}}
\multiput(30.16,41.76)(0.1,0.5){1}{\line(0,1){0.5}}
\multiput(30.08,41.26)(0.08,0.5){1}{\line(0,1){0.5}}
\multiput(30.03,40.76)(0.05,0.5){1}{\line(0,1){0.5}}
\multiput(30,40.25)(0.03,0.51){1}{\line(0,1){0.51}}
\put(30,39.75){\line(0,1){0.51}}
\multiput(30,39.75)(0.03,-0.51){1}{\line(0,-1){0.51}}
\multiput(30.03,39.24)(0.05,-0.5){1}{\line(0,-1){0.5}}
\multiput(30.08,38.74)(0.08,-0.5){1}{\line(0,-1){0.5}}
\multiput(30.16,38.24)(0.1,-0.5){1}{\line(0,-1){0.5}}
\multiput(30.26,37.74)(0.13,-0.49){1}{\line(0,-1){0.49}}
\multiput(30.39,37.25)(0.15,-0.48){1}{\line(0,-1){0.48}}
\multiput(30.54,36.77)(0.18,-0.48){1}{\line(0,-1){0.48}}
\multiput(30.71,36.29)(0.1,-0.23){2}{\line(0,-1){0.23}}
\multiput(30.91,35.82)(0.11,-0.23){2}{\line(0,-1){0.23}}
\multiput(31.14,35.37)(0.12,-0.22){2}{\line(0,-1){0.22}}
\multiput(31.38,34.93)(0.13,-0.21){2}{\line(0,-1){0.21}}
\multiput(31.65,34.5)(0.14,-0.21){2}{\line(0,-1){0.21}}
\multiput(31.94,34.08)(0.1,-0.13){3}{\line(0,-1){0.13}}
\multiput(32.25,33.68)(0.11,-0.13){3}{\line(0,-1){0.13}}
\multiput(32.58,33.3)(0.12,-0.12){3}{\line(0,-1){0.12}}
\multiput(32.93,32.93)(0.12,-0.12){3}{\line(1,0){0.12}}
\multiput(33.3,32.58)(0.13,-0.11){3}{\line(1,0){0.13}}
\multiput(33.68,32.25)(0.13,-0.1){3}{\line(1,0){0.13}}
\multiput(34.08,31.94)(0.21,-0.14){2}{\line(1,0){0.21}}
\multiput(34.5,31.65)(0.21,-0.13){2}{\line(1,0){0.21}}
\multiput(34.93,31.38)(0.22,-0.12){2}{\line(1,0){0.22}}
\multiput(35.37,31.14)(0.23,-0.11){2}{\line(1,0){0.23}}
\multiput(35.82,30.91)(0.23,-0.1){2}{\line(1,0){0.23}}
\multiput(36.29,30.71)(0.48,-0.18){1}{\line(1,0){0.48}}
\multiput(36.77,30.54)(0.48,-0.15){1}{\line(1,0){0.48}}
\multiput(37.25,30.39)(0.49,-0.13){1}{\line(1,0){0.49}}
\multiput(37.74,30.26)(0.5,-0.1){1}{\line(1,0){0.5}}
\multiput(38.24,30.16)(0.5,-0.08){1}{\line(1,0){0.5}}
\multiput(38.74,30.08)(0.5,-0.05){1}{\line(1,0){0.5}}
\multiput(39.24,30.03)(0.51,-0.03){1}{\line(1,0){0.51}}
\put(39.75,30){\line(1,0){0.51}}
\multiput(40.25,30)(0.51,0.03){1}{\line(1,0){0.51}}
\multiput(40.76,30.03)(0.5,0.05){1}{\line(1,0){0.5}}
\multiput(41.26,30.08)(0.5,0.08){1}{\line(1,0){0.5}}
\multiput(41.76,30.16)(0.5,0.1){1}{\line(1,0){0.5}}
\multiput(42.26,30.26)(0.49,0.13){1}{\line(1,0){0.49}}
\multiput(42.75,30.39)(0.48,0.15){1}{\line(1,0){0.48}}
\multiput(43.23,30.54)(0.48,0.18){1}{\line(1,0){0.48}}
\multiput(43.71,30.71)(0.23,0.1){2}{\line(1,0){0.23}}
\multiput(44.18,30.91)(0.23,0.11){2}{\line(1,0){0.23}}
\multiput(44.63,31.14)(0.22,0.12){2}{\line(1,0){0.22}}
\multiput(45.07,31.38)(0.21,0.13){2}{\line(1,0){0.21}}
\multiput(45.5,31.65)(0.21,0.14){2}{\line(1,0){0.21}}
\multiput(45.92,31.94)(0.13,0.1){3}{\line(1,0){0.13}}
\multiput(46.32,32.25)(0.13,0.11){3}{\line(1,0){0.13}}
\multiput(46.7,32.58)(0.12,0.12){3}{\line(1,0){0.12}}
\multiput(47.07,32.93)(0.12,0.12){3}{\line(0,1){0.12}}
\multiput(47.42,33.3)(0.11,0.13){3}{\line(0,1){0.13}}
\multiput(47.75,33.68)(0.1,0.13){3}{\line(0,1){0.13}}
\multiput(48.06,34.08)(0.14,0.21){2}{\line(0,1){0.21}}
\multiput(48.35,34.5)(0.13,0.21){2}{\line(0,1){0.21}}
\multiput(48.62,34.93)(0.12,0.22){2}{\line(0,1){0.22}}
\multiput(48.86,35.37)(0.11,0.23){2}{\line(0,1){0.23}}
\multiput(49.09,35.82)(0.1,0.23){2}{\line(0,1){0.23}}
\multiput(49.29,36.29)(0.18,0.48){1}{\line(0,1){0.48}}
\multiput(49.46,36.77)(0.15,0.48){1}{\line(0,1){0.48}}
\multiput(49.61,37.25)(0.13,0.49){1}{\line(0,1){0.49}}
\multiput(49.74,37.74)(0.1,0.5){1}{\line(0,1){0.5}}
\multiput(49.84,38.24)(0.08,0.5){1}{\line(0,1){0.5}}
\multiput(49.92,38.74)(0.05,0.5){1}{\line(0,1){0.5}}
\multiput(49.97,39.24)(0.03,0.51){1}{\line(0,1){0.51}}

\linethickness{1mm}
\put(10,40){\line(1,0){20}}
\linethickness{1mm}
\put(50,40){\line(1,0){30}}
\linethickness{0.3mm}
\put(31,10){\line(0,1){25}}
\linethickness{0.3mm}
\put(49,10){\line(0,1){25}}
\put(15,45){\makebox(0,0)[cc]{$p$}}

\put(15,35){\makebox(0,0)[cc]{$\alpha$}}

\put(64,45){\makebox(0,0)[cc]{$-p-k_1-k_2$}}

\put(62,35){\makebox(0,0)[cc]{$\beta$}}

\put(40,40){\makebox(0,0)[cc]{$\Gamma^{(4)}$}}

\put(27,20){\makebox(0,0)[cc]{$k_1$}}

\put(35,20){\makebox(0,0)[cc]{$\ve_1$}}

\put(45,20){\makebox(0,0)[cc]{$k_2$}}

\put(54,20){\makebox(0,0)[cc]{$\ve_2$}}

\end{picture}

}

\def\figsix{

\def\JPicScale{0.8}
\ifx\JPicScale\undefined\def\JPicScale{1}\fi
\unitlength \JPicScale mm

}

\def\fignine{

\def\JPicScale{0.6}
\ifx\JPicScale\undefined\def\JPicScale{1}\fi
\unitlength \JPicScale mm
\begin{picture}(90,80)(0,0)
\linethickness{0.3mm}
\put(80,39.75){\line(0,1){0.5}}
\multiput(79.99,40.75)(0.01,-0.5){1}{\line(0,-1){0.5}}
\multiput(79.96,41.25)(0.02,-0.5){1}{\line(0,-1){0.5}}
\multiput(79.92,41.74)(0.04,-0.5){1}{\line(0,-1){0.5}}
\multiput(79.87,42.24)(0.05,-0.5){1}{\line(0,-1){0.5}}
\multiput(79.81,42.73)(0.06,-0.49){1}{\line(0,-1){0.49}}
\multiput(79.74,43.23)(0.07,-0.49){1}{\line(0,-1){0.49}}
\multiput(79.65,43.72)(0.09,-0.49){1}{\line(0,-1){0.49}}
\multiput(79.55,44.21)(0.1,-0.49){1}{\line(0,-1){0.49}}
\multiput(79.44,44.69)(0.11,-0.49){1}{\line(0,-1){0.49}}
\multiput(79.32,45.18)(0.12,-0.48){1}{\line(0,-1){0.48}}
\multiput(79.18,45.66)(0.14,-0.48){1}{\line(0,-1){0.48}}
\multiput(79.04,46.13)(0.15,-0.48){1}{\line(0,-1){0.48}}
\multiput(78.88,46.61)(0.16,-0.47){1}{\line(0,-1){0.47}}
\multiput(78.71,47.07)(0.17,-0.47){1}{\line(0,-1){0.47}}
\multiput(78.52,47.54)(0.09,-0.23){2}{\line(0,-1){0.23}}
\multiput(78.33,48)(0.1,-0.23){2}{\line(0,-1){0.23}}
\multiput(78.13,48.45)(0.1,-0.23){2}{\line(0,-1){0.23}}
\multiput(77.91,48.9)(0.11,-0.22){2}{\line(0,-1){0.22}}
\multiput(77.68,49.35)(0.11,-0.22){2}{\line(0,-1){0.22}}
\multiput(77.44,49.78)(0.12,-0.22){2}{\line(0,-1){0.22}}
\multiput(77.19,50.22)(0.12,-0.22){2}{\line(0,-1){0.22}}
\multiput(76.93,50.64)(0.13,-0.21){2}{\line(0,-1){0.21}}
\multiput(76.66,51.06)(0.14,-0.21){2}{\line(0,-1){0.21}}
\multiput(76.38,51.47)(0.14,-0.21){2}{\line(0,-1){0.21}}
\multiput(76.09,51.88)(0.15,-0.2){2}{\line(0,-1){0.2}}
\multiput(75.79,52.27)(0.1,-0.13){3}{\line(0,-1){0.13}}
\multiput(75.48,52.66)(0.1,-0.13){3}{\line(0,-1){0.13}}
\multiput(75.16,53.05)(0.11,-0.13){3}{\line(0,-1){0.13}}
\multiput(74.83,53.42)(0.11,-0.12){3}{\line(0,-1){0.12}}
\multiput(74.49,53.79)(0.11,-0.12){3}{\line(0,-1){0.12}}
\multiput(74.14,54.14)(0.12,-0.12){3}{\line(0,-1){0.12}}
\multiput(73.79,54.49)(0.12,-0.12){3}{\line(1,0){0.12}}
\multiput(73.42,54.83)(0.12,-0.11){3}{\line(1,0){0.12}}
\multiput(73.05,55.16)(0.12,-0.11){3}{\line(1,0){0.12}}
\multiput(72.66,55.48)(0.13,-0.11){3}{\line(1,0){0.13}}
\multiput(72.27,55.79)(0.13,-0.1){3}{\line(1,0){0.13}}
\multiput(71.88,56.09)(0.13,-0.1){3}{\line(1,0){0.13}}
\multiput(71.47,56.38)(0.2,-0.15){2}{\line(1,0){0.2}}
\multiput(71.06,56.66)(0.21,-0.14){2}{\line(1,0){0.21}}
\multiput(70.64,56.93)(0.21,-0.14){2}{\line(1,0){0.21}}
\multiput(70.22,57.19)(0.21,-0.13){2}{\line(1,0){0.21}}
\multiput(69.78,57.44)(0.22,-0.12){2}{\line(1,0){0.22}}
\multiput(69.35,57.68)(0.22,-0.12){2}{\line(1,0){0.22}}
\multiput(68.9,57.91)(0.22,-0.11){2}{\line(1,0){0.22}}
\multiput(68.45,58.13)(0.22,-0.11){2}{\line(1,0){0.22}}
\multiput(68,58.33)(0.23,-0.1){2}{\line(1,0){0.23}}
\multiput(67.54,58.52)(0.23,-0.1){2}{\line(1,0){0.23}}
\multiput(67.07,58.71)(0.23,-0.09){2}{\line(1,0){0.23}}
\multiput(66.61,58.88)(0.47,-0.17){1}{\line(1,0){0.47}}
\multiput(66.13,59.04)(0.47,-0.16){1}{\line(1,0){0.47}}
\multiput(65.66,59.18)(0.48,-0.15){1}{\line(1,0){0.48}}
\multiput(65.18,59.32)(0.48,-0.14){1}{\line(1,0){0.48}}
\multiput(64.69,59.44)(0.48,-0.12){1}{\line(1,0){0.48}}
\multiput(64.21,59.55)(0.49,-0.11){1}{\line(1,0){0.49}}
\multiput(63.72,59.65)(0.49,-0.1){1}{\line(1,0){0.49}}
\multiput(63.23,59.74)(0.49,-0.09){1}{\line(1,0){0.49}}
\multiput(62.73,59.81)(0.49,-0.07){1}{\line(1,0){0.49}}
\multiput(62.24,59.87)(0.49,-0.06){1}{\line(1,0){0.49}}
\multiput(61.74,59.92)(0.5,-0.05){1}{\line(1,0){0.5}}
\multiput(61.25,59.96)(0.5,-0.04){1}{\line(1,0){0.5}}
\multiput(60.75,59.99)(0.5,-0.02){1}{\line(1,0){0.5}}
\multiput(60.25,60)(0.5,-0.01){1}{\line(1,0){0.5}}
\put(59.75,60){\line(1,0){0.5}}
\multiput(59.25,59.99)(0.5,0.01){1}{\line(1,0){0.5}}
\multiput(58.75,59.96)(0.5,0.02){1}{\line(1,0){0.5}}
\multiput(58.26,59.92)(0.5,0.04){1}{\line(1,0){0.5}}
\multiput(57.76,59.87)(0.5,0.05){1}{\line(1,0){0.5}}
\multiput(57.27,59.81)(0.49,0.06){1}{\line(1,0){0.49}}
\multiput(56.77,59.74)(0.49,0.07){1}{\line(1,0){0.49}}
\multiput(56.28,59.65)(0.49,0.09){1}{\line(1,0){0.49}}
\multiput(55.79,59.55)(0.49,0.1){1}{\line(1,0){0.49}}
\multiput(55.31,59.44)(0.49,0.11){1}{\line(1,0){0.49}}
\multiput(54.82,59.32)(0.48,0.12){1}{\line(1,0){0.48}}
\multiput(54.34,59.18)(0.48,0.14){1}{\line(1,0){0.48}}
\multiput(53.87,59.04)(0.48,0.15){1}{\line(1,0){0.48}}
\multiput(53.39,58.88)(0.47,0.16){1}{\line(1,0){0.47}}
\multiput(52.93,58.71)(0.47,0.17){1}{\line(1,0){0.47}}
\multiput(52.46,58.52)(0.23,0.09){2}{\line(1,0){0.23}}
\multiput(52,58.33)(0.23,0.1){2}{\line(1,0){0.23}}
\multiput(51.55,58.13)(0.23,0.1){2}{\line(1,0){0.23}}
\multiput(51.1,57.91)(0.22,0.11){2}{\line(1,0){0.22}}
\multiput(50.65,57.68)(0.22,0.11){2}{\line(1,0){0.22}}
\multiput(50.22,57.44)(0.22,0.12){2}{\line(1,0){0.22}}
\multiput(49.78,57.19)(0.22,0.12){2}{\line(1,0){0.22}}
\multiput(49.36,56.93)(0.21,0.13){2}{\line(1,0){0.21}}
\multiput(48.94,56.66)(0.21,0.14){2}{\line(1,0){0.21}}
\multiput(48.53,56.38)(0.21,0.14){2}{\line(1,0){0.21}}
\multiput(48.12,56.09)(0.2,0.15){2}{\line(1,0){0.2}}
\multiput(47.73,55.79)(0.13,0.1){3}{\line(1,0){0.13}}
\multiput(47.34,55.48)(0.13,0.1){3}{\line(1,0){0.13}}
\multiput(46.95,55.16)(0.13,0.11){3}{\line(1,0){0.13}}
\multiput(46.58,54.83)(0.12,0.11){3}{\line(1,0){0.12}}
\multiput(46.21,54.49)(0.12,0.11){3}{\line(1,0){0.12}}
\multiput(45.86,54.14)(0.12,0.12){3}{\line(1,0){0.12}}
\multiput(45.51,53.79)(0.12,0.12){3}{\line(0,1){0.12}}
\multiput(45.17,53.42)(0.11,0.12){3}{\line(0,1){0.12}}
\multiput(44.84,53.05)(0.11,0.12){3}{\line(0,1){0.12}}
\multiput(44.52,52.66)(0.11,0.13){3}{\line(0,1){0.13}}
\multiput(44.21,52.27)(0.1,0.13){3}{\line(0,1){0.13}}
\multiput(43.91,51.88)(0.1,0.13){3}{\line(0,1){0.13}}
\multiput(43.62,51.47)(0.15,0.2){2}{\line(0,1){0.2}}
\multiput(43.34,51.06)(0.14,0.21){2}{\line(0,1){0.21}}
\multiput(43.07,50.64)(0.14,0.21){2}{\line(0,1){0.21}}
\multiput(42.81,50.22)(0.13,0.21){2}{\line(0,1){0.21}}
\multiput(42.56,49.78)(0.12,0.22){2}{\line(0,1){0.22}}
\multiput(42.32,49.35)(0.12,0.22){2}{\line(0,1){0.22}}
\multiput(42.09,48.9)(0.11,0.22){2}{\line(0,1){0.22}}
\multiput(41.87,48.45)(0.11,0.22){2}{\line(0,1){0.22}}
\multiput(41.67,48)(0.1,0.23){2}{\line(0,1){0.23}}
\multiput(41.48,47.54)(0.1,0.23){2}{\line(0,1){0.23}}
\multiput(41.29,47.07)(0.09,0.23){2}{\line(0,1){0.23}}
\multiput(41.12,46.61)(0.17,0.47){1}{\line(0,1){0.47}}
\multiput(40.96,46.13)(0.16,0.47){1}{\line(0,1){0.47}}
\multiput(40.82,45.66)(0.15,0.48){1}{\line(0,1){0.48}}
\multiput(40.68,45.18)(0.14,0.48){1}{\line(0,1){0.48}}
\multiput(40.56,44.69)(0.12,0.48){1}{\line(0,1){0.48}}
\multiput(40.45,44.21)(0.11,0.49){1}{\line(0,1){0.49}}
\multiput(40.35,43.72)(0.1,0.49){1}{\line(0,1){0.49}}
\multiput(40.26,43.23)(0.09,0.49){1}{\line(0,1){0.49}}
\multiput(40.19,42.73)(0.07,0.49){1}{\line(0,1){0.49}}
\multiput(40.13,42.24)(0.06,0.49){1}{\line(0,1){0.49}}
\multiput(40.08,41.74)(0.05,0.5){1}{\line(0,1){0.5}}
\multiput(40.04,41.25)(0.04,0.5){1}{\line(0,1){0.5}}
\multiput(40.01,40.75)(0.02,0.5){1}{\line(0,1){0.5}}
\multiput(40,40.25)(0.01,0.5){1}{\line(0,1){0.5}}
\put(40,39.75){\line(0,1){0.5}}
\multiput(40,39.75)(0.01,-0.5){1}{\line(0,-1){0.5}}
\multiput(40.01,39.25)(0.02,-0.5){1}{\line(0,-1){0.5}}
\multiput(40.04,38.75)(0.04,-0.5){1}{\line(0,-1){0.5}}
\multiput(40.08,38.26)(0.05,-0.5){1}{\line(0,-1){0.5}}
\multiput(40.13,37.76)(0.06,-0.49){1}{\line(0,-1){0.49}}
\multiput(40.19,37.27)(0.07,-0.49){1}{\line(0,-1){0.49}}
\multiput(40.26,36.77)(0.09,-0.49){1}{\line(0,-1){0.49}}
\multiput(40.35,36.28)(0.1,-0.49){1}{\line(0,-1){0.49}}
\multiput(40.45,35.79)(0.11,-0.49){1}{\line(0,-1){0.49}}
\multiput(40.56,35.31)(0.12,-0.48){1}{\line(0,-1){0.48}}
\multiput(40.68,34.82)(0.14,-0.48){1}{\line(0,-1){0.48}}
\multiput(40.82,34.34)(0.15,-0.48){1}{\line(0,-1){0.48}}
\multiput(40.96,33.87)(0.16,-0.47){1}{\line(0,-1){0.47}}
\multiput(41.12,33.39)(0.17,-0.47){1}{\line(0,-1){0.47}}
\multiput(41.29,32.93)(0.09,-0.23){2}{\line(0,-1){0.23}}
\multiput(41.48,32.46)(0.1,-0.23){2}{\line(0,-1){0.23}}
\multiput(41.67,32)(0.1,-0.23){2}{\line(0,-1){0.23}}
\multiput(41.87,31.55)(0.11,-0.22){2}{\line(0,-1){0.22}}
\multiput(42.09,31.1)(0.11,-0.22){2}{\line(0,-1){0.22}}
\multiput(42.32,30.65)(0.12,-0.22){2}{\line(0,-1){0.22}}
\multiput(42.56,30.22)(0.12,-0.22){2}{\line(0,-1){0.22}}
\multiput(42.81,29.78)(0.13,-0.21){2}{\line(0,-1){0.21}}
\multiput(43.07,29.36)(0.14,-0.21){2}{\line(0,-1){0.21}}
\multiput(43.34,28.94)(0.14,-0.21){2}{\line(0,-1){0.21}}
\multiput(43.62,28.53)(0.15,-0.2){2}{\line(0,-1){0.2}}
\multiput(43.91,28.12)(0.1,-0.13){3}{\line(0,-1){0.13}}
\multiput(44.21,27.73)(0.1,-0.13){3}{\line(0,-1){0.13}}
\multiput(44.52,27.34)(0.11,-0.13){3}{\line(0,-1){0.13}}
\multiput(44.84,26.95)(0.11,-0.12){3}{\line(0,-1){0.12}}
\multiput(45.17,26.58)(0.11,-0.12){3}{\line(0,-1){0.12}}
\multiput(45.51,26.21)(0.12,-0.12){3}{\line(0,-1){0.12}}
\multiput(45.86,25.86)(0.12,-0.12){3}{\line(1,0){0.12}}
\multiput(46.21,25.51)(0.12,-0.11){3}{\line(1,0){0.12}}
\multiput(46.58,25.17)(0.12,-0.11){3}{\line(1,0){0.12}}
\multiput(46.95,24.84)(0.13,-0.11){3}{\line(1,0){0.13}}
\multiput(47.34,24.52)(0.13,-0.1){3}{\line(1,0){0.13}}
\multiput(47.73,24.21)(0.13,-0.1){3}{\line(1,0){0.13}}
\multiput(48.12,23.91)(0.2,-0.15){2}{\line(1,0){0.2}}
\multiput(48.53,23.62)(0.21,-0.14){2}{\line(1,0){0.21}}
\multiput(48.94,23.34)(0.21,-0.14){2}{\line(1,0){0.21}}
\multiput(49.36,23.07)(0.21,-0.13){2}{\line(1,0){0.21}}
\multiput(49.78,22.81)(0.22,-0.12){2}{\line(1,0){0.22}}
\multiput(50.22,22.56)(0.22,-0.12){2}{\line(1,0){0.22}}
\multiput(50.65,22.32)(0.22,-0.11){2}{\line(1,0){0.22}}
\multiput(51.1,22.09)(0.22,-0.11){2}{\line(1,0){0.22}}
\multiput(51.55,21.87)(0.23,-0.1){2}{\line(1,0){0.23}}
\multiput(52,21.67)(0.23,-0.1){2}{\line(1,0){0.23}}
\multiput(52.46,21.48)(0.23,-0.09){2}{\line(1,0){0.23}}
\multiput(52.93,21.29)(0.47,-0.17){1}{\line(1,0){0.47}}
\multiput(53.39,21.12)(0.47,-0.16){1}{\line(1,0){0.47}}
\multiput(53.87,20.96)(0.48,-0.15){1}{\line(1,0){0.48}}
\multiput(54.34,20.82)(0.48,-0.14){1}{\line(1,0){0.48}}
\multiput(54.82,20.68)(0.48,-0.12){1}{\line(1,0){0.48}}
\multiput(55.31,20.56)(0.49,-0.11){1}{\line(1,0){0.49}}
\multiput(55.79,20.45)(0.49,-0.1){1}{\line(1,0){0.49}}
\multiput(56.28,20.35)(0.49,-0.09){1}{\line(1,0){0.49}}
\multiput(56.77,20.26)(0.49,-0.07){1}{\line(1,0){0.49}}
\multiput(57.27,20.19)(0.49,-0.06){1}{\line(1,0){0.49}}
\multiput(57.76,20.13)(0.5,-0.05){1}{\line(1,0){0.5}}
\multiput(58.26,20.08)(0.5,-0.04){1}{\line(1,0){0.5}}
\multiput(58.75,20.04)(0.5,-0.02){1}{\line(1,0){0.5}}
\multiput(59.25,20.01)(0.5,-0.01){1}{\line(1,0){0.5}}
\put(59.75,20){\line(1,0){0.5}}
\multiput(60.25,20)(0.5,0.01){1}{\line(1,0){0.5}}
\multiput(60.75,20.01)(0.5,0.02){1}{\line(1,0){0.5}}
\multiput(61.25,20.04)(0.5,0.04){1}{\line(1,0){0.5}}
\multiput(61.74,20.08)(0.5,0.05){1}{\line(1,0){0.5}}
\multiput(62.24,20.13)(0.49,0.06){1}{\line(1,0){0.49}}
\multiput(62.73,20.19)(0.49,0.07){1}{\line(1,0){0.49}}
\multiput(63.23,20.26)(0.49,0.09){1}{\line(1,0){0.49}}
\multiput(63.72,20.35)(0.49,0.1){1}{\line(1,0){0.49}}
\multiput(64.21,20.45)(0.49,0.11){1}{\line(1,0){0.49}}
\multiput(64.69,20.56)(0.48,0.12){1}{\line(1,0){0.48}}
\multiput(65.18,20.68)(0.48,0.14){1}{\line(1,0){0.48}}
\multiput(65.66,20.82)(0.48,0.15){1}{\line(1,0){0.48}}
\multiput(66.13,20.96)(0.47,0.16){1}{\line(1,0){0.47}}
\multiput(66.61,21.12)(0.47,0.17){1}{\line(1,0){0.47}}
\multiput(67.07,21.29)(0.23,0.09){2}{\line(1,0){0.23}}
\multiput(67.54,21.48)(0.23,0.1){2}{\line(1,0){0.23}}
\multiput(68,21.67)(0.23,0.1){2}{\line(1,0){0.23}}
\multiput(68.45,21.87)(0.22,0.11){2}{\line(1,0){0.22}}
\multiput(68.9,22.09)(0.22,0.11){2}{\line(1,0){0.22}}
\multiput(69.35,22.32)(0.22,0.12){2}{\line(1,0){0.22}}
\multiput(69.78,22.56)(0.22,0.12){2}{\line(1,0){0.22}}
\multiput(70.22,22.81)(0.21,0.13){2}{\line(1,0){0.21}}
\multiput(70.64,23.07)(0.21,0.14){2}{\line(1,0){0.21}}
\multiput(71.06,23.34)(0.21,0.14){2}{\line(1,0){0.21}}
\multiput(71.47,23.62)(0.2,0.15){2}{\line(1,0){0.2}}
\multiput(71.88,23.91)(0.13,0.1){3}{\line(1,0){0.13}}
\multiput(72.27,24.21)(0.13,0.1){3}{\line(1,0){0.13}}
\multiput(72.66,24.52)(0.13,0.11){3}{\line(1,0){0.13}}
\multiput(73.05,24.84)(0.12,0.11){3}{\line(1,0){0.12}}
\multiput(73.42,25.17)(0.12,0.11){3}{\line(1,0){0.12}}
\multiput(73.79,25.51)(0.12,0.12){3}{\line(1,0){0.12}}
\multiput(74.14,25.86)(0.12,0.12){3}{\line(0,1){0.12}}
\multiput(74.49,26.21)(0.11,0.12){3}{\line(0,1){0.12}}
\multiput(74.83,26.58)(0.11,0.12){3}{\line(0,1){0.12}}
\multiput(75.16,26.95)(0.11,0.13){3}{\line(0,1){0.13}}
\multiput(75.48,27.34)(0.1,0.13){3}{\line(0,1){0.13}}
\multiput(75.79,27.73)(0.1,0.13){3}{\line(0,1){0.13}}
\multiput(76.09,28.12)(0.15,0.2){2}{\line(0,1){0.2}}
\multiput(76.38,28.53)(0.14,0.21){2}{\line(0,1){0.21}}
\multiput(76.66,28.94)(0.14,0.21){2}{\line(0,1){0.21}}
\multiput(76.93,29.36)(0.13,0.21){2}{\line(0,1){0.21}}
\multiput(77.19,29.78)(0.12,0.22){2}{\line(0,1){0.22}}
\multiput(77.44,30.22)(0.12,0.22){2}{\line(0,1){0.22}}
\multiput(77.68,30.65)(0.11,0.22){2}{\line(0,1){0.22}}
\multiput(77.91,31.1)(0.11,0.22){2}{\line(0,1){0.22}}
\multiput(78.13,31.55)(0.1,0.23){2}{\line(0,1){0.23}}
\multiput(78.33,32)(0.1,0.23){2}{\line(0,1){0.23}}
\multiput(78.52,32.46)(0.09,0.23){2}{\line(0,1){0.23}}
\multiput(78.71,32.93)(0.17,0.47){1}{\line(0,1){0.47}}
\multiput(78.88,33.39)(0.16,0.47){1}{\line(0,1){0.47}}
\multiput(79.04,33.87)(0.15,0.48){1}{\line(0,1){0.48}}
\multiput(79.18,34.34)(0.14,0.48){1}{\line(0,1){0.48}}
\multiput(79.32,34.82)(0.12,0.48){1}{\line(0,1){0.48}}
\multiput(79.44,35.31)(0.11,0.49){1}{\line(0,1){0.49}}
\multiput(79.55,35.79)(0.1,0.49){1}{\line(0,1){0.49}}
\multiput(79.65,36.28)(0.09,0.49){1}{\line(0,1){0.49}}
\multiput(79.74,36.77)(0.07,0.49){1}{\line(0,1){0.49}}
\multiput(79.81,37.27)(0.06,0.49){1}{\line(0,1){0.49}}
\multiput(79.87,37.76)(0.05,0.5){1}{\line(0,1){0.5}}
\multiput(79.92,38.26)(0.04,0.5){1}{\line(0,1){0.5}}
\multiput(79.96,38.75)(0.02,0.5){1}{\line(0,1){0.5}}
\multiput(79.99,39.25)(0.01,0.5){1}{\line(0,1){0.5}}

\linethickness{0.3mm}
\multiput(26,6)(0.12,0.12){167}{\line(1,0){0.12}}
\linethickness{0.3mm}
\multiput(74,26)(0.12,-0.12){167}{\line(1,0){0.12}}
\linethickness{0.3mm}
\put(60,60){\line(0,1){20}}
\put(59,40){\makebox(0,0)[cc]{$V^{(3)}$}}

\put(53,70){\makebox(0,0)[cc]{$\mu,\nu$}}

\put(75,70){\makebox(0,0)[cc]{$-k_1-k_2$}}

\put(29,20){\makebox(0,0)[cc]{$\ve_1$}}

\put(40,11){\makebox(0,0)[cc]{$k_1$}}

\put(75,15){\makebox(0,0)[cc]{$\ve_2$}}

\put(87,20){\makebox(0,0)[cc]{$k_2$}}

\end{picture}

}

\def\figten{

\def\JPicScale{0.5}
\ifx\JPicScale\undefined\def\JPicScale{1}\fi
\unitlength \JPicScale mm

}

\begin{document}

\baselineskip 24pt

\begin{center}
{\Large \bf  Subleading Soft Theorem for  Multiple Soft Gravitons}

\end{center}

\vskip .6cm
\medskip

\vspace*{4.0ex}

\baselineskip=18pt

\centerline{\rm Subhroneel Chakrabarti$^{a}$, Sitender Pratap 
Kashyap$^{a}$,  Biswajit Sahoo$^{a}$, Ashoke Sen$^{a}$, 
Mritunjay Verma$^{a,b}$}

\vspace*{4.0ex}

\centerline{\large \it $^a$Harish-Chandra Research Institute, HBNI}
\centerline{\large \it  Chhatnag Road, Jhusi,
Allahabad 211019, India}



\medskip

\centerline{\large \it  $^b$International Centre for Theoretical Sciences}
\centerline{\large \it  
Hesarghatta,
Bengaluru - 560 089, India.}

\vspace*{1.0ex}
\centerline{\small E-mail:  subhroneelchak, sitenderpratap, biswajitsahoo, sen, mritunjayverma @hri.res.in}

\vspace*{5.0ex}

\centerline{\bf Abstract} \bigskip

We derive the subleading soft graviton theorem in a generic quantum theory of
gravity for arbitrary number of soft external gravitons and arbitrary number of finite
energy external states carrying arbitrary mass and spin. Our results are valid to all
orders in perturbation theory when the number of non-compact space-time dimensions 
is six or more, but only for tree amplitudes for five or less non-compact space-time
dimensions due to enhanced contribution to loop amplitudes from the infrared region.

\vspace*{5.0ex}

\vfill \eject

\baselineskip=18pt

\tableofcontents

\sectiono{Introduction} \label{s1}

Soft graviton theorem expresses the scattering amplitude of finite energy
external states and low energy gravitons in terms of the amplitude without
the low energy gravitons\cite{weinberg1,weinberg2,jackiw1,jackiw2}.  They
have been investigated intensively during the last few 
years\cite{1103.2981,1404.4091,1404.7749,1405.1015,
1405.1410,1405.2346,
1405.3413,1405.3533,1406.6574,1406.6987,1406.7184,1407.5936,
1407.5982,1408.4179,1410.6406,1412.3699,1504.01364,1507.08882,
1509.07840,1604.00650,1604.03893,1607.02700,1611.02172,1611.07534,1611.03137,
1702.02350}
due to their connection to asymptotic symmetries\cite{1312.2229,1401.7026,
1411.5745,1506.05789,1509.01406,1605.09094,1608.00685,1612.08294,1701.00496,
1612.05886,1703.05448}. 
They have also been investigated in string theory\cite{ademollo,shapiro,
1406.4172,1406.5155,1411.6661,1502.05258,1505.05854,1507.08829,
1511.04921,
1512.00803,1601.03457,1604.03355,1610.03481,1702.03934,1703.00024}.
In particular in 
specific quantum field theories and string theories, amplitudes with several finite
energy external states and one soft graviton have been analyzed to subsubleading
order, leading to the subsubleading soft graviton theorem in these theories.
A general proof of the soft graviton theorem in a generic quantum theory of gravity
was given in \cite{1702.03934,1703.00024,1706.00759} for 
one external soft graviton and arbitrary number of other finite energy
external states carrying arbitrary mass and spin.  

For specific theories, soft graviton
amplitudes with two soft gravitons have also been investigated in 
\cite{1503.04816,1504.05558,1504.05559,1507.00938,1604.02834,1607.02700,1702.02350,
1705.06175}.
Our goal in this paper will be to derive,  in a generic quantum theory of gravity, 
the form of the soft graviton theorem to the first
subleading order in soft momentum for arbitrary number of soft gravitons and
for arbitrary number of finite energy external states
carrying arbitrary mass and spin. The limit we
consider is when all the soft momenta become small at the same rate.
As discussed in section \ref{sIR}, in order to avoid 
enhanced contribution to loop diagrams from the infrared region, 
we shall restrict our analysis to the case where
the number of non-compact space-time dimensions $D$ is six or more. For $D\le 5$ our
analysis will be valid for tree amplitudes. We expect that even in D=5, where the
amplitudes are infrared finite, the enhanced infrared contributions of the type described
in section \ref{sIR} will cancel in the sum over graphs 
and our result will be valid also for $D=5$ to all loop orders. However, we have not
proved this yet.

Our final result for an amplitude with $N$ external finite energy particles carrying  polarizations and
momenta $(\eps_i, p_i)$ for $i=1,\cdots, N$, and $M$ soft gravitons carrying polarizations
and momenta $(\ve_r, k_r)$ for $r=1,\cdots, M$, takes the form
\ben \label{efullgenintro}
A &=& \Bigg\{ \prod_{i=1}^N \eps_{i,\alpha_i}(p_i) \Bigg\} \ \Bigg[
\Bigg\{\prod_{r=1}^M \, S^{(0)}_r \Bigg\} \ 
\ \Gamma^{\alpha_1\cdots \alpha_N}
+ \sum_{s=1}^M \Bigg\{\prod_{r=1\atop r\ne s}^M \, S^{(0)}_r \Bigg\}\ 
\left[S^{(1)}_s \Gamma\right]^{\alpha_1\cdots \alpha_N} 
\nonumber \\ &&  \hskip -.5in
+ \sum_{r,u=1\atop r<u}^M 
\Bigg\{ \prod_{s=1\atop s\ne r,u}^M \, S^{(0)}_s \Bigg\} 
\ \Bigg\{\sum_{j=1}^N \ \{p_j\cdot (k_r+k_u)\}^{-1}\ 
\ \MM(p_j; \ve_r, k_r, \ve_u, k_u) \Bigg\}  \
\Gamma^{\alpha_1\cdots  \alpha_N} \Bigg]\,  , 
\een
where
\be \label{defs0}
S^{(0)}_r = \sum_{\ell=1}^N (p_\ell\cdot k_r)^{-1}
\ \ve_{r,\mu\nu} \, p_\ell^\mu  p_{\ell}^\nu  \, ,
\ee
\be \label{edefs1}
[S^{(1)}_s \Gamma]^{\alpha_1\cdots \alpha_N}=
\sum_{j=1}^N (p_j\cdot k_s)^{-1} 
\, \ve_{s,b\mu}  \,  k_{s a} \, 
p_j^\mu\
\left[ p_j^b 
{\p \Gamma^{\alpha_1\cdots\alpha_N} \over \p p_{ja}} - p_j^a 
{\p \Gamma^{\alpha_1\cdots\alpha_N} \over \p p_{jb}}
+   (J^{ab})_{\beta_j}^{~\alpha_j}
\Gamma^{\alpha_1\cdots \alpha_{j-1} \beta_j \alpha_{j+1} \cdots \alpha_N}\right],
\ee
\ben \label{edefmiintro}
&& \MM(p_i; \ve_1, k_1, \ve_2, k_2) \nonumber \\
&=&
(p_i\cdot k_1)^{-1}  (p_i\cdot k_2)^{-1} 
 \ \Bigg\{-k_1\cdot k_2 \ p_i\cdot \ve_1\cdot p_i \ 
 p_i\cdot \ve_2\cdot p_i
 \nonumber \\ && \hskip -.3in
 + \ 2 \ p_i\cdot k_2 \ p_i\cdot \ve_1\cdot p_i \ p_i\cdot \ve_2\cdot k_1 
  + 2 \ p_i\cdot k_1 \ p_i\cdot \ve_2\cdot p_i \ p_i\cdot \ve_1\cdot k_2
  - 2  \ p_i\cdot k_1 \ p_i\cdot k_2 \ p_i\cdot \ve_1\cdot \ve_2\cdot p_i\Bigg\}
 \nonumber \\ &&  \hskip -.3in
 +\ (k_1\cdot k_2)^{-1}
 \Bigg\{-(k_2\cdot\ve_{1}\cdot\ve_{2}\cdot p_i)(k_2\cdot p_i) 
 -(k_1\cdot\ve_{2}\cdot\ve_{1}\cdot p_i)(k_1\cdot p_i)  \nonumber\\
&&+\ (k_2\cdot\ve_{1}\cdot\ve_{2}\cdot p_i)(k_1\cdot p_i)
+(k_1\cdot\ve_{2}\cdot\ve_{1}\cdot p_i)(k_2\cdot p_i)  - \ve_1^{\gamma\delta}\ve_{2\gamma\delta}(k_{1} \cdot p_i)( k_{2}\cdot p_i)  \nonumber\\
&& -\ 2(p_i\cdot\ve_{1}\cdot k_2)(p_i\cdot\ve_{2}\cdot k_1) 
+ (p_i\cdot\ve_{2}\cdot p_i)(k_2\cdot\ve_{1}\cdot k_2)  
+ (p_i\cdot\ve_{1}\cdot p_i)(k_1\cdot\ve_{2}\cdot k_1)\Bigg\}\, ,
\nonumber\\
\een
and $\Gamma^{\alpha_1\cdots\alpha_N}$ is defined such that 
\be 
\Gamma(\eps_1, p_1, \cdots , \eps_N, p_N) \equiv 
\left\{\prod_{i=1}^N \eps_{i,\alpha_i}\right\}
\Gamma^{\alpha_1\cdots\alpha_N}\, ,
\ee 
gives the amplitude without the soft gravitons,
including the momentum conserving delta function. The indices $\alpha,\beta,\gamma,
\delta$ run
over all the fields of the theory and $J^{ab}$ is the (reducible) representation of the spin angular
momentum generator on the fields. The indices $a,b$ as well as $\mu,\nu,\rho$ are
space-time coordinate / momentum labels. We shall use Einstein summation convention
for the indices $\alpha,\beta,\cdots$ carried by the fields and also for the space-time
coordinate labels $a,b\cdots$ and $\mu,\nu,\cdots$, but not for the indices $r,s,\cdots$
labelling the external soft gravitons and $i,j,\cdots$ labelling the external finite energy
particles. For the signature of the space-time metric we shall use mostly + sign 
convention.

The rest of the paper is organized as follows. In section \ref{s2} we prove the subleading 
soft graviton theorem for two external soft gravitons and arbitrary number of external
states of arbitrary mass and spin. In section \ref{s3} we carry out various consistency
checks of this formula. These include test of gauge invariance and also comparison 
with existing results. In particular we find that neither the first nor the second line of
\refb{efullgenintro} is gauge invariant by itself but their sum is gauge invariant.
We generalize the result to the case of multiple soft gravitons in
section \ref{s4}. 

Derivation of double soft theorem from asymptotic symmetries has been pursued in \cite{progress}.

\sectiono{Amplitudes with two soft gravitons} \label{s2}

\begin{figure}
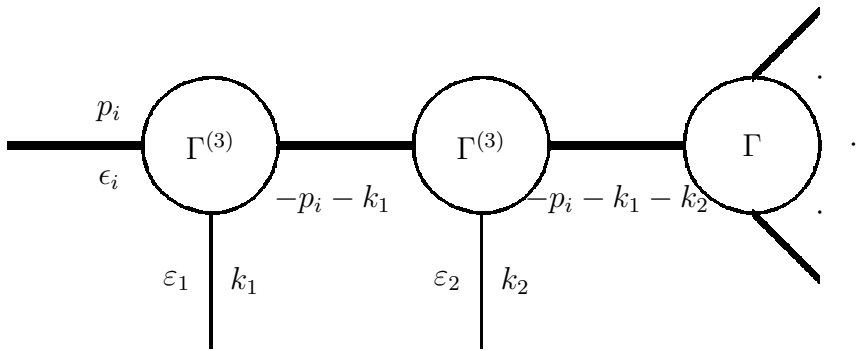


\begin{center}

\figtwo

\end{center}

\vskip -.5in

\caption{A leading contribution to the amplitude with two soft gravitons.
\label{fig2}}

\end{figure}

\begin{figure}

\begin{center}

\figseven

\end{center}

\caption{A leading contribution to the amplitude with two soft gravitons.
\label{fig7}}

\end{figure}

\begin{figure}
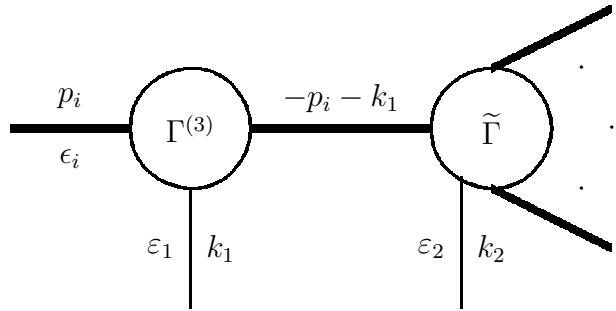


\begin{center}

\figthree

\end{center}

\vskip -.5in

\caption{A subleading contribution to the amplitude with two soft gravitons. The subamplitude 
$\wt\Gamma$ excludes all diagrams where the soft particle carrying momentum $k_2$ 
gets attached to one of
external lines of $\wt\Gamma$.
\label{fig3}}
\end{figure}

In this section we shall analyze an amplitude with arbitrary number of finite energy external
states and two soft gravitons in the limit when the momenta carried by the soft gravitons
become soft at the same rate. 
The relevant diagrams are shown in Figs.~\ref{fig2}-\ref{fig6}. 
We use the convention that all external momenta are ingoing, 
thick lines represent finite energy propagators and thin
lines represent soft propagators. 
$\ve_r,k_r$ for $r=1,2$ represent the polarizations and momenta carried by the soft
gravitons subject to the constraint
\be \label{epol}
\eta^{\mu\nu} \, \ve_{r,\mu\nu}=0, \quad k_r^\mu \, \ve_{r,\mu\nu}=0\, .
\ee
$\Gamma^{(3)}$ and $\Gamma^{(4)}$ denote one particle irreducible (1PI) three and
four point functions and $\Gamma$ denotes full amputated Green's function. 
In Fig.~\ref{fig3},
$\wt\Gamma$ denotes sum of all amputated Feynman diagrams in which the soft graviton
is not attached to an external leg via a 1PI three point function. The internal thick lines of
the diagrams represent full quantum corrected propagators carrying finite momentum.
For Figs.~\ref{fig2}
and \ref{fig3} we also have to consider diagrams where the two soft gravitons are 
exchanged.

Among these diagrams the contributions from Fig.~\ref{fig2} and Fig.~\ref{fig7} 
have two nearly on-shell propagators giving two powers of soft momentum in the
denominators. For example in Fig.~\ref{fig2} the line carrying momentum $p_i+k_1$
is proportional to
\be 
\{ (p_i+k_1)^2 + M_i^2\}^{-1} = (2p_i\cdot k_1)^{-1}\, ,
\ee
using the on-shell condition $k_1^2=0$, $p_i^2+M_i^2=0$ if the mass of the internal
state is the same as the mass of the $i$-th external state.
Therefore the contribution from these diagrams 
begins at the leading order.
The rest of the diagrams have only one nearly on-shell propagator and therefore their
contribution begins at the subleading order. The contribution from Fig.~\ref{fig6} is
somewhat deceptive -- it appears to have one nearly on-shell propagator carrying finite 
energy giving one power of soft momentum in the denominator and a soft internal 
propagator giving two powers of soft momentum in the denominator. However the
three graviton vertex  has two powers of soft momentum in the numerator. Therefore
the contribution from this diagram begins with one inverse power of soft momentum and
is subleading.

\begin{figure}

\begin{center}

\figfive

\end{center}

\vskip -.5in

\caption{A subleading contribution to the amplitude with two soft gravitons.
\label{fig5}}

\end{figure}

\begin{figure}

\begin{center}

\figsix

\end{center}

\caption{A subleading contribution to the amplitude with two soft gravitons.
\label{fig6}}

\end{figure}

\subsection{Expressions for the vertices and propagators}

Our strategy for deriving the vertices will be the same as that 
in \cite{1702.03934,1703.00024,1706.00759}. We begin with
the 1PI effective action of the theory and use Lorentz covariant gauge fixing conditions
such that the propagators computed from this gauge fixed action do not have double poles.
We now find the coupling of the soft graviton to the rest of the fields by covariantizing
this action. As in \cite{1703.00024,1706.00759} 
we shall assume that all the fields carry tangent space indices so that
covariantization corresponds to replacing ordinary derivatives by covariant derivatives and
then converting the tensor indices arising from derivatives to tangent space indices by
contraction with inverse vielbeins. For simplicity we shall choose a gauge in which 
the metric always has determinant $-1$ so that we do not need to worry about the 
multiplicative factor of $\sqrt{-\det g}$ while covariantizing the action. This is done by
parametrizing the metric as
\be 
g_{\mu\nu} = \left(e^{2S\eta}\, \eta\right)_{\mu\nu} = \eta_{\mu\nu} + 2\, S_{\mu\nu} + 2\, 
S_{\mu\rho} S^{\rho}_{~\nu} + \cdots, \quad S_{\mu\nu} = S_{\nu\mu}, \quad 
S_\mu^{~\mu}=0 \, ,
\ee
where all indices are raised and lowered by the flat metric $\eta$. We also introduce the
vielbein fields
\be
e_\mu^{~a} = \left(e^{S\eta}\right)_{\mu}^{~a} = \delta_\mu^{~a}
+ S_\mu^{~a} + {1\over 2}
S_{\mu}^{~b} S_{b}^{~a} +\cdots, \quad E_a^{~\mu} = 
\left(e^{-S\eta}\right)_{a}^{~\mu} = \delta_a^{~\mu}
- S_a^{~\mu} + {1\over 2}
S_{a}^{~b} S_{b}^{~\mu} +\cdots \, .
\ee
Covariantization of the action now involves the following step. 
Let $\{\phi_\alpha\}$ denote the set of all the fields of the theory.
We replace a chain of ordinary derivatives $\p_{a_1}\cdots \p_{a_n}$ acting
on a field $\phi_\alpha$ by
\be \label{ecov}
E_{a_1}^{~\mu_1}\cdots E_{a_n}^{~\mu_n} D_{\mu_1}\cdots D_{\mu_n}
\ee
where 
\be \label{ecov2}
D_\mu \phi_\alpha = \p_\mu \phi_\alpha + {1\over 2} 
\omega_\mu^{ab} (J_{ab})_\alpha^{~\beta} \phi_\beta\, ,
\ee
with $(J_{ab})_\alpha^{~\beta}$ representing the action of 
spin angular momentum generator on all the fields, normalized so  that acting on a covariant vector field
$\phi_c$, we have
\be 
(J^{ab})_c^{~d} = \delta^a_{~c} \eta^{bd} - \delta^b_{~c} \eta^{ad}\, .
\ee
 For our analysis we shall only need
the expression for $\omega_{\mu}^{ab}$ to first order in $S_{\mu\nu}$. This is given 
by\footnote{Terms involving higher powers of $S$ will give rise to vertices that 
have two or more soft gravitons, {\it and} a power of soft momentum. Such vertices will not
contribute to the amplitude to subleading order in soft momentum.}
\be \label{ecov3}
\omega_\mu^{ab} = \p^b S_\mu^{~a} - \p^a S_\mu^{~b}\, .
\ee
For each pair of covariant derivatives acting on the field $\phi_\alpha$, we also have
a contribution from the Christoffel symbol
\be  \label{ecov4}
D_{\mu} D_\nu \phi_\alpha = \cdots -\left\{ {\rho\atop \mu ~ \nu} \right\} D_\rho\, 
\phi_\alpha\, ,
\ee
where 
\be
\left\{ {\rho\atop \mu ~ \nu} \right\}= \p_\mu S_{\nu}^{~\rho} + \p_\nu S_{\mu}^{~\rho}
- \p^\rho S_{\mu\nu} + \hbox{terms involving quadratic and higher powers of $S$}\, ,
\ee
and $\cdots$ terms represent the usual derivatives and spin connection term.
Since we shall compute subleading soft graviton amplitudes we shall only keep terms
up to first order in the derivatives of soft gravitons. Also for amplitudes with two soft
gravitons
we only need to keep up to terms with two powers of soft graviton field $S_{\mu\nu}$.
As we shall see, for specific vertices we can make further truncation of the action.

\begin{figure}
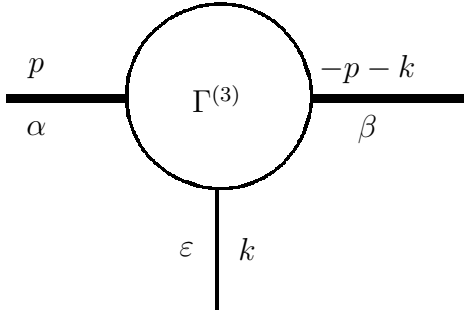


\begin{center}

\figone

\end{center}

\vskip -.7in

\caption{A 1PI vertex involving two finite energy particles and one soft particle.
\label{fig1}}

\end{figure}

Let us now derive the form of the three point vertex involving one soft graviton and
two finite energy fields, as shown in Fig.~\ref{fig1}. 
For this we first express the quadratic part of the 1PI action as
\be \label{equad}
{1\over 2} \ 
\int {d^D q_1\over (2\pi)^D} \, {d^D q_2\over (2\pi)^D}\, (2\pi)^D \delta^{(D)}(q_1+q_2)
\phi_\alpha(q_1) \, \KK^{\alpha\beta}(q_2) \, \phi_\beta(q_2)\, ,
\ee
where we take
\be \label{esym}
\KK^{\alpha\beta}(q)=\KK^{\beta\alpha}(-q)\, .
\ee
For grassmann odd fields there will be an extra minus sign in this equation, but it does
not affect the final results.
If the soft graviton carries polarization $\ve$ and momentum $k$, then the coupling
of single soft graviton to the fields $\phi_\alpha$, obtained by covariantizing \refb{equad},
takes the form\cite{1706.00759}
\ben \label{ethreept}
S^{(3)} &=& {1\over 2} \int {d^D q_1\over (2\pi)^D} \, {d^D q_2\over (2\pi)^D} \,  (2\pi)^D \delta^{(D)}(q_1+q_2+k)
\nonumber \\ && \times \Phi_\alpha(q_1) 
\Bigg[ - \ve_{\mu \nu} q_2^\nu  {\p\over \p q_{2\mu}}\KK^{\alpha\beta} (q_2)
+ {1\over 2} (k_b \, \ve_{a\mu} - k_a \, \ve_{b\mu}) {\p\over \p q_{2\mu}}\KK^{\alpha\gamma}(q_2) 
\left(J^{ab}\right)_{\gamma}^{~\beta}  \nonumber \\ &&
- {1\over 2} {\p^2 \KK^{\alpha\beta}(q_2)\over \p q_{2\mu} \p q_{2\nu}}
q_{2\rho} \left(k_\mu \ve_\nu^{~\rho} + 
k_\nu \ve_\mu^{~\rho} - k^\rho \ve_{\mu\nu}\right) \Bigg] \Phi_\beta(q_2)\, .
\een
In this equation the first term inside the square bracket represents the effect of 
multiplication by $E_a^{~\mu} = \delta_a^{~\mu} - S_a^{~\mu}$ in \refb{ecov}. The
second term is the effect of the spin connection \refb{ecov3} 
appearing in the definition of the
covariant derivative in \refb{ecov2} and the third term is the effect of the Christoffel
symbol appearing in \refb{ecov4}. From this we can derive an expression for the
soft graviton vertex shown in Fig.~\ref{fig1} to order $k$:
\ben \label{evertex}
&& \Gamma^{(3)\alpha\beta}(\ve, k; p, -p-k) \nonumber \\
&=& {i\over 2} \Bigg[- \ve_{\mu \nu} (p+k)^\nu  {\p\over \p p_{\mu}}\KK^{\alpha\beta} (-p-k)
- \ve_{\mu \nu} p^\nu  {\p\over \p p_{\mu}}\KK^{\beta\alpha} (p)
\nonumber \\ &&
+ {1\over 2} (k_a \, \ve_{b\mu} - k_b \, \ve_{a\mu}) {\p\over \p p_{\mu}}\KK^{\alpha\gamma}(-p-k) 
\left(J^{ab}\right)_{\gamma}^{~\beta} 
- {1\over 2} (k_a \, \ve_{b\mu} - k_b \, \ve_{a\mu}) {\p\over \p p_{\mu}}\KK^{\beta\gamma}(p) 
\left(J^{ab}\right)_{\gamma}^{~\alpha}  \nonumber \\ &&
-{1\over 2} {\p^2 \KK^{\alpha\beta}(-p-k)\over \p p_{\mu} \p p_{\nu}}
(-p_{\rho}-k_\rho) \left(k_\mu \ve_\nu^{~\rho} + 
k_\nu \ve_\mu^{~\rho} - k^\rho \ve_{\mu\nu}\right) 
\nonumber \\ &&
-{1\over 2} {\p^2 \KK^{\beta\alpha}(p)\over \p p_{\mu} \p p_{\nu}}
p_{\rho} \left(k_\mu \ve_\nu^{~\rho} + 
k_\nu \ve_\mu^{~\rho} - k^\rho \ve_{\mu\nu}\right) 
\Bigg] \, .
\een
Using \refb{esym},
\refb{epol}, and  expanding each term in Taylor series in the soft momentum $k$, we arrive
at the following expression for the vertex $\Gamma^{(3)}$ in Fig.~\ref{fig1} to order $k$:
\ben \label{egam3}
&& \Gamma^{(3)}(\ve, k; p, p-k) \nonumber \\
&=& i\, \bigg[ -\ve_{\mu\nu} p^\nu {\p \KK(-p)\over \p p_\mu}
-{1\over 2} \ve_{\mu\nu} p^\nu k_\rho {\p^2 \KK(-p) \over \p p_\mu \p p_\rho} 
+ {1\over 2} k_a \ve_{b\mu} {\p \KK(-p)\over \p p_\mu} J^{ab} 
- {1\over 2} k_a \ve_{b\mu} (J^{ab})^T {\p \KK(-p)\over \p p_\mu}\bigg] \, ,
\nonumber \\
\een
where we have used a matrix notation and 
$(J^{ab})^T$ denotes the transpose of $J^{ab}$, i.e. $((J^{ab})^T)^\alpha_{~\gamma}
= (J^{ab})_\gamma^{~\alpha}$. 

\begin{figure}
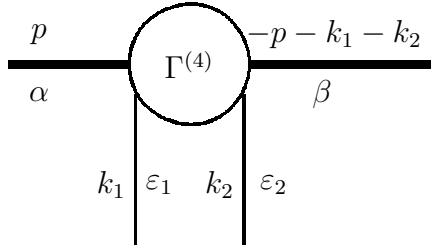


\begin{center}

\figeight

\end{center}

\vskip -.5in

\caption{A 1PI vertex involving two finite energy particles and two soft particles.
\label{fig8}}

\end{figure}

Next we consider the four point vertex containing two soft gravitons and two finite
energy particles as shown in Fig.~\ref{fig8}. Since this vertex appears in Fig.~\ref{fig5}
which begins contributing at the subleading order, we need to evaluate this to leading
power in the soft momentum. Therefore we can ignore the spin connection and Christoffel
symbol terms in the expression for the covariant derivatives appearing in \refb{ecov}, and
only focus on the contribution from the $E_a^{~\mu}$ terms. Since we have two soft gravitons,
we need to keep terms quadratic in the soft graviton field $S_{\mu\nu}$.  These can come
from two sources -- either one power of $S$ from two $E_a^{~\mu}$'s  or two powers of
$S$ from a single $E_a^{~\mu}$. The resulting action is given by
\ben
&& {1\over 2} \int {d^D q_1\over (2\pi)^D} \, {d^D q_2\over (2\pi)^D} \, 
{d^D \ell_1\over (2\pi)^D} \, {d^D \ell_2\over (2\pi)^D} \,
(2\pi)^D \delta^{(D)}(q_1+q_2+\ell_1+\ell_2) \Phi_\alpha(q_1)  \Phi_\beta(q_2) 
\nonumber \\ && \times 
\Bigg[ {1\over 2} S_{\mu \nu}(\ell_1) S_{\rho\sigma}(\ell_2) 
q_2^\nu  q_2^\sigma {\p^2 \KK^{\alpha\beta} (q_2)\over \p q_{2\mu}\p q_{2\rho}}
+{1\over 2} S_{\mu}^{~b} S_{b\nu} q_2^\nu {\p \KK^{\alpha\beta} (q_2)\over \p q_{2\mu}}
 \Bigg]\, .
\een
Using this and the symmetry \refb{esym}, 
we get the following form of the vertex shown in Fig.~\ref{fig8} to leading order in soft momenta,
written in the matrix notation:
\ben \label{eexpgam4}
&& \Gamma^{(4)}(\ve_1, k_1, \ve_2, k_2; p, -p-k_1-k_2)
\nonumber \\
&=& 
{i}\, \left[\ve_{1,\mu \nu}\ve_{2,\rho\sigma}
p^\nu  p^\sigma {\p^2 \KK (-p)\over \p p_{\mu}\p p_{\rho}}
+ {1\over 2} \left(\ve_{1,\mu}^{~~~b} \, \ve_{2,b\nu}
+ \ve_{2,\mu}^{~~~b} \, \ve_{1,b\nu}\right) \, p^\nu {\p \KK(-p)\over \p p_{\mu}}
\right]\, .
\een

\begin{figure}
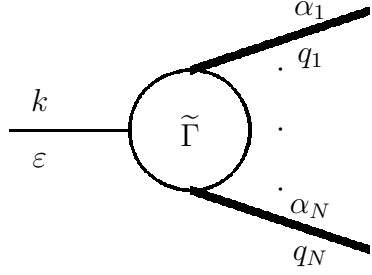


\begin{center}

\figfour

\end{center}

\vskip -.7in

\caption{An amputated amplitude with one external soft particle and many external finite 
energy particles. We exclude from this any diagram where the soft particle gets
attached to one of the external lines.
\label{fig4}}

\end{figure}

Next let us consider the contribution from the amplitude in
Fig.~\ref{fig4} for off-shell external momenta $q_1,\cdots , q_N$. 
This can be obtained by
covariantizing the truncated Green's function 
$\Gamma^{\alpha_1\cdots \alpha_N}(q_1,\cdots q_N)$ without the soft graviton.
Since this amplitude appears inside Fig.~\ref{fig3} which begins contributing at the
subleading order, we only need the leading contribution from this amplitude. This is easily
computed using the covariantization procedure, giving the result\cite{1703.00024}
\be \label{ecompver}
\wt\Gamma^{\alpha_1\cdots \alpha_N}(\ve, k; q_1,\cdots q_N) 
= - \sum_{i=1}^N \ve_{\mu\nu} q_i^\mu {\p\over \p q_{i\nu}} 
\Gamma^{\alpha_1\cdots \alpha_N}(q_1,\cdots , q_N)\, ,
\ee
reflecting the effect of having to multiply every factor of momentum (derivative with
respect to space-time coordinates) by inverse vielbeins as in \refb{ecov}.

\begin{figure}
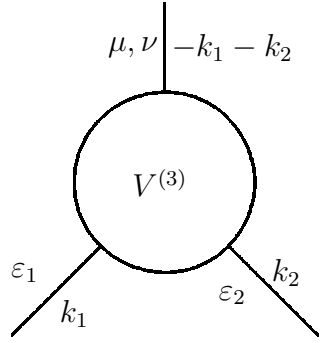


\begin{center}

\fignine

\end{center}

\vskip -.2in

\caption{A 1PI vertex involving three soft gravitons.
\label{fig9}}

\end{figure}

The next vertex to be evaluated is the three point vertex of three soft gravitons as shown in
Fig.~\ref{fig9}, involving external on-shell soft gravitons carrying momenta $k_1, k_2$ and 
polarizations $\ve_1, \ve_2$ respectively and internal soft graviton carrying momenta $-k_1-k_2$
and polarization labelled by the pair of indices $(\mu,\nu)$. 
This vertex appears in Fig.~\ref{fig6} which begins contributing at
the subleading order. Therefore we need to evaluate this vertex to 
leading order in soft momenta -- given by the Einstein-Hilbert action.
This is best done by regarding the external soft gravitons as 
background field $S_{\mu\nu}$ so that
the vertex can be regarded as the one point function of the 
internal graviton in the presence of soft
graviton background. This is proportional to $R_{\mu\nu} - {1\over 2} R \, g_{\mu\nu}$ computed from the
soft graviton metric. Evaluating this to quadratic order in $S_{\rho\sigma}$ 
we can read out the vertex. Using standard results on the expansion of connection and curvature in powers
of fluctuations in the metric (see {\it e.g.} \cite{9405057,9411092}) we find that the 
vertex takes the form:
\ben\label{esoftvert}
&& V^{(3)}_{\mu\nu}(\ve_1, k_1, \ve_2, k_2) \nonumber \\
&=&{i\over 2} \, \ve_{1,a b }\ve_{2,c d }\Biggl[\Bigg\{\eta_{\mu\nu} 
\eta^{a c }\eta^{b d }k_1^\rho k_{2\rho}-2\eta^{a d }
\eta^{c }_{\;\;\nu}k_2^b  k_{2\mu} -2\eta^{c  b }\eta^{a }_{\;\;\nu}k_1^d  k_{1\mu}  
+2\eta^{a d }\eta^c _{\;\;\nu} k_{1\mu} k_2^b  \nonumber\\
&&+2\eta^{c b }\eta^a _{\;\;\mu}  k_1^d  k_{2\nu}  
-2\eta^{a c }\eta^{b d }k_{1\mu}  k_{2\nu}   
-4\eta^{a }_{\;\;\nu}\eta^{c }_{\;\;\mu}k_1^d   k_2^b 
+ 2\eta^{c }_{\;\;\mu}\eta^{d }_{\;\;\nu}k_2^b  k_2^a   
+ 2\eta^{a }_{\;\;\mu}\eta^{b }_{\;\;\nu}k_1^d  k_1^c  \Bigg\}
\nonumber \\ && + \{\mu\leftrightarrow \nu\}\Biggl]\, . 
\een

We now turn to the computation of the propagators. In the normalization in which the three point
vertex of Fig.~\ref{fig9} is given by \refb{esoftvert}, the soft graviton propagator in the de Donder gauge takes the
form:
\be \label{egravprop}
G_{\mu\nu,\rho\sigma}(k)=- \frac{1}{2}\left(\eta_{\mu\rho}\eta_{\nu\sigma}+\eta_{\nu\rho}
\eta_{\mu\sigma}- \frac{2}{d-2}\eta_{\mu\nu}\eta_{\rho\sigma}\right)\frac{i}{k^2}\, ,
\ee
where $\mu,\nu$ are the indices carried by one of the gravitons and $\rho,\sigma$ are the indices
carried by the other graviton.

The final ingredient is the propagator for an internal finite energy line
carrying momentum $q$.
This is given by $i\, \KK^{-1}_{\alpha\beta}(q)$.  We define
\be \label{edefxi}
\Xi^j_{\alpha\beta}(q) = i\, \KK^{-1}_{\alpha\beta}(q) \, (q^2 + M_j^2)\, ,
\ee
where $M_j$ is the mass of the $j$-th external state. Then the propagator can be
expressed as
\be \label{eprop}
\Delta(q) = (q^2 + M_j^2)^{-1} \, \Xi^j(q)\, ,
\ee
where we have adopted the matrix notation dropping the indices $\alpha,\beta$.

Now from \refb{edefxi} we have
\be 
\KK(q) \ \Xi^i (q) = i\, (q^2 + M_i^2) \, .
\ee
Taking derivatives of this with respect to momenta we arrive at the 
following relations:
\ben \label{eexchange}
{\p \KK(-p)\over \p p_\mu} \Xi^i(-p) &=& -  \KK(-p) {\p \Xi^i(-p)\over \p p_\mu} 
+ 2\, i\, p^\mu\, ,
\nonumber \\ 
{\p^2 \KK(-p)\over \p p_\mu \p p_\nu} \Xi^i(-p) &\hskip -.2in=& \hskip -.2in - {\p \KK(-p)\over \p p_\mu}
{\p \Xi^i(-p)\over \p p_\nu} - {\p \KK(-p)\over \p p_\nu}
{\p \Xi^i(-p)\over \p p_\mu} - \KK(-p) {\p^2 \Xi^i(-p)\over \p p_\mu \p p_\nu}
+2 \, i\, \eta^{\mu\nu}\, , 
\nonumber \\
\een
Finally
rotational invariant of $\KK$ implies the following relations:
\ben \label{ejmove}
(J^{ab})^T \KK(-p) &=& - \KK(-p) J^{ab} + p^a {\p \KK(-p)\over \p p_b} 
- p^b {\p \KK(-p)\over \p p_a} \, ,\nonumber \\
J^{ab}\,  \Xi^i(-p) &=&-\Xi^i(-p) (J^{ab})^T  - p^a {\p \Xi^i(-p)\over \p p_b} 
+ p^b {\p \Xi^i(-p)\over \p p_a}\, .
\een

\subsection{Evaluation of the diagrams}

We begin with the evaluation of Fig.~\ref{fig2}. Even though we can use the form
\refb{eprop} for the internal propagator for any $j$, \refb{eprop} being independent of
$j$ due to \refb{edefxi}, we shall use the form \refb{eprop} with $j=i$ when the soft
gravitons attach to the $i$-th external line. In this case the propagator carrying momentum
$-p_i-k$ for some soft momentum $k$ takes the form
\be 
\Delta(-p_i-k) = \{ (p_i+k)^2 + M_i^2\}^{-1} \, \Xi^i (-p_i-k) = (2 p_i\cdot k+k^2)^{-1} 
\, \Xi^i(-p_i-k)\, .
\ee
We now define
\be 
\Gamma_{(i)}^{\alpha_i}(p_i)= \Bigg\{ \prod_{j=1\atop j\ne i}^N \eps_{j,\alpha_j} \Bigg\} \, 
\Gamma^{\alpha_1 \cdots \alpha_N}(p_1,\cdots , p_N)\, ,
\ee
with the understanding that $\Gamma_{(i)}^{\alpha_i}(p_i)$ also 
implicitly depends on the
$p_j$'s and $\eps_j$'s for $j\ne i$.
Using this we can express the contribution from Fig.~\ref{fig2} as
\ben \label{ea1f}
A_1&\equiv& \sum_{i=1}^N 
(2p_i\cdot k_1)^{-1} (2 p_i\cdot (k_1+k_2)+2k_1\cdot k_2)^{-1} \, \eps_i^T \, 
\Gamma^{(3)}(\ve_1, k_1; p_i, -p_i-k_1)\, \Xi^i(-p_i-k_1) \nonumber \\ 
&&
\Gamma^{(3)}(\ve_2, k_2; p_i+k_1, -p_i-k_1-k_2) \, \Xi^i(-p_i-k_1-k_2) \,
\Gamma_{(i)}
(p_i+k_1+k_2) \, ,
\een
where we have summed over soft graviton insertion on different external legs.
We now use the expression \refb{egam3} for $\Gamma^{(3)}$ and manipulate this
expression as follows:
\begin{enumerate}
\item Take all the $J^{ab}$ factors to the extreme right using \refb{ejmove}
and their derivatives with respect to $p^\mu$.
\item Expand $\KK$, $\Xi^i$ and $\Gamma_{(i)}$ in Taylor series expansion in
$k_1$, $k_2$, and keep up to the first subleading terms in soft momenta.
\item Use the relations \refb{eexchange} 
to move all momentum derivatives to the extreme right to the extent possible.
\item Finally use the on-shell condition 
\be \label{eepK}
\eps_i^T \KK(-p)=0\, ,
\ee
to set all terms in which the left-most $\KK$ does not have a derivative acting on it to zero.
\end{enumerate}
While these steps are sufficient to  arrive at the final result given in \refb{edefa1},
for the analysis of section \ref{s4} we shall need some of the results that appear
in the intermediate stages. For example, Taylor series expansion in $k$, together
with the use of \refb{eexchange}, \refb{ejmove} leads to
the result
\ben \label{emodule} 
\Gamma^{(3)}( \ve, k; p, -p-k) \ \Xi^i(-p-k)&=& \Bigl[ 2 \ \ve^{\mu\nu} p_\mu p_\nu
+ i\, \ve_{\mu\nu} p^\nu \KK(-p) {\p \Xi^i(-p)\over \p p_\mu}
+ 2\ \ve_{b\mu}
 k_{a}p^\mu  (J^{ab})^T \nonumber \\ && \hskip 1in
 +\KK(-p) \ \QQ(p, k) \Bigl]\
\een
to subleading order. Here 
\be \label{edefqpk}
\QQ(p,k)\equiv {i\over 2} \, k\cdot p\, \ve_{b\mu} 
\, {\p^2\Xi(-p)\over \p p_\mu \p p_b} + i \, \ve_{b\mu} \, k_a\, {\p \Xi(-p) \over \p p_\mu} \ (J^{ab})^T\, ,
\ee 
denotes a term that receives contribution from subleading order in
soft momentum. We shall see that its contribution to the amplitude
vanishes due to \refb{eepK}.
Using \refb{emodule} we can express the
amplitude \refb{ea1f} as
\ben
A_1 &=& \sum_{i=1}^N 
(2p_i\cdot k_1)^{-1} (2 p_i\cdot (k_1+k_2)+2k_1\cdot k_2)^{-1} \nonumber \\ &&
\hskip .2in \eps_i^T  \Bigl[ 2 \ \ve_1^{\mu\nu} p_{i\mu} p_{i\nu} 
+ i\, \ve_{1,\mu\nu} p_i^\nu \KK(-p_i) {\p \Xi^i(-p_i)\over \p p_{i\mu}}
+ 2\ \ve_{1, b\mu}
 k_{1a}p_i^\mu  (J^{ab})^T +\KK(-p_i) \ \QQ(p_i, k_1) \Bigl] \nonumber \\ &&
\hskip .2in  \Bigl[ 2 \ \ve_2^{\rho\sigma} (p_{i\rho}+k_{1\rho}) (p_{i\sigma}+k_{1\sigma})
+ i\, \ve_{2,\rho\sigma} (p_i^\sigma+k_1^\sigma) \KK(-p_i-k_1) {\p \Xi^i(-p_i-k_1)\over \p p_{i\rho}}
\nonumber \\ &&  \hskip .3in
+ 2\ \ve_{2, d\rho}\,
 k_{2c}\, (p_i^\rho+k_1^\rho)\   (J^{cd})^T +\KK(-p_i) \ \QQ(p_i, k_2 ) \Bigl] 
\Gamma_{(i)}(p_i+k_1+k_2)\, , 
\een
to first subleading order. Expanding the terms inside the second square bracket and $\Gamma_{(i)}$
in a Taylor series expansion in
$k_1$ and $k_2$, and using \refb{eepK}, \refb{ejmove}, we get, up to subleading order,
\ben \label{edefa1}
A_1
&=&\sum\limits_{i=1}^N (p_i\cdot k_1)^{-1} \{p_i\cdot (k_1+k_2)+k_1\cdot k_2\}^{-1}\eps_i^T \Biggl[
\ve_1^{\sigma\tau} p_{i\sigma} p_{i\tau}\ \ve_{2\mu\nu}p_i^\nu\, p_i^\mu \ \Gamma_{(i)}
(p_i) \nonumber\\
&&+2 \ve_1^{\sigma\tau} p_{i\sigma} p_{i\tau}\ \ve_{2,\mu\nu}k_1^\nu\, p_i^\mu\ \Gamma_{(i)}
(p_i)+ \ve_1^{\sigma\tau} p_{i\sigma} p_{i\tau} k_{2a}\ve_{2,b\mu} p_i^\mu (J^{ab})^T\Gamma_{(i)}
(p_i)\nonumber\\
&&+ \ve_{1,b\sigma}
 k_{1a}p_i^\sigma   \, \ve_{2,\mu\nu}p_i^\nu\, p_i^\mu (J^{ab})^T\Gamma_{(i)}
(p_i)+ \ve_1^{\sigma\tau} p_{i\sigma} p_{i\tau}\ \ve_{2,\mu\nu}p_i^\nu\, p_i^\mu
\ (k_1+k_2)_\rho\;\frac{\p\Gamma_{(i)}
(p_i)}{\p p_{i\rho}} \nonumber\\
&&+{1\over 2} i \,  (k_{1}\cdot p_i)  \, \ve_{1,\mu\sigma}\,
\ve_{2,\rho\nu} \, p_i^\sigma \, p_{i}^{\nu}\,
 {\p \KK(-p_i)\over \p p_{i\mu}}\, 
{\p \Xi^i(-p_i)\over \p p_{i\rho}}\, \Gamma_{(i)}
(p_i)\Biggl]\, .
\een
To this, we also need to add an expression in which we interchange $(k_1,\ve_1)\leftrightarrow (k_2,\ve_2)$. This gives the amplitude
\ben \label{edefa1p}
A_1'
&=&\sum\limits_{i=1}^N (p_i\cdot k_2)^{-1} (p_i\cdot (k_2+k_1)+2k_2\cdot k_1)^{-1}\eps_i^T \Biggl[
\ve_2^{\sigma\tau} p_{i\sigma} p_{i\tau}\ \ve_{1,\mu\nu}p_i^\nu\, p_i^\mu\ \Gamma_{(i)}
(p_i) \nonumber\\
&&+2 \ve_2^{\sigma\tau} p_{i\sigma} p_{i\tau}\ \ve_{1,\mu\nu}k_2^\nu\, p_i^\mu\ \Gamma_{(i)}
(p_i)+\ve_2^{\sigma\tau} p_{i\sigma} p_{i\tau} k_{1a}\ve_{1,b\mu} p_i^\mu (J^{ab})^T\Gamma_{(i)}
(p_i)\nonumber\\
&&+ \ve_{2,b\sigma}
 k_{2a}p_i^\sigma   \, \ve_{1,\mu\nu}p_i^\nu\, p_i^\mu (J^{ab})^T\Gamma_{(i)}
(p_i)+\ve_2^{\sigma\tau} p_{i\sigma} p_{i\tau}\ \ve_{1,\mu\nu}p_i^\nu\, p_i^\mu
\ (k_2+k_1)_\rho
\;\frac{\p\Gamma_{(i)}
(p_i)}{\p p_{i\rho}} \nonumber\\
&&+{1\over 2} i \,  (k_{2}\cdot p_i)  \, \ve_{2,\mu\sigma}
 \ve_{1,\rho\nu} p_i^\sigma p_{i}^{\nu}
 {\p \KK(-p_i)\over \p p_{i\mu}}
{\p \Xi^i(-p_i)\over \p p_{i\rho}}\Gamma_{(i)}
(p_i)\Biggl] \, .
\een

The contribution from Fig.~\ref{fig7} can be evaluated by knowing the result for single
soft graviton insertion since the two parts of the diagram on which the two soft gravitons
are inserted can be evaluated independently.  We shall express this as
\ben \label{euse}
&& (2p_i\cdot k_1)^{-1} \, (2p_j\cdot k_2)^{-1} 
\, \{\ve_i^T \, \Gamma^{(3)}(\ve_1, k_1; p_i, -p_i-k_1) \, \Xi^i(-p_i-k_1)\} \nonumber \\
&&
\otimes \,  \{\ve_j^T \, \Gamma^{(3)}(\ve_2, k_2; p_j, -p_j-k_2) \, \Xi^j(-p_j-k_2)\}
\Gamma_{(i,j)}(p_i+k_1, p_j+k_2)\, ,
\een
where $\Gamma_{(i,j)}^{\alpha_i\alpha_j}$ is defined in the same way as 
$\Gamma_{(i)}$ except that we now strip off both the polarization tensors of 
the $i$-th and the $j$-th leg:
\be \label{egamij}
\Gamma_{(i,j)}^{\alpha_i\alpha_j}(p_i, p_j) \equiv \Bigg\{ \prod_{\ell=1\atop\ell \ne i,j}^N 
\eps_{\ell, \alpha_\ell}\Bigg\} \
\Gamma^{\alpha_1\cdots \alpha_N}(p_1,\cdots , p_N)\, .
\ee
It is understood that in \refb{euse} the terms inside the first curly bracket contracts with
the first index $\alpha_i$ of $\Gamma_{(i,j)}$ and the terms inside the second bracket contracts with
the second index $\alpha_j$ of $\Gamma_{(i,j)}$.  
By manipulating the matrices acting on the $i$-th and the $j$-th leg independently 
in the
same way as before, using the results
\be\label{egamid}
\eps_{i,\alpha} \Gamma_{(i,j)}^{\alpha\beta}(p_i,p_j) = \Gamma_{(j)}^\beta(p_j),
\quad \eps_{j,\beta} \Gamma_{(i,j)}^{\alpha\beta}(p_i,p_j) = \Gamma_{(i)}^\alpha(p_i),
\ee
and summing over insertions on all external legs, we arrive at the following result
for the amplitude up to first subleading order:
\ben \label{edefa2}
A_2 &=& \sum_{i,j=1\atop i\ne j}^N (p_i\cdot k_1)^{-1}\, (p_j\cdot k_2)^{-1} 
\ \ve_{1,\mu\nu} \, p_i^\mu  p_{i}^\nu \, \ve_{2,\rho\sigma} \, p_j^\rho  p_{j}^\sigma
\ \Gamma(\ve_1, k_1, \ve_2, k_2; \eps_1, p_1,\cdots , \eps_N, p_N)
\nonumber \\ && \hskip -.3in
+ \sum_{i,j=1\atop i\ne j}^N (p_i\cdot k_1)^{-1}\, (p_j\cdot k_2)^{-1} 
\, \ve_{2,\rho\sigma} \, p_j^\rho  p_{j}^\sigma \, \eps_i^T\,  \left[
\ve_{1,\mu\nu} \, p_i^\mu  p_{i}^\nu k_{1\tau} {\p \Gamma_{(i)}(p_i) \over \p p_{i\tau}}
+  k_{1a} \, \ve_{1,b\mu}  \, 
p_i^\mu\,  (J^{ab})^T
\Gamma_{(i)}(p_i)\right]
\nonumber \\ && \hskip -.3in
+ \sum_{i,j=1\atop i\ne j}^N (p_i\cdot k_2)^{-1}\, (p_j\cdot k_1)^{-1} 
\, \ve_{1,\rho\sigma} \, p_j^\rho  p_{j}^\sigma \, \eps_i^T\,  \left[
\ve_{2,\mu\nu} \, p_i^\mu  p_{i}^\nu k_{2\tau} {\p \Gamma_{(i)}(p_i) \over \p p_{i\tau}}
+  k_{2a} \, \ve_{2,b\mu}  \, 
p_i^\mu\,  (J^{ab})^T
\Gamma_{(i)}(p_i)\right]\, . \nonumber \\
\een

Next we consider the contribution from Fig.~\ref{fig3}.  The contribution from this term has
at most one pole in the soft momentum and therefore begins at subleading order.
Therefore we only need the leading contribution from this diagram. 
For this we use the result \refb{ecompver} for the off-shell amplitude shown in
Fig.~\ref{fig4}. 
This gives
the following expression for the contribution from Fig.~\ref{fig3}:
\be 
A_3=- \sum_{i=1}^N
(2p_i\cdot k_1)^{-1} \eps_i^T \Gamma^{(3)}(\ve_1, k_1; p_i, -p_i-k_1) \, \Xi^i(-p_1-k_1) 
\sum_{j=1}^N \eps_j^T \ve_2^{\mu\nu} p_{j\mu} {\p\over \p p_j^\nu} 
\Gamma_{(i,j)}(p_i, p_j)\, ,
\ee
where again we have summed over the insertion of the first soft graviton on all external
finite energy states.
We can now manipulate this using the form of $\Gamma^{(3)}$ given earlier. 
This leads to
\ben\label{edefa3}
A_3&=&- \sum_{i=1}^N
(p_i\cdot k_1)^{-1}  \, 
\ve_1^{\rho\sigma} p_{i\rho} p_{i\sigma}\, \sum_{j=1}^N \eps_j^T \ve_2^{\mu\nu} p_{j\mu} {\p\over \p p_j^\nu} 
\Gamma_{(j)}\, .
\een
The diagram obtained by interchanging $(k_1,\ve_1)\leftrightarrow (k_2,\ve_2)$ 
gives
\ben \label{edefa3p}
A_3'&=&-\sum_{i=1}^N(p_i\cdot k_2)^{-1}  \,  \ve_2^{\rho\sigma} p_{i\rho} p_{i\sigma}\, \sum_{j=1}^N
 \eps_j^T \ve_1^{\mu\nu} p_{j\mu} {\p\over \p p_j^\nu} 
\Gamma_{(j)}\, .
\een

Fig.~\ref{fig5} also begins contributing at the subleading order. Therefore we only need its 
leading contribution, which is given by
\be 
A_4 = \sum_{i=1}^N \, \{2p_i\cdot (k_1+k_2)\}^{-1} \eps_i^T 
\Gamma^{(4)}(\ve_1,k_1,\ve_2, k_2;
p_i, -p_i-k_1-k_2) \, \Xi^i(-p_i-k_1-k_2) \, \Gamma_{(i)}(p_i)\, .
\ee
This can be evaluated using the expression 
\refb{eexpgam4}
for the vertex 
$\Gamma^{(4)}$ shown in Fig.~\ref{fig8} and manipulating the resulting expression  in the
same way as the previous diagrams. The result is
\ben \label{edefa4}
A_4 &=& \sum\limits_{i=1}^N \{p_i\cdot (k_1+k_2)\}^{-1} \eps_i^T 
\biggl[ -2\varepsilon_{1,\mu}^{\,\;\;\;\,\nu}\varepsilon_{2,\nu\rho}p_{i}^{\rho}p_i^\mu
\nonumber \\  && \hskip .3in - {i\over 2} 
\Bigl(\varepsilon_{1,\mu\sigma}\varepsilon_{2,\rho\nu}p_i^{\sigma}p_i^{\nu}
 +\varepsilon_{1,\rho\sigma}\varepsilon_{2,\mu\nu}
 p_i^{\sigma}p_i^{\nu}\Bigl) {\p \KK(-p_i)\over \p p_{i\mu}}
{\p \Xi^i(-p_i)\over \p p_{i\rho}}\biggl] \Gamma_{(i)}(p_i)\, .
\een

Finally we turn to the computation of the diagram shown in Fig.~\ref{fig6}. Its contribution is given by
\be \label{eina5}
A_5 = V^{(3)\mu\nu}(\ve_1, k_1, \ve_2, k_2) G_{\mu\nu,\rho\sigma}(k_1+k_2)\
\sum_{i=1}^N \eps_i^T \, \Gamma^{(3)(\rho\sigma)} \left(k_1+k_2; p_i, -p_i - k_1-k_2
\right)\ 
\Gamma_{(i)}(p_i)\, ,
\ee
where $V^{(3)}$ and $G_{\mu\nu,\rho\sigma}$ have been defined in \refb{esoftvert} and 
\refb{egravprop} respectively, and $\Gamma^{(3)(\rho\sigma)}$ is defined via the equation
\be
\Gamma^{(3)}\left(\ve, k; p, -p - k
\right) = \ve_{\rho\sigma}  
\Gamma^{(3)(\rho\sigma)}
\left(k; p, -p - k
\right)\, .
\ee
Using the leading order expression for $\Gamma^{(3)}$ given in \refb{egam3},
and the relations \refb{eexchange}, \refb{ejmove}, \refb{eepK} this can  be brought to the form
\ben\label{edefa5}
A_5&=& \sum\limits_{i=1}^N\{p_i\cdot(k_1+k_2)\}^{-1}(k_1\cdot k_2)^{-1}\epsilon_i^T\Bigl[-(k_2\cdot\ve_{1}\cdot\ve_{2}\cdot p_i)(k_2\cdot p_i) -(k_1\cdot\ve_{2}\cdot\ve_{1}\cdot p_i)(k_1\cdot p_i)  \nonumber\\
&&+(k_2\cdot\ve_{1}\cdot\ve_{2}\cdot p_i)(k_1\cdot p_i)+(k_1\cdot\ve_{2}\cdot\ve_{1}\cdot p_i)(k_2\cdot p_i)  - \ve_1^{cd}\ve_{2,cd}\ (k_{1} \cdot p_i)\ ( k_{2}\cdot p_i)  \nonumber\\
&& -2(p_i\cdot\ve_{1}\cdot k_2)(p_i\cdot\ve_{2}\cdot k_1) + (p_i\cdot\ve_{2}\cdot p_i)(k_2\cdot\ve_{1}\cdot k_2)  + (p_i\cdot\ve_{1}\cdot p_i)(k_1\cdot\ve_{2}\cdot k_1)\Bigl]\Gamma_{(i)}(p_i)\nonumber\\
\een

The full amplitude is given by
\ben \label{efull}
A &=& A_1+A_1'+A_2 + A_3 + A_3' + A_4 + A_5 \nonumber \\
&=& \Bigg\{\sum_{i=1}^N (p_i\cdot k_1)^{-1} \ \ve_{1,\mu\nu} \, p_i^\mu  p_{i}^\nu\Bigg\}
\ \Bigg\{ \sum_{j=1}^N (p_j\cdot k_2)^{-1} 
 \, \ve_{2,\rho\sigma} \, p_j^\rho  p_{j}^\sigma\Bigg\}
\ \Gamma(\eps_1, p_1,\cdots , \eps_N, p_N) 
\nonumber \\ && \hskip -.5in
+  \Bigg\{\sum_{j=1}^N (p_j\cdot k_2)^{-1} 
\, \ve_{2,\rho\sigma} \, p_j^\rho  p_{j}^\sigma\Bigg\} \ 
\sum_{i=1}^N (p_i\cdot k_1)^{-1}\,
\ve_{1,b\mu}  \,  k_{1a} \, 
p_i^\mu\  \eps_i^T\, \left[ p_i^b 
{\p \Gamma_{(i)}(p_i) \over \p p_{ia}} - p_i^a 
{\p \Gamma_{(i)}(p_i) \over \p p_{ib}}
+   (J^{ab})^T
\Gamma_{(i)}(p_i)\right]
\nonumber \\ && \hskip -.5in
+ \Bigg\{ \sum_{j=1}^N (p_j\cdot k_1)^{-1} 
\, \ve_{1,\rho\sigma} \, p_j^\rho  p_{j}^\sigma\Bigg\} \ 
\sum_{i=1}^N (p_i\cdot k_2)^{-1}\,  \ve_{2,b\mu}  \,  k_{2a} \, 
p_i^\mu\  \eps_i^T\, \left[ p_i^b 
{\p \Gamma_{(i)}(p_i) \over \p p_{ia}} - p_i^a 
{\p \Gamma_{(i)}(p_i) \over \p p_{ib}}
+   (J^{ab})^T
\Gamma_{(i)}(p_i)\right]
\nonumber \\ && \hskip -.5in
+ \Bigg\{ \sum_{i=1}^N   \,  \{p_i\cdot (k_1+k_2)\}^{-1} \ 
\MM (p_i; \ve_1, k_1, \ve_2, k_2) \Bigg\} \ \Gamma(\eps_1, p_1,
\cdots , \eps_N, p_N)\, ,
\een 
where
\ben \label{edefmi}
\MM(p_i; \ve_1, k_1, \ve_2, k_2) &=&
(p_i\cdot k_1)^{-1}  (p_i\cdot k_2)^{-1} 
 \ \Bigg\{- (k_1\cdot k_2) \ (p_i\cdot \ve_1\cdot p_i) \ 
( p_i\cdot \ve_2\cdot p_i)
 \nonumber \\ && \hskip -.3in
 + \ 2 \  (p_i\cdot k_2) \ (p_i\cdot \ve_1\cdot p_i) \ (p_i\cdot \ve_2\cdot k_1) 
  + 2 \ (p_i\cdot k_1) \ (p_i\cdot \ve_2\cdot p_i) \ (p_i\cdot \ve_1\cdot k_2)
 \nonumber \\ &&  \hskip -.3in 
 - \ 2  \ (p_i\cdot k_1) \ (p_i\cdot k_2) \ (p_i\cdot \ve_1\cdot \ve_2\cdot p_i)\Bigg\}
 \nonumber \\ &&  \hskip -1.6in
 +\ (k_1\cdot k_2)^{-1}
 \Bigg\{-(k_2\cdot\ve_{1}\cdot\ve_{2}\cdot p_i)\ (k_2\cdot p_i) 
 -(k_1\cdot\ve_{2}\cdot\ve_{1}\cdot p_i)\ (k_1\cdot p_i)  \nonumber\\
&& \hskip  -1in +\ (k_2\cdot\ve_{1}\cdot\ve_{2}\cdot p_i)\ (k_1\cdot p_i)
+(k_1\cdot\ve_{2}\cdot\ve_{1}\cdot p_i)\ (k_2\cdot p_i)  - \ve_1^{cd}\ve_{2,cd}\ 
(k_{1} \cdot p_i) \ ( k_{2}\cdot p_i)  \nonumber\\
&& \hskip -1in -\ 2(p_i\cdot\ve_{1}\cdot k_2)\ (p_i\cdot\ve_{2}\cdot k_1) 
+ (p_i\cdot\ve_{2}\cdot p_i)\ (k_2\cdot\ve_{1}\cdot k_2)  
+ (p_i\cdot\ve_{1}\cdot p_i)\ (k_1\cdot\ve_{2}\cdot k_1)\Bigg\}\, .
\nonumber\\
\een
Here we have used the shorthand notation $p_i\cdot \ve_1\cdot p_i\equiv
\ve_{1,\mu\nu} p_i^\mu p_i^\nu$ etc. $\MM$ receives contributions from the first two
terms in \refb{edefa1} and \refb{edefa1p} and also from \refb{edefa4} and \refb{edefa5}.

\subsection{Infrared issues} \label{sIR}

\begin{figure}
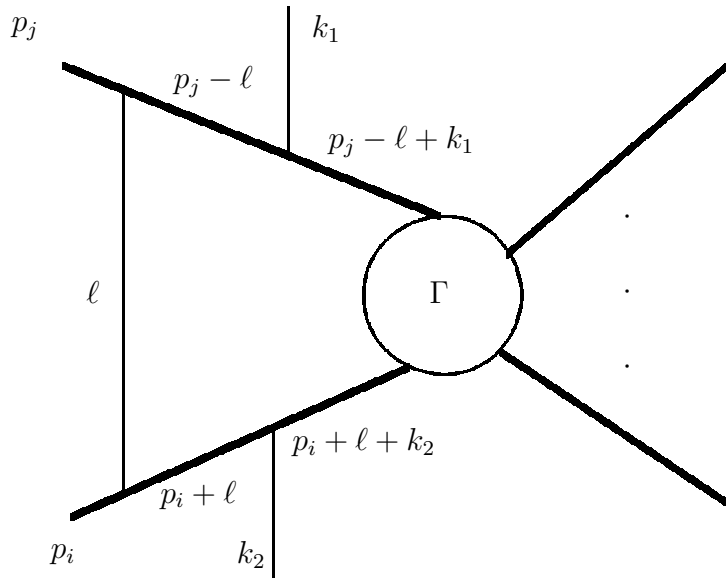


\begin{center}

\figextra

\end{center}

\vskip -.2in

\caption{A possible subleading contribution in five non-compact dimensions.
\label{figex}}

\end{figure}

In our analysis we have assumed that possible soft factors in the denominator arise
from propagators but not from the 1PI vertices. This holds when the
number of non-compact space-time dimensions $D$ is sufficiently high. However we shall
now show that for $D\le 5$, individual contributions violate this condition due to
infrared effects in
the loop.  Let us consider for example the diagram shown in Fig.~\ref{figex}.
In the 1PI effective field theory, this corresponds to a graph similar to one shown in
Fig.~\ref{fig3}, but with both soft gravitons connected to the vertex $\wt\Gamma$. 
If there is
no inverse power of soft momenta from $\wt\Gamma$ then this contribution is
subsubleading and can be ignored. 
However let us consider the limit in which the loop momentum $\ell$ 
in Fig.~\ref{figex} becomes soft -- of the same order as the external
soft momenta. In this limit each of the propagators carrying momenta $p_i+\ell$,
$p_i+\ell+k_2$, $p_j-\ell$ and $p_j-\ell+k_1$ gives one power of soft momentum in the
denominator and the soft propagator carrying momentum $\ell$ gives two powers of
soft momentum in the denominator. On the other hand in $D$ non-compact space-time
dimensions the loop momentum integration measure goes as 
$D$ powers of soft momentum. Therefore the net power of soft momentum that we
get from this graph for soft $\ell$ is $D-6$,
and in $D=5$ this integral can give a term containing one power of soft momentum in the 
denominator, giving a subleading contribution. Since we have not included these diagrams
in our analysis we conclude that for loop amplitudes our result is valid for $D\ge 6$. 
It is easy to see by simple power counting 
that higher loop amplitudes do not lead to any additional enhancement
from the infrared region of loop momenta.

Similar
analysis can be carried out for multiple soft graviton amplitudes of the kind described
in section \ref{s4}. As we connect each external soft graviton to an internal nearly
on-shell line carrying finite energy, 
the number of powers of soft momentum in the denominator goes up by one.
However the required number of powers of soft momentum in the denominator of
the subleading 
contribution also goes up by one. Therefore the result of section \ref{s4} continues to be
valid for loop amplitudes for $D\ge 6$, irrespective of the number of external soft gravitons.

Even though this analysis shows that individual diagrams can give contributions
beyond what we have included in our analysis for $D\le 5$, we expect that 
for $D=5$ such contributions
will cancel when we sum over all diagrams. This expectation arises out of standard
results on factorization of soft loops\cite{yennie,grammer} 
that tells us that after summing over graphs,
the contribution from the region of soft loop momentum takes the form of a product
of an amplitude without soft loop and a soft factor that arises from graphs like
Fig.~\ref{figex} without the external soft lines.  Since the graphs like Fig.~\ref{figex} 
without soft external lines do not receive large contribution from the small $\ell$ region,
and are furthermore independent of the external soft momenta, their contribution may be
absorbed into the definition of the amplitude without the soft gravitons. Therefore
we conclude that the contribution from the loop momentum integration region for small
$\ell$ in graphs like Fig.~\ref{figex} 
must cancel in the sum over graphs. Nevertheless since our general analysis
relies on the analysis of individual contributions of different graphs of the type shown in
Fig.~\ref{fig2}-\ref{fig6}, 
and since the coefficients of Taylor series expansion of
these individual contributions as well as those not included in Fig.~\ref{fig2}-\ref{fig6} 
(like Fig.~\ref{figex}) do receive large contribution from
small loop momentum region, we cannot give a foolproof argument that our general
result is not affected by infrared contributions of the type described above.

Note that similar infrared enhancement also occurs for amplitudes with 
single soft graviton, but by
analyzing the tensor structure of these contributions it was argued in \cite{1706.00759} 
that gauge
invariance prevents corrections to the soft theorem from such effects 
to subsubleading order for $D\ge 5$. Similar argument has not been developed for
multiple soft graviton amplitudes.

This problem of course does not arise for
tree amplitudes where the vertices are always polynomial in momenta. Therefore for
tree amplitudes our results hold in all dimensions.

\sectiono{Consistency checks} \label{s3}

In this section we shall carry out various consistency checks on our result. 
First we shall check the
internal consistency of our result. Then we shall compare our results with the previous
results derived for specific theories.

\subsection{Internal consistency}

The first internal consistency check of our result comes from the 
requirement of gauge invariance.
This means that if we make the transformation
\be
\ve_{r,\mu\nu} \to k_{r\mu} \, \xi_{r\nu} +  k_{r\nu} \, \xi_{r\mu}\, , \qquad r=1,2\, ,
\ee 
for any vector $\xi_r$ satisfying $k_r\cdot \xi_r=0$, the result \refb{efull} 
does not  change. Checking this involves tedious but straightforward algebra, and needs
use of the equations
\be \label{econmom}
\sum_{i=1}^N p_{i\mu} \ \Gamma^{\alpha_1\cdots \alpha_N} = 0\,  ,
\ee
and 
\be \label{econang}
\sum_{i=1}^N \left[ p_i^b \ {\p \Gamma^{\alpha_1\cdots \alpha_N}\over \p p_{ia}}
- p_i^a \ {\p \Gamma^{\alpha_1\cdots \alpha_N}\over \p p_{ib}}
+ (J^{ab})_{\beta_i}^{~~\alpha_i}  \Gamma^{\alpha_1\cdots \alpha_{i-1}\beta_i
\alpha_{i+1}\cdots \alpha_N}
\right] = 0\, ,
\ee
reflecting respectively  translational and rotational invariance of the amplitude without the
soft graviton.  While making this analysis we also need to be careful to ensure that while
passing $p_{i\mu} $  through $\p / \p p_{j\nu}$ 
in order to make use of \refb{econmom}, we have to take into account the extra terms
proportional to $\delta_{ij} \delta_\mu^\nu$. For this reason the terms in the third and fourth lines
of \refb{efull} are not gauge invariant by themselves -- their gauge variation cancels against the
variation of the term in the last line of \refb{efull}. More specifically if we denote by $\delta_r$ the
gauge variation:
\be \label{evardef}
\delta_r\ : \qquad \ve_{r,\mu\nu} \to \ve_{r,\mu\nu} + k_{r\mu}\xi_{r\nu} +  k_{r\nu}\xi_{r\mu}\, ,
\ee
for some vector $\xi_r$ satisfying $k_r\cdot \xi_r=0$, then under $\delta_1$ the term in the
third line of \refb{efull} remains unchanged, but the term in the fourth line changes by
\be \label{evar1}
- 2\ \sum_{i=1}^N \ (p_i\cdot k_2)^{-1} \ \ve_{2,b\mu} \ p_i^b p_i^\mu \ k_2\cdot \xi_1 \ \eps_i^T 
\Gamma_{(i)}\, .
\ee
On the other hand we get, after using momentum conservation equation $\sum_{j=1}^N p_j
\Gamma_{(i)}=0$,
\be \label{evar2}
\sum_{i=1}^N \ \{p_i\cdot (k_1+k_2)\}^{-1} \
\delta_1 \MM(p_i; \ve_1, k_1, \ve_2, k_2) \ \eps_i^T 
\Gamma_{(i)}=   2\ \sum_{i=1}^N (p_i\cdot k_2)^{-1} \  
\ve_{2,b\mu} \ p_i^b p_i^\mu \ k_2\cdot \xi_1 \ \eps_i^T 
\Gamma_{(i)}\, .
\ee
Using this one can easily verify that the $\delta_1$ variation of the fourth line and the last line of
\refb{efull} cancel. A similar analysis shows that the $\delta_2$ variation of the third and the last
lines of \refb{efull} cancel, and that the fourth line of \refb{efull} is invariant under $\delta_2$.

The second consistency requirement arises from the fact that individual terms in \refb{efull} depend
on the off-shell data on $\Gamma^{\alpha_1\cdots \alpha_N}$ while  the actual result should be
insensitive to such off-shell extension. For example if we add to $\Gamma^{\alpha_1\cdots \alpha_N}$
any term proportional to $p_i^2+M_i^2$, it does not affect the on-shell amplitude without the 
soft gravitons since it vanishes on-shell. However $\p\Gamma^{\alpha_1\cdots \alpha_N}/\p p_{i_\mu}$
receives a contribution proportional to $p_i^\mu$ that does not vanish on-shell. We note however 
that in \refb{efull} the derivatives of $\Gamma^{\alpha_1\cdots \alpha_N}$ come in a very special
combination that vanishes under addition of any term to $\p\Gamma/\p p_{i\mu}$ proportional to
$p_i^\mu$. Therefore \refb{efull} is not sensitive to such additional terms in $\Gamma$.

More generally we can add to $\Gamma^{\alpha_1\cdots \alpha_N}$ any term proportional to
$\KK^{\alpha_i\beta}(-p_i) \GG_{\beta}^{\alpha_1\cdots \alpha_{i-1}\alpha_{i+1}\cdots \alpha_N}$ 
for any function $\GG$, since its contribution to on-shell
amplitudes without the soft gravitons vanishes due to \refb{eepK}. Using \refb{eepK} and 
the rotational invariance
of $\KK$ described in \refb{ejmove}, is easy to see however that
the addition of such terms to $\Gamma$ does not affect \refb{efull}.

\subsection{Comparison with known results}

In order to compare the amplitude with known results, it is convenient to rewrite the amplitude
\refb{efull} as a sum of two terms $\AAA_1+\AAA_2$ by adding and subtracting a specific term
given in the last two lines of \refb{esplit1}:
\ben \label{esplit1}
&& \AAA_1 = \Bigg\{\sum_{i=1}^N (p_i\cdot k_1)^{-1} \ \ve_{1,\mu\nu} \, p_i^\mu  p_{i}^\nu\Bigg\}
\ \Bigg\{ \sum_{j=1}^N (p_j\cdot k_2)^{-1} 
 \, \ve_{2,\rho\sigma} \, p_j^\rho  p_{j}^\sigma\Bigg\}
\ \Gamma(\eps_1, p_1,\cdots , \eps_N, p_N) 
\nonumber \\ && \hskip -.5in
+  \Bigg\{\sum_{j=1}^N (p_j\cdot k_2)^{-1} 
\, \ve_{2,\rho\sigma} \, p_j^\rho  p_{j}^\sigma\Bigg\} \ 
\sum_{i=1}^N (p_i\cdot k_1)^{-1}\,
\ve_{1,b\mu}  \,  k_{1a} \, 
p_i^\mu\  \eps_i^T\, \left[ p_i^b 
{\p \Gamma_{(i)}(p_i) \over \p p_{ia}} - p_i^a 
{\p \Gamma_{(i)}(p_i) \over \p p_{ib}}
+   (J^{ab})^T
\Gamma_{(i)}(p_i)\right]
\nonumber \\ && \hskip -.5in
+ \Bigg\{ \sum_{j=1}^N (p_j\cdot k_1)^{-1} 
\, \ve_{1,\rho\sigma} \, p_j^\rho  p_{j}^\sigma\Bigg\} \ 
\sum_{i=1}^N (p_i\cdot k_2)^{-1}\,  \ve_{2,b\mu}  \,  k_{2a} \, 
p_i^\mu\  \eps_i^T\, \left[ p_i^b 
{\p \Gamma_{(i)}(p_i) \over \p p_{ia}} - p_i^a 
{\p \Gamma_{(i)}(p_i) \over \p p_{ib}}
+   (J^{ab})^T
\Gamma_{(i)}(p_i)\right]
\nonumber \\ 
&& 
 + \  (k_1\cdot k_2)^{-1} 
 \sum_{i=1}^N \ (p_i\cdot k_1)^{-1} \ (p_i\cdot k_2)^{-1} \  \bigg\{
(k_1\cdot \ve_2 \cdot k_1)(p_i\cdot \ve_1\cdot p_i) \ (p_i\cdot k_2)
\nonumber \\ && \hskip 2in +\ (k_2\cdot \ve_1 \cdot k_2) \ (p_i\cdot \ve_2\cdot p_i) \ (p_i\cdot k_1) 
\bigg\} \ \eps_i^T \Gamma_{(i)}(p_i)\, ,
\een
\be \label{eplit2}
\AAA_2 = \Bigg\{ \sum_{i=1}^N 
\NN(p _i; \ve_1, k_1, \ve_2, k_2) \Bigg\} \, \Gamma(\eps_1, p_1, \cdots , \eps_N, p_N)
\, ,
\ee 
where
\ben \label{edefwtmi}
\NN(p_i; \ve_1, k_1, \ve_2, k_2)
&=&  \{p_i\cdot (k_1+k_2)\}^{-1}  \ \MM(p_i; \ve_1, k_1, \ve_2, k_2)  \nonumber \\ 
&&\hskip -.3in  - \ (k_1\cdot k_2)^{-1} \ (p_i\cdot k_1)^{-1} \ (p_i\cdot k_2)^{-1}
\nonumber \\ && \hskip -.3in  \times \ \bigg\{
(k_1\cdot \ve_2 \cdot k_1) \ (p_i\cdot \ve_1\cdot p_i) \ (p_i\cdot k_2)
+ (k_2\cdot \ve_1 \cdot k_2) \ (p_i\cdot \ve_2\cdot p_i) \ (p_i\cdot k_1)
\bigg\}
\, , \nonumber \\
\een
$\MM$ being given in \refb{edefmi}.
With this definition $\AAA_1$ and $\AAA_2$ can be shown to be separately gauge invariant.

Refs.~\cite{1607.02700,1702.02350} computed the double soft limit for scattering of gravitons in 
Einstein gravity using
CHY scattering equations\cite{1306.6575,1307.2199,1309.0885,1409.8256,1412.3479}. 
Since our result is valid for general finite energy external states in any theory, 
it must also be valid for scattering
of gravitons. Therefore we can compare the two results.
The contribution in \cite{1607.02700,1702.02350} 
comes from two separate terms, the degenerate
solutions and non-degenerate solutions. The contribution from the degenerate solutions
agrees 
with our amplitude $\AAA_2$ given in \refb{eplit2} up to a 
sign after using momentum
conservation rules \refb{econmom}.
The contribution from the non-degenerate 
solutions were evaluated in \cite{1702.02350} to 
give only the first three lines of \refb{esplit1}.
However the analysis was carried out in a gauge in which $k_1\cdot \ve_2=0$ and $k_2\cdot \ve_1=0$.
For this choice of gauge the contribution from the last two lines of \refb{esplit1} vanishes. Therefore,  up to the issue with signs mentioned above, there
is agreement between our results and the results in pure gravity derived from CHY equations
in \cite{1607.02700,1702.02350}, 
with \refb{esplit1} giving the full gauge invariant 
version of the contribution
from non-degenerate solutions of CHY equations. 
By carefully reanalyzing the double soft limit of the CHY formula for the scattering
amplitudes we have been able to show that the result obtained from the CHY formula
actually agrees with ours including the sign\cite{appear}.

Ref.~\cite{1504.05558} computed the double soft limit of graviton scattering amplitude in four
space-time dimensions  using BCFW recursion relations\cite{0501052}. This analysis
was also carried out in the gauge $k_1\cdot \ve_2=0$ and $k_2\cdot \ve_1=0$.
In this gauge the subleading contribution to $\AAA_1$ comes only from the second and the third 
lines which, written in the spinor helicity notation, has the standard form involving derivatives with
respect to the spinor helicity variables, called `non-contact terms' in \cite{1504.05558}. 
Therefore we focus on the $\AAA_2$ term. 
Ref.~\cite{1702.02350} showed that the contribution from the degenerate solution to the CHY equations
agrees with the `contact terms' computed in \cite{1504.05558} using BCFW recursion relations.
Therefore our result for $\AAA_2$ agrees with the contact terms of \cite{1504.05558} up to the sign factor discussed earlier. 
We have also verified this independently by noting that in the gauge 
$k_1\cdot \ve_2=0$ and $k_2\cdot \ve_1=0$ many of the terms in $\AAA_2$ vanish and the remaining
terms take the form
\ben \label{esimpli}
&& \sum_{i=1}^N   \,  \{p_i\cdot (k_1+k_2)\}^{-1} \ 
\Bigg[ - \ (p_i\cdot k_1)^{-1}  (p_i\cdot k_2)^{-1} 
 (k_1\cdot k_2) \ (p_i\cdot \ve_1\cdot p_i) \ 
( p_i\cdot \ve_2\cdot p_i) \nonumber \\ && 
  - \ 2  \  (p_i\cdot \ve_1\cdot \ve_2\cdot p_i)
 -\ (k_1\cdot k_2)^{-1}\, 
  \ve_1^{cd}\ve_{2,cd}\ 
(k_{1} \cdot p_i) \ ( k_{2}\cdot p_i)  \Bigg] \Gamma(\eps_1, p_1,
\cdots , \eps_N, p_N)\, 
\, .
\een
By expressing this in the spinor helicity notation we find that when the
two soft gravitons carry the same helicity \refb{esimpli} vanishes. This is 
in agreement
with the result of \cite{1504.05558}.  On the other hand when the two soft gravitons carry
opposite helicities, 
$\AAA_2$ gives a non-zero result that agrees with the `contact terms'
of \cite{1504.05558} up to a sign. We have not tried to resolve this discrepancy in sign
between our results and that of \cite{1504.05558}. However given that we have now verified 
that the CHY result for contact terms actually comes with a sign opposite to that found
in \cite{1607.02700,1702.02350} and agrees with our
amplitude $\AAA_2$\cite{appear}, it seems that the difference in sign 
between our results and the BCFW results may be due to some differences in convention,
{\it e.g.} the difference in the choice of sign of the graviton polarization 
tensor.\footnote{We have used the convention that the graviton 
polarization tensors in four dimensions are given by squares of the gauge field polarization
tensors without any extra sign.}

\sectiono{Amplitudes with arbitrary number of soft gravitons} \label{s4}

The method described in the earlier sections can now be generalized to derive the
expression for the amplitude with multiple soft gravitons when the momenta carried by
all the soft gravitons become small at the same rate. We shall first write down the
result and then explain how we arrive at it. The subleading 
soft graviton amplitude with $M$
soft gravitons carrying momenta $k_1,\cdots , k_M$ and polarizations $\ve_1,\cdots ,\ve_M$
and $N$ finite energy particles carrying momenta $p_1,\cdots , p_N$ and polarizations
$\eps_1,\cdots , \eps_N$ is given by
\ben \label{efullgen}
A &=& \prod_{r=1}^M \, \left\{\sum_{i=1}^N (p_i\cdot k_r)^{-1}
\ \ve_{r,\mu\nu} \, p_i^\mu  p_{i}^\nu \right\}
\ \Gamma(\eps_1, p_1,\cdots , \eps_N, p_N) 
\nonumber \\ && \hskip -.6in
+ \sum_{s=1}^M \sum_{j=1}^N (p_j\cdot k_s)^{-1} 
\, \ve_{s,b\mu}  \,  k_{s a} \, 
p_j^\mu\  \eps_j^T\, \left[ p_j^b 
{\p \Gamma_{(j)}(p_j) \over \p p_{ja}} - p_j^a 
{\p \Gamma_{(j)}(p_j) \over \p p_{jb}}
+   (J^{ab})^T
\Gamma_{(j)}(p_j)\right]
\nonumber \\ &&
\times \prod_{r=1\atop r\ne s}^M \, \left\{\sum_{i=1}^N (p_i\cdot k_r)^{-1}
\ \ve_{r,\mu\nu} \, p_i^\mu  p_{i}^\nu \right\}
\nonumber \\ &&  \hskip -.6in
+ \sum_{r,u=1\atop r<u}^M \Bigg\{
\sum_{j=1}^N \ \{p_j\cdot (k_r+k_u)\}^{-1}
\ \MM (p_j; \ve_r, k_r, \ve_u, k_u) \
\eps_j^T \ \Gamma_{(j)}(p_j)\Bigg\} \, \prod_{s=1\atop s\ne r,u}^M \, \left\{\sum_{i=1}^N (p_i\cdot k_s)^{-1}
\ \ve_{s,\mu\nu} \, p_i^\mu  p_{i}^\nu \right\}\,  , \nonumber \\
\een
where $\MM(p_j; \ve_r, k_r, \ve_u, k_u)$ has been defined in \refb{edefmi}. Independently
of the general argument given below, we have used Cadabra\cite{0608005,0701238} and
Mathematica\cite{math} to check 
\refb{efullgen} explicitly for amplitudes
with three soft gravitons.

\begin{figure}

\begin{center}

\figten

\end{center}

\vskip -.3in

\caption{A leading contribution to the amplitude with multiple soft gravitons.
\label{fig10}}

\end{figure}

We begin by reviewing the derivation of the leading term given in the first line of
\refb{efullgen}.  For this note that this term may be rearranged as
\be\label{erear}
\sum_{A_1,\cdots A_N; \ A_i\subset \{1,\cdots , M\}
\atop A_i\cap A_j = \emptyset \, {\rm for} \, i\ne j; \
A_1\cup A_2\cup \cdots \cup A_N=\{1,\cdots M\}}
\prod_{i=1}^N \left\{ \prod_{r\in A_i}  \ve_{r,\mu\nu} \, p_i^\mu  p_{i}^\nu\right\} \
\left\{ \prod_{r\in A_i} (p_i\cdot k_r)^{-1} \right\} \ 
\Gamma(\eps_1, p_1,\cdots , \eps_N, p_N) \, .
\ee
Physically the  $i$-th term in the product 
represents the contribution from the soft gravitons in the set $A_i$
attached to the $i$-th finite energy external line. To see how we get this factor, let us
denote the momenta of the soft gravitons attached from the outermost end to the
innermost end of the $i$-th line in a given graph 
by $\tilde k_1,\cdots \tilde k_n$. The corresponding polarizations are denoted by
$\tilde\ve_1,\cdots \tilde \ve_n$. This is shown in Fig.~\ref{fig10}.
The unordered
set $\{\tilde k_1,\cdots , \tilde k_n\}$ coincides with the set $\{k_s; s\in A_i\}$. 
A similar statement holds for the polarizations.
The leading contribution from the products of three point vertices and propagators
associated with the $i$-th line of the graph may be computed using \refb{emodule}, \refb{eepK} 
and is given by
\be
\left\{ \prod_{r=1}^n  \tilde\ve_{r,\mu\nu} \, p_i^\mu  p_{i}^\nu\right\} \
\
\{p_i\cdot \tilde k_1\}^{-1} \{p_i\cdot (\tilde k_1 + \tilde k_2)\}^{-1} \cdots
\{p_i\cdot (\tilde k_1+\cdots +\tilde k_n)\}^{-1}\, .
\ee
The total contribution obtained after summing
over all permutations of the
momenta $\tilde k_1,\cdots , \tilde k_n$  using \refb{eap1} is given by
\ben
&&\hskip -.4in
\left\{ \prod_{r=1}^n  \tilde\ve_{r,\mu\nu} \, p_i^\mu  p_{i}^\nu\right\} \
\sum_{{\rm permutations \, of}\, \tilde k_1,\cdots \tilde k_n} 
\{p_i\cdot \tilde k_1\}^{-1} \{p_i\cdot (\tilde k_1 + \tilde k_2)\}^{-1} \cdots
\{p_i\cdot (\tilde k_1+\cdots +\tilde k_n)\}^{-1}
\nonumber \\ &=& 
\left\{ \prod_{r=1}^n  \tilde\ve_{r,\mu\nu} \, p_i^\mu  p_{i}^\nu\right\} \
\left\{ \prod_{s=1}^n (p_i\cdot \tilde k_s)^{-1} \right\}
=\left\{ \prod_{r\in A_i}  \ve_{r,\mu\nu} \, p_i^\mu  p_{i}^\nu\right\} \
\left\{ \prod_{s\in A_i} (p_i\cdot k_s)^{-1} \right\} \, . 
\een
This reproduces \refb{erear}.

We now turn to the analysis of the subleading terms. For this let us first analyze the contribution
from the products of the propagators and vertices in Fig.~\ref{fig10} to subleading order. Using
\refb{emodule}  this may be expressed as
\ben \label{emoduleprod} 
&& \{2p_i\cdot \tilde k_1\}^{-1}  \{2 p_i\cdot (\tilde k_1 + \tilde k_2) + 2 \ \tilde k_1\cdot \tilde k_2
\}^{-1} \cdots
\left\{2p_i\cdot (\tilde k_1+\cdots +\tilde k_n) + 2\sum_{r,u=1\atop 
r<u}^n \tilde k_r\cdot \tilde k_u
\right\}^{-1} \nonumber \\
&&\eps_i^T \, \Bigl[ 2 \ \tilde\ve_1^{\mu\nu}\ p_{i\mu} p_{i\nu}
+ i\, \tilde\ve_{1,\mu\nu} p_i^\nu \KK(-p_i) {\p \Xi^i(-p_i)\over \p p_{i\mu}}
+ 2\ \tilde\ve_{1,b\mu}
 \tilde k_{1a}p_i^\mu  (J^{ab})^T +\KK(-p_i) \ \QQ(p_i, \tilde k_1)\Bigl] \nonumber \\ &&
 \Bigl[ 2 \ \tilde\ve_2^{\mu\nu}  \{p_{i\mu}+\tilde k_{1\mu}\} \{p_{i\nu}+\tilde k_{1\nu}\}
+ i\, \tilde\ve_{2,\mu\nu} (p_i^\nu+\tilde k_1^\nu) \KK(-p_i-\tilde k_1) {\p \Xi^i(-p_i-\tilde k_1)\over \p p_{i\mu}}
\nonumber \\ && \hskip 1in 
+ 2\ \tilde\ve_{2,b\mu}
 \tilde k_{2a}p_i^\mu  (J^{ab})^T +\KK(-p_i) \ \QQ(p_i, \tilde k_2)\Bigl]
\nonumber \\ &&
\cdots 
 \Bigl[ 2 \ \tilde\ve_n^{\mu\nu}  \{p_{i\mu}+\tilde k_{1\mu}+\cdots + \tilde k_{(n-1)\mu}\}
  \{p_{i\nu}+\tilde k_{1\nu}+\cdots + \tilde k_{(n-1)\nu}\} \nonumber \\ &&
+ i\, \tilde\ve_{n,\mu\nu} (p_i^\nu+\tilde k_1^\nu +\cdots +\tilde k_{n-1}^\nu)\, 
\KK(-p_i-\tilde k_1-\tilde k_2-\cdots \tilde k_{n-1}) 
{\p \Xi^i(-p_i-\tilde k_1-\tilde k_2-\cdots -\tilde k_{n-1})\over \p p_{i\mu}}
\nonumber \\ && \hskip .5in + 2\ \tilde\ve_{n,b\mu}
 \tilde k_{na}p_i^\mu  (J^{ab})^T +\KK(-p_i) \ \QQ(p_i, \tilde k_n)\Bigl] \,
  \Gamma_{(i)}(p_i+\tilde k_1+\cdots +\tilde k_n)\, .
\een

First let us analyze the contribution from the $\tilde k_r\cdot \tilde k_u$ terms in the
denominator. Since this is subleading, we need to expand one of the denominators 
to first order in  $\tilde k_r\cdot \tilde k_u$, set  $\tilde k_r\cdot \tilde k_u=0$ in
the rest of the denominators, and pick the leading contribution from all other factors.
This leads to
\be 
-\Bigg\{\sum_{m=2}^n \sum_{r,u=1\atop r<u}^m {\tilde k_r\cdot \tilde k_u\over p_i\cdot
(\tilde k_1+\cdots + \tilde k_m)} \Bigg\} \, \Bigg\{\prod_{\ell=1}^n
{1\over p_i\cdot (\tilde k_1+\cdots +\tilde k_\ell)} \Bigg\} 
\Bigg\{ \prod_{s=1}^n \tilde\ve_{s,\mu\nu} p_i^\mu p_i^\nu \Bigg\}
\eps_i^T \ \Gamma_{(i)}(p_i)\, .
\ee
After performing the sum over all permutations of $\tilde k_1,\cdots, \tilde k_n$ 
using \refb{eap2} this gives
\be \label{ekk1}
-\prod_{s=1}^n \, \Bigg\{ (p_i\cdot \tilde k_s)^{-1} \ 
\tilde
\ve_{s,\mu\nu} \, p_i^\mu p_i^\nu \Bigg\} \sum_{r,u=1\atop r< u}^n \tilde k_r\cdot \tilde k_u \, 
\{p_i\cdot (\tilde k_r+\tilde k_u)\}^{-1}\, .
\ee

Next we consider the terms involving the contraction of $\tilde\ve_u$ with 
$\tilde k_r$ for $r<u$,
coming from the first term inside each square bracket in \eqref{emoduleprod}. Since this term is subleading, once we pick one of these 
factors we must pick the leading terms from all the other factors. Again using \refb{eepK} we can express the
sum of all such contributions as
\be 
2\ \sum_{r,u=1\atop r<u}^n \  \Bigg\{ 
\prod_{s=1\atop s\ne u}^n \tilde\ve_{s,\mu\nu} p_i^\mu p_i^\nu \Bigg\}
\ \tilde\ve_{u,\mu\nu} \, p_i^\mu \ \tilde k_r^\nu \, \Bigg\{\prod_{m=1}^n
{1\over p_i\cdot (\tilde k_1+\cdots +\tilde k_m)} \Bigg\} 
\eps_i^T \ \Gamma_{(i)}(p_i)\, .
\ee
After summing over all permutations of $(\tilde k_1, \tilde \ve_1), \cdots , (\tilde k_n, \tilde \ve_n)$ using \refb{eap3}  this gives
\ben \label{ekk2} 
&&  2\ \Bigg\{\prod_{s=1}^n \, (p_i\cdot \tilde k_s)^{-1}\Bigg\}  
\ \sum_{r,u=1\atop r< u}^n \ \{ p_i\cdot (\tilde k_r+\tilde k_u)\}^{-1}\
\Bigg\{\prod_{s=1\atop s\ne r,u}^n \, \tilde\ve_{s,\mu\nu}
p_i^\mu p_i^\nu
\Bigg\}  \nonumber \\ &&  \hskip .3in 
  \Bigg\{ (p_i\cdot \tilde k_u) \
(p_i\cdot \tilde\ve_r\cdot p_i) \ (p_i\cdot \tilde\ve_u\cdot \tilde k_r) 
+ (p_i\cdot \tilde k_r) \ (p_i\cdot \tilde\ve_u\cdot p_i) \ (p_i\cdot \tilde\ve_r\cdot \tilde k_u)
  \Bigg\} \ \eps_i^T \ \Gamma_{(i)}(p_i)\, . \nonumber \\
\een

We now turn to the rest of the contribution from \refb{emoduleprod}  in which we drop the
$\tilde k_r\cdot \tilde k_u$ factors in the denominator and also the 
terms involving contraction of 
$\tilde k_r$ with $\tilde \ve_u$ in the first term inside each square bracket.
Our first task will be to expand the factors of $\KK$ and $\Xi^i$ in Taylor series expansion in powers
of the soft momenta. It is easy to see however that to the first subleading order, the order $\tilde{k}^\mu$ terms
in the expansion of $\Xi^i$ do not contribute to the amplitude. This is due to the fact that once we have
picked a subleading term proportional to $\tilde{k}_s^\rho \p^2 \Xi^i /\p p_i^\mu \p  p_i^\rho$, 
we must replace the argument of $\KK$ by 
$-p_i$ in the accompanying factor and in all other factors we must pick the leading term.
In this case repeated use of \refb{eepK} shows that the corresponding contribution vanishes. Therefore we
can replace all factors of $\p\Xi^i (-p_i-\tilde k_1-\cdots)/\p p_i^\mu$ by $\p\Xi^i(-p_i)/\p p_i^\mu$. 
Similar argument shows that all the $\KK(-p_i) \QQ$ terms,
and the terms involving contraction of $\tilde\ve_u$ with $\tilde k_r$
in the second term inside each square bracket in \refb{emoduleprod}, give vanishing
contribution at the subleading order. This allows us to
express the rest of the contribution from \refb{emoduleprod} as
\ben \label{enew1} 
&& (2p_i\cdot\tilde  k_1)^{-1}  \{2 p_i\cdot (\tilde k_1 + \tilde k_2)\}^{-1} \cdots
\{2p_i\cdot (\tilde k_1+\cdots +\tilde k_n)\}^{-1} \nonumber \\
&&\eps_i^T \, \Bigl[ 2 \ \EE_1 + 2 \ \LL_1
+ 2\ \tilde\ve_{1,b\mu}
 \tilde k_{1a}p_i^\mu  (J^{ab})^T \Bigl] \nonumber \\ &&
 \Bigl[ 2 \ \EE_2 + 2\ \LL_2
+ i\, \tilde\ve_{2,\mu\nu} p_i^\nu \tilde k_{1\rho} {\p\KK(-p_i)\over \p p_{i\rho}} {\p \Xi^i(-p_i)\over \p p_{i\mu}}
+ 2\ \tilde\ve_{2,b\mu}
 \tilde k_{2a}p_i^\mu  (J^{ab})^T \Bigl] \ \cdots 
\nonumber \\ && \hskip -.5in
 \Bigl[ 2 \ \EE_n + 2\ \LL_n 
 + i\, \tilde\ve_{n,\mu\nu} p_i^\nu \, (\tilde k_{1\rho}+\cdots 
+\tilde k_{n-1,\rho}) {\p \KK(-p_i) \over \p p_{i\rho}}\
{\p \Xi^i(-p_i)\over \p p_{i\mu}} 
+ 2\ \tilde\ve_{n,b\mu}
 \tilde k_{na}p_i^\mu  (J^{ab})^T \Bigl] \, 
 \nonumber \\ && \hskip .5in 
\Gamma_{(i)}(p_i+\tilde k_1+\cdots +\tilde k_n)\, ,
\een
where,
\be \label{edeflspi}
\EE_s = \tilde\ve_s^{\mu\nu} p_{i\mu} p_{i\nu}, \qquad \LL_s= 
{i\over 2} \tilde\ve_s^{\mu\nu} p_{i\nu}
\KK(-p_i) \, {\p\Xi^i(-p_i)\over \p p_{i\mu}}\, .
\ee
We now expand \refb{enew1} in powers of soft momenta. Even though $\LL_s$ is 
leading order, its contribution to the amplitude vanishes by \refb{eepK} unless
there is some other matrix sitting between $\eps_i^T$ and $\LL_s$. The possible
terms come from picking up either the term proportional to $\p \KK/\p p_{i\rho}
\ \p \Xi/\p p_{i\mu}$ or $(J^{ab})^T$ from one of the factors. Both these terms are
subleading and therefore we can pick at most one such term in the product, with
the other factors being given by $\EE_s+\LL_s$.
Therefore if we expand \refb{enew1} and pick the subleading factor from the $r$-th
term in the product, then in the product of $\EE_s+\LL_s$,
we can drop all factors of $\LL_s$ for
$s<r$ since they sit to the left of the subleading factor and will  vanish due to 
\refb{eepK}. This gives the following expression for the subleading contribution to
\refb{enew1}:
\ben \label{enew1.5}
&& (p_i\cdot \tilde k_1)^{-1}  \{p_i\cdot (\tilde k_1 + \tilde k_2)\}^{-1} \cdots
\{p_i\cdot (\tilde k_1+\cdots +\tilde k_n)\}^{-1} \nonumber \\
&&\eps_i^T \,  \Bigg[\sum_{r=1}^n \bigg\{\prod_{s=1}^{r-1}
\EE_s\bigg\} \  \Bigl[
 {i\over 2} \, \tilde\ve_{r,\mu\nu} p_i^\nu \, (\tilde k_{1\rho}+\cdots 
+\tilde k_{r-1,\rho}) {\p \KK(-p_i) \over \p p_{i\rho}}\
{\p \Xi^i(-p_i)\over \p p_{i\mu}} 
+ \ \tilde\ve_{r,b\mu}
 \tilde k_{ra}p_i^\mu  (J^{ab})^T \Bigl] \nonumber \\ &&
\Bigg\{\prod_{s=r+1}^n (\EE_s+\LL_s)
\Bigg\} \ \Gamma_{(i)}(p_i)
 \nonumber \\ &&
+\,  (p_i\cdot \tilde k_1)^{-1}  \{p_i\cdot (\tilde k_1 + \tilde k_2)\}^{-1} \cdots
\{p_i\cdot (\tilde k_1+\cdots +\tilde k_n)\}^{-1} \  \bigg\{\prod_{s=1}^{n}
\EE_s\bigg\} \  \sum_{r=1}^n \tilde k_{r\rho} \ {\p \Gamma_{(i)}(p_i)\over \p p_{i\rho}}\, .
\nonumber \\
\een
The last term comes from the Taylor series expansion of $\Gamma_{(i)}$ in powers of soft momenta.
In the product the $(\EE_s+\LL_s)$'s 
are ordered from left to right in the order of increasing $s$.
We now manipulate the product $\prod_{s=r+1}^n (\EE_s+\LL_s)$ as follows. 
If the subleading factor is the one proportional to
$\p \KK/\p p_{i\rho}
\ \p \Xi/\p p_{i\mu}$ then we leave the product of the factors $(\EE_s+\LL_s)$ for $s>r$
unchanged. However if the subleading factor is the one proportional to $(J^{ab})^T$, then
we expand the
product of the factors $(\EE_s+\LL_s)$ for $s>r$ as
\be \label{eexp22}
(\EE_{r+1}+\LL_{r+1}) \cdots (\EE_n +\LL_n)
=\EE_{r+1}\cdots \EE_n + \sum_{u=r+1}^n \EE_{r+1}\cdots \EE_{u-1} \, \LL_u \, (\EE_{u+1}+\LL_{u+1})
\cdots (\EE_n+\LL_n)\, .
\ee
Using this, and combining the contribution from the first term on the right hand side of  \refb{eexp22}
with the last term in \refb{enew1.5}, we can express \refb{enew1.5} as
\ben \label{elast}
&& (p_i\cdot \tilde k_1)^{-1}  \{p_i\cdot (\tilde k_1 + \tilde k_2)\}^{-1} \cdots
\{p_i\cdot (\tilde k_1+\cdots +\tilde k_n)\}^{-1} \nonumber \\
&&\eps_i^T \,  \Bigg[\sum_{r=1}^n \bigg\{\prod_{s=1\atop s\ne r}^n 
\EE_s\bigg\}
\left\{\tilde\ve_{r,b\mu}
 \tilde k_{ra}p_i^\mu  (J^{ab})^T \Gamma_{(i)}(p_i)+ \tilde\ve_r^{\mu\nu} p_{i\mu} p_{i\nu} \ \tilde k_{r\rho}
 {\p\Gamma_{(i)}\over
 \p p_{i\rho}}\right\} \nonumber \\ &&
\hskip -.3in  +  {i\over 2} \sum_{r=1}^n \bigg\{\prod_{s=1}^{r-1}
 \EE_s \bigg\}
 \tilde\ve_{r,\mu\nu} p_i^\nu \, (\tilde k_{1\rho}+\cdots 
+\tilde k_{r-1,\rho}) {\p \KK(-p_i) \over \p p_{i\rho}}
{\p \Xi^i(-p_i)\over \p p_{i\mu}}  \Bigg\{\prod_{s=r+1}^n (\EE_s+\LL_s)\Bigg\} \ \Gamma_{(i)} \nonumber \\ &&
\hskip -.3in +  {i\over 2} \sum_{r=1}^n \sum_{u=r+1}^n 
\bigg\{\prod_{s=1\atop s\ne r}^{u-1} 
\EE_s\bigg\} \
\tilde\ve_{r,b\mu}
 \tilde k_{ra}p_i^\mu  (J^{ab})^T \
 \tilde\ve_{u,\rho\sigma} p_i^\sigma \KK(-p_i) 
{\p \Xi^i(-p_i)\over \p p_{i\rho}} \Bigg\{\prod_{s=u+1}^n (\EE_s+\LL_s)
\Bigg\} \ \Gamma_{(i)}
 \Bigg]\, . \nonumber \\
\een
We now use \refb{ejmove} to move the $\KK(-p_i)$ factor in the last term to the left of
$(J^{ab})^T$ and use \refb{eepK}. 
This allows us to express \refb{elast}  as
\ben\label{enew2}
&& (p_i\cdot \tilde k_1)^{-1}  \{p_i\cdot (\tilde k_1 + \tilde k_2)\}^{-1} \cdots
\{p_i\cdot (\tilde k_1+\cdots +\tilde k_n)\}^{-1} \nonumber \\
&& \eps_i^T \,  \Bigg[\sum_{r=1}^n \bigg\{\prod_{s=1\atop s\ne r}^n 
\EE_s \bigg\}
\left\{\tilde\ve_{r,b\mu}
 \tilde k_{ra}p_i^\mu  (J^{ab})^T \Gamma_{(i)}(p_i)+ \tilde\ve_r^{\mu\nu} p_{i\mu} p_{i\nu} \tilde k_{r\rho}
 {\p\Gamma_{(i)}\over
 \p p_{i\rho}}\right\} \nonumber \\ &&
 +  {i\over 2} \sum_{r=1}^n \bigg\{\prod_{s=1}^{r-1} 
\EE_s\bigg\}
 \tilde\ve_{r,\mu\nu} p_i^\nu \, (\tilde k_{1\rho}+\cdots 
+\tilde k_{r-1,\rho}) {\p \KK(-p_i) \over \p p_{i\rho}}
{\p \Xi^i(-p_i)\over \p p_{i\mu}} 
\Bigg\{\prod_{s=r+1}^n (\EE_s+\LL_s)
\Bigg\} \ \Gamma_{(i)} \nonumber \\ &&
+  {i\over 2} \sum_{r,u=1\atop r<u}^n \bigg\{\prod_{s=1\atop s\ne r}^{u-1} 
\EE_s \bigg\}
\ \tilde\ve_{r,b\mu}
 \tilde k_{ra}p_i^\mu  
\  \tilde\ve_{u,\rho\sigma} p_i^\sigma \left(p_i^a {\p \KK(-p_i)\over \p p_{ib}} -  
 p_i^b {\p \KK(-p_i)\over \p p_{ia}}\right)
{\p \Xi^i(-p_i)\over \p p_{i\rho}} \nonumber \\ && \hskip 2in  
\Bigg\{\prod_{s=u+1}^n (\EE_s+\LL_s) \Bigg\} \ \Gamma_{(i)}
 \Bigg]\, . 
\een
It is easy to see that terms proportional to $p_i^b \p \KK/\p p_{ia}$ in the fourth 
line of \refb{enew2}
cancels the terms in the third line of \refb{enew2}. Therefore we are left with
\ben\label{enew3}
&& (p_i\cdot \tilde k_1)^{-1}  \{p_i\cdot (\tilde k_1 + \tilde k_2)\}^{-1} \cdots
\{p_i\cdot (\tilde k_1+\cdots +\tilde k_n)\}^{-1} \nonumber \\
&&\eps_i^T \,  \Bigg[\sum_{r=1}^n \Bigg\{\prod_{s=1\atop s\ne r}^n \EE_s
\Bigg\}
\left\{\tilde\ve_{r,b\mu}
 \tilde k_{ra}p_i^\mu  (J^{ab})^T \Gamma_{(i)}(p_i)+ \tilde\ve_r^{\mu\nu} p_{i\mu} p_{i\nu} \tilde k_{r\rho}
 {\p\Gamma_{(i)}\over
 \p p_{i\rho}}\right\} \nonumber \\ &&
 +  {i\over 2} \sum_{r,u=1\atop r<u}^n \Bigg\{\prod_{s=1\atop s\ne r}^{u-1}
\EE_s\Bigg\}
\ p_i\cdot \tilde k_{r}  \
  \tilde\ve_{r,b\mu} \, p_i^\mu \ \tilde\ve_{u,\rho\sigma} \, p_i^\sigma\ 
{\p \KK(-p_i)\over \p p_{ib}} {\p \Xi^i(-p_i)\over \p p_{i\rho}} 
\Bigg\{\prod_{s=u+1}^n 
(\EE_s+\LL_s)\Bigg\} \ \Gamma_{(i)}
 \Bigg]\, . \nonumber \\
 \een

First consider
the term in the second line of \refb{enew3}. 
We  sum over all permutations of $(\tilde \ve_1, \tilde k_1),\cdots , (\tilde\ve_n,
\tilde k_n)$. 
After the sum over $r$ is performed, this
expression is already invariant under the permutations of the soft gravitons inserted on
the $i$-th line. Therefore we simply have to sum the expression in the first line
over all permutations using \refb{eap1}, producing the result:
\be 
\bigg\{\prod_{s=1}^n (p_i\cdot \tilde k_s)^{-1} \bigg\} \
\eps_i^T \,  \Bigg[\sum_{r=1}^n \prod_{s=1\atop s\ne r}^n \{\tilde\ve_s^{\mu\nu} p_{i\mu} p_{i\nu}\}
\left\{\tilde\ve_{r,b\mu}
 \tilde k_{ra}p_i^\mu  (J^{ab})^T \Gamma_{(i)}(p_i)+ \tilde\ve_r^{\mu\nu} p_{i\mu} p_{i\nu} \tilde k_{r\rho}
 {\p\Gamma_{(i)}\over
 \p p_{i\rho}}\right\} \, .
 \ee
Since this is already subleading, we have to pick the leading contribution from all the
other external legs, producing factors of $\prod_{s\in A_j}\big\{
(p_j\cdot k_s)^{-1} \ve_{s,\mu\nu} p_j^\mu p_j^\nu\big\}$ 
after summing over
permutations of the soft gravitons. Finally we sum over all ways of distributing the soft gravitons on the
external lines. The net contribution from these terms is given by
\be \label{ess1}
\sum_{r=1}^M \sum_{i=1}^N (p_i\cdot k_r)^{-1} 
\, \ve_{r,b\rho}  \,  k_{r a} \, 
p_i^\rho\  \eps_i^T\, \left[ p_i^b 
{\p \Gamma_{(i)}(p_i) \over \p p_{ia}} 
+   (J^{ab})^T
\Gamma_{(i)}(p_i)\right]
\prod_{s=1\atop s\ne r}^M \, \bigg\{\sum_{j=1}^N (p_j\cdot k_s)^{-1}
\ \ve_{s,\mu\nu} \, p_j^\mu  p_{j}^\nu \bigg\}\, .
 \ee
We now combine this with the contribution from the sum of graphs where one soft graviton
attaches to the amplitude via the vertex $\wt\Gamma$ shown in Fig.~\ref{fig4} and the
other soft gravitons attach to the external lines. Using 
\refb{ecompver} we get the contribution from these graphs to be 
\be \label{ess2}
- \sum_{r=1}^M \Bigg[\prod_{s=1\atop s\ne r}^M \, \bigg\{\sum_{j=1}^N (p_j\cdot k_s)^{-1}
\ \ve_{s,\mu\nu} \, p_j^\mu  p_{j}^\nu \bigg\}\Bigg] \  \sum_{i=1}^N 
\, \ve_{r,ab}  \, 
p_i^a\  \eps_i^T\, 
{\p \Gamma_{(i)}(p_i) \over \p p_{ib}} 
\, .
 \ee
The sum of \refb{ess1} and \refb{ess2} reproduces the terms in the second
and third line of \refb{efullgen}.

Let us now turn to the contribution from the last line 
of \refb{enew3}. We express
$p_i\cdot \tilde k_r$ factor as
\be \label{ex1}
p_i \cdot (\tilde k_1+\cdots + \tilde k_r) - p_i\cdot (\tilde k_1+\cdots \tilde  k_{r-1})
\ee
so that  each term in \refb{ex1} cancels one of the denominator factors in
the first line of \refb{enew3}. Now we are supposed to sum over all permutations of
the soft gravitons carrying the labels $1,\cdots\, n$. However
instead of summing over all permutations of $\tilde k_1,\cdots , \tilde k_n$ in one
step, let us first fix the positions of all soft gravitons except the one carrying momentum $\tilde k_r$,
and sum over all insertions of the soft graviton carrying momentum $\tilde k_r$
to the left of the one carrying momentum $\tilde k_u$. Using \refb{ex1} at each step, it is easy to
see that the contributions from the terms cancel pairwise.
For example for three soft gravitons, with 1 fixed to the left of 3, and the position of 2
summed over on all positions to the left of 3, we have
\ben
&& \{p_i\cdot \tilde k_1\}^{-1} \{p_i\cdot (\tilde k_1+\tilde k_2)\}^{-1} 
\{p_i\cdot (\tilde k_1+\tilde k_2+\tilde k_3)\}^{-1} \ \{p_i \cdot (\tilde k_1+\tilde k_2)- p_i\cdot \tilde k_1\}
\nonumber \\
&& +  \ \{p_i\cdot \tilde k_2\}^{-1} \{p_i\cdot (\tilde k_1+\tilde k_2)\}^{-1} 
\{p_i\cdot (\tilde k_1+\tilde k_2+\tilde k_3)\}^{-1} \{p_i\cdot  \tilde k_2\} \nonumber \\
&=&\{p_i\cdot \tilde k_1\}^{-1}\ \{p_i\cdot (\tilde k_1+\tilde k_2+\tilde k_3)\}^{-1}\, .
\een
As a result of this pairwise cancellation, at the end we are left with only one term
arising from the insertion of $\tilde k_r$ just to the left of $\tilde k_u$. In order to express the result in a 
convenient form
we relabel the gravitons attached to the $i$-th line from left to right, other than the one carrying
momentum $\tilde k_r$, as 
\be (\hat \ve_1, \hat k_1),
\cdots , (\hat \ve_{u-2}, \hat k_{u-2}),  (\tilde\ve_u, \tilde k_u), 
(\hat \ve_{u+1}, \hat k_{u+1}), \cdots , (\hat\ve_n, \hat k_n)\, .
\ee
and sum over all insertions of
the graviton carrying the quantum numbers $(\tilde \ve_r, \tilde k_r)$ to the left of
$(\tilde\ve_u, \tilde k_u)$. Then for fixed $r,s$ the result
is given by
\ben\label{enew4}
&& 
(p_i\cdot \hat k_1)^{-1}  \{p_i\cdot (\hat k_1 + \hat k_2)\}^{-1} \cdots
\{p_i\cdot (\hat k_1+\cdots +\hat k_{u-2})\}^{-1} \nonumber \\ &&
\{p_i\cdot (\hat k_1+\cdots + \hat k_{u-2} + \tilde k_r + \tilde k_u)\}^{-1}  \cdots
\{p_i\cdot (\hat k_1+\cdots+ \hat k_{u-2} + \tilde k_r + \tilde k_u + \hat k_{u+1} +\cdots 
+\hat k_n)\}^{-1} \nonumber \\
&&\eps_i^T \,  \Bigg[ {i\over 2} \Bigg\{\prod_{s=1}^{u-2}\
\wh\EE_s \Bigg\}  \
  \tilde\ve_{r,b\mu} \, p_i^\mu \ \tilde\ve_{u,\rho\sigma} \, p_i^\sigma\ 
{\p \KK(-p_i)\over \p p_{ib}} {\p \Xi^i(-p_i)\over \p p_{i\rho}} 
\Bigg\{\prod_{s=u+1}^n (\wh\EE_s+\wh\LL_s)\Bigg\} \
 \Bigg] \Gamma_{(i)}(p_i)\, , 
 \een
where
\be
\wh\EE_s = \hat\ve_s^{\mu\nu} p_{i\mu} p_{i\nu}, \qquad
\wh\LL_s = {i\over 2} \hat\ve_s^{\mu\nu} p_{i\nu}
\KK(-p_i) \, {\p\Xi^i(-p_i)\over \p p_{i\mu}}\, .
\ee

Next we add to this a term obtained by exchanging the positions of $r$ and $u$. 
This is equivalent to exchanging the $\rho$ and $b$ indices in 
$(\p\KK/\p p_{ib}) (\p\Xi/\p p_{i\rho})$ and gives
\ben\label{enew5}
&& 
(p_i\cdot \hat k_1)^{-1}  \{p_i\cdot (\hat k_1 + \hat k_2)\}^{-1} \cdots
\{p_i\cdot (\hat k_1+\cdots +\hat k_{u-2})\}^{-1} \nonumber \\ &&
\{p_i\cdot (\hat k_1+\cdots + \hat k_{u-2} + \tilde k_r + \tilde k_u)\}^{-1}  \cdots
\{p_i\cdot (\hat k_1+\cdots+ \hat k_{u-2} + \tilde k_r + \tilde k_u + \hat k_{u+1} +\cdots 
+\hat k_n)\}^{-1} \nonumber \\
&&\eps_i^T \,  \Bigg[ {i\over 2} \Bigg\{\prod_{s=1}^{u-2}\
\wh\EE_s \Bigg\}\
  \tilde\ve_{r,b\mu} \, p_i^\mu \ \tilde\ve_{u,\rho\sigma} \, p_i^\sigma\ 
{\p \KK(-p_i)\over \p p_{i\rho}} {\p \Xi^i(-p_i)\over \p p_{ib}} 
\Bigg\{\prod_{s=u+1}^n (\wh\EE_s+\wh\LL_s)\Bigg\} \
 \Bigg] \Gamma_{(i)}(p_i)\, . 
 \nonumber \\
 \een

\begin{figure}
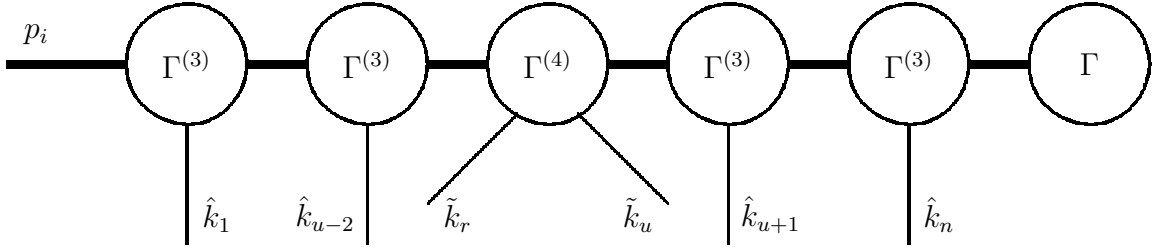


\begin{center}

\figeleven

\end{center}

\vskip -.4in

\caption{A subleading contribution to the amplitude with multiple soft gravitons.
\label{fig11}}

\end{figure}

\begin{figure}
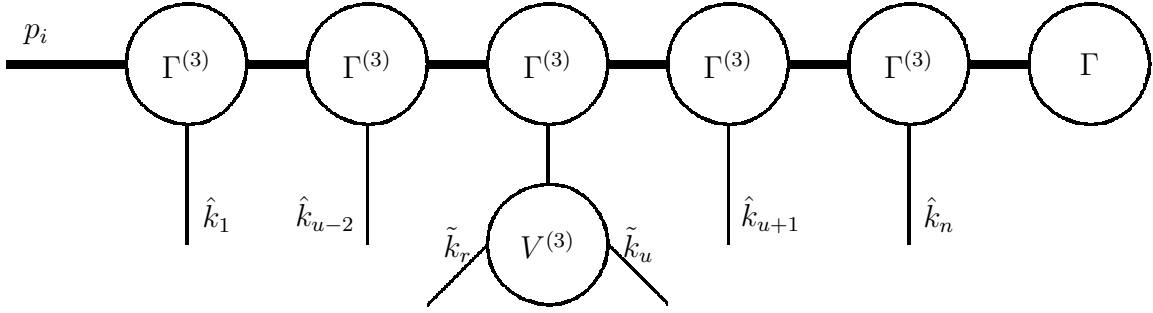


\begin{center}

\figtwelve

\end{center}

\vskip -.2in

\caption{Another subleading contribution to the amplitude with multiple soft gravitons.
\label{fig12}}

\end{figure}

Summing over the remaining permutations corresponds to treating the $r$-th and
$u$-th graviton as one unit sitting together, and summing over all permutations of the $n-1$ objects
generated this way. However it will be more convenient for us to first add to this the contribution from
the diagrams shown in in Fig.~\ref{fig11} and
Fig.~\ref{fig12}.  Since these diagrams are subleading, we need to pick the leading
contribution from all the vertices and propagators. For the product of any of 
the $\Gamma^{(3)}$
vertices and the propagator to the right of this vertex, we can use \refb{emodule} 
to generate
factors of $(\wh\EE_s+\wh\LL_s)$ in the numerator. Using \refb{eepK} we can argue that
in all such factors to the left of the vertex, where momentum $\tilde k_r+\tilde k_u$
enters the finite energy line, we can drop the $\wh\LL_s$ factors. 
Following analysis similar to the ones leading to \refb{edefa4} and \refb{edefa5}, we
arrive at the following results for
Figs.\ref{fig11} and \ref{fig12} respectively,
\ben\label{enew11}
&& 
(p_i\cdot \hat k_1)^{-1}  \{p_i\cdot (\hat k_1 + \hat k_2)\}^{-1} \cdots
\{p_i\cdot (\hat k_1+\cdots +\hat k_{u-2})\}^{-1} \nonumber \\ &&
\{p_i\cdot (\hat k_1+\cdots + \hat k_{u-2} + \tilde k_r + \tilde k_u)\}^{-1}  \cdots
\{p_i\cdot (\hat k_1+\cdots+ \hat k_{u-2} + \tilde k_r + \tilde k_u + \hat k_{u+1} +\cdots 
+\hat k_n)\}^{-1} \nonumber \\
&&\eps_i^T \,  \Bigg[ \Bigg\{\prod_{s=1}^{u-2}\
\wh\EE_s \Bigg\}
\biggl\{ -2\varepsilon_{r\mu}^{\;\;\;\,\nu}\varepsilon_{u,\nu\rho}p_{i}^{\rho}p_i^\mu- {i\over 2} 
\Bigl(\varepsilon_{r,\mu\sigma}\varepsilon_{u,\rho\nu}p_i^{\sigma}p_i^{\nu}
 +\varepsilon_{r,\rho\sigma}\varepsilon_{u,\mu\nu}p_i^{\sigma}p_i^{\nu}\Bigl) {\p \KK(-p_i)\over \p p_{i\mu}}
{\p \Xi^i(-p_i)\over \p p_{i\rho}}\biggl\} \nonumber \\
&& \hskip 1in \Bigg\{\prod_{s=u+1}^n (\wh\EE_s+\wh\LL_s)\Bigg\} \
 \Bigg] \Gamma_{(i)}(p_i)\, , 
 \een
and\footnote{We could have dropped the $\wh\LL_s$ factors from \refb{enew12} using
\refb{eepK}, but will postpone this till the next step.}
\ben\label{enew12}
&& 
(\tilde k_r\cdot \tilde k_u)^{-1}\ 
(p_i\cdot \hat k_1)^{-1}  \{p_i\cdot (\hat k_1 + \hat k_2)\}^{-1} \cdots
\{p_i\cdot (\hat k_1+\cdots +\hat k_{u-2})\}^{-1} \nonumber \\ &&
\{p_i\cdot (\hat k_1+\cdots + \hat k_{u-2} + \tilde k_r + \tilde k_u)\}^{-1}  \cdots
\{p_i\cdot (\hat k_1+\cdots+ \hat k_{u-2} + \tilde k_r + \tilde k_u + \hat k_{u+1} +\cdots 
+\hat k_n)\}^{-1} \nonumber \\
&&\eps_i^T \,  \Bigg[\Bigg\{\prod_{s=1}^{u-2}\
\wh\EE_s \Bigg\}
\  \Bigl\{-(\tilde k_u\cdot\tilde\ve_r\cdot\tilde\ve_u\cdot p_i)(\tilde k_u\cdot p_i) -(\tilde k_r\cdot\tilde\ve_u\cdot\tilde\ve_r\cdot p_i)(\tilde k_r\cdot p_i)  \nonumber\\
&&+(\tilde k_u\cdot\tilde\ve_r\cdot\tilde\ve_u\cdot p_i)(\tilde k_r\cdot p_i)+(\tilde k_r\cdot\tilde\ve_u\cdot\tilde\ve_r\cdot p_i)(\tilde k_u\cdot p_i)  - \tilde\ve_r^{cd}\ve_{u,cd}
(\tilde k_{r} \cdot p_i)( \tilde k_{u}\cdot p_i)  \nonumber\\
&& -2(p_i\cdot\tilde\ve_r\cdot \tilde k_u)(p_i\cdot\tilde\ve_u\cdot \tilde k_r) + (p_i\cdot\tilde\ve_u\cdot p_i)(\tilde k_u\cdot\tilde\ve_r\cdot \tilde k_u)  + (p_i\cdot\tilde\ve_r\cdot p_i)(\tilde k_r\cdot\tilde\ve_u\cdot \tilde k_r)\Bigl\}
\nonumber \\ && \hskip 1in 
\Bigg\{\prod_{s=u+1}^n (\wh\EE_s+\wh\LL_s)\Bigg\} \
 \Bigg] \Gamma_{(i)}(p_i)\, .
 \een
After adding these to \refb{enew4}, \refb{enew5} the 
terms involving derivatives of $\KK$ and $\Xi$ get canceled. Once these terms cancel, we can drop 
the terms proportional to $\wh\LL_s$. The result takes the form
\ben \label{enewpart}
&& 
(p_i\cdot \hat k_1)^{-1}  \{p_i\cdot (\hat k_1 + \hat k_2)\}^{-1} \cdots
\{p_i\cdot (\hat k_1+\cdots +\hat k_{u-2})\}^{-1} \nonumber \\ &&
\{p_i\cdot (\hat k_1+\cdots + \hat k_{u-2} + \tilde k_r + \tilde k_u)\}^{-1}  \cdots
\{p_i\cdot (\hat k_1+\cdots+ \hat k_{u-2} + \tilde k_r + \tilde k_u + \hat k_{u+1} +\cdots 
+\hat k_n)\}^{-1} \nonumber \\
&&(\tilde k_r\cdot \tilde k_u)^{-1}\ 
\eps_i^T \,  \Bigg[\Bigg\{\prod_{s=1}^{u-2}\
\wh\EE_s \Bigg\}\ \Bigg\{\prod_{s=u+1}^n \wh\EE_s\Bigg\} 
\  \Bigl\{-2 \ \tilde k_r\cdot \tilde k_u \ \varepsilon_{r\mu}^{\;\;\;\,\nu}\varepsilon_{u,\nu\rho}p_{i}^{\rho}
p_i^\mu-(\tilde k_u\cdot\tilde\ve_r\cdot\tilde\ve_u\cdot p_i)(\tilde k_u\cdot p_i) 
\nonumber\\&&
-(\tilde k_r\cdot\tilde\ve_u\cdot\tilde\ve_r\cdot p_i)(\tilde k_r\cdot p_i) 
+(\tilde k_u\cdot\tilde\ve_r\cdot\tilde\ve_u\cdot p_i)(\tilde k_r\cdot p_i) 
+(\tilde k_r\cdot\tilde\ve_u\cdot\tilde\ve_r\cdot p_i)(\tilde k_u\cdot p_i) \nonumber\\ &&
 - \tilde\ve_r^{cd}\tilde\ve_{u,cd}
(\tilde k_{r} \cdot p_i)( \tilde k_{u}\cdot p_i)  -2(p_i\cdot\tilde\ve_r\cdot \tilde k_u)(p_i\cdot\tilde\ve_u\cdot \tilde k_r) 
+ (p_i\cdot\tilde\ve_u\cdot p_i)(\tilde k_u\cdot\tilde\ve_r\cdot \tilde k_u)  
\nonumber\\
&&+ (p_i\cdot\tilde\ve_r\cdot p_i)(\tilde k_r\cdot\tilde\ve_u\cdot \tilde k_r)\Bigl\}
 \Bigg] \Gamma_{(i)}(p_i)\, .
 \een

We can now 
sum over all permutations of the soft gravitons carrying 
momenta $\hat k_1,\cdots , \hat k_{u-2}$,
$\hat k_{u+1},\cdots , \hat k_n$ and the relative position of the 
unit carrying momentum $\tilde k_r + \tilde k_u$ among these.
The only factors that differ for different permutations 
are the factors in the first two lines of \refb{enewpart}.
Sum over permutations using \refb{eap1} converts these to
\be
\Bigg\{\prod_{s=1}^{u-1} (p_i\cdot \hat k_s)^{-1}\Bigg\} \ 
\Bigg\{\prod_{s=u+1}^{n} (p_i\cdot \hat k_s)^{-1} \Bigg\}
\ \{p_i\cdot (\tilde k_r+\tilde k_u)\}^{-1}
=\{p_i\cdot (\tilde k_r+\tilde k_u)\}^{-1}\
 \Bigg\{\prod_{s=1\atop s\ne r,u}^{n} (p_i\cdot \tilde k_s)^{-1}\Bigg\} \ 
\, .
\ee
where we have used the fact that the unordered set 
$\{\tilde k_r, \tilde k_u, \hat k_1,\cdots , \hat k_{u-2},
\hat k_{u+1},\cdots \hat k_{n}\}$ corresponds to the set $\{\tilde k_1,\cdots, \tilde k_n\}$.
Using a similar relation for the polarizations we can express the product of $\wh\EE_s$
factors in \refb{enewpart} as $\prod_{s\ne r,u}\EE_s$.
We now sum over
all possible choices of $r,u$ from the set $\{1,\cdots, n\}$, and
add to this 
the contribution \refb{ekk1}, \refb{ekk2}.
This gives
\ben\label{enew7}
&& \sum_{r,u=1\atop r<u}^n \Bigg\{\prod_{s=1\atop s\ne r,u}^n\
(\tilde\ve_s^{\mu\nu} p_{i\mu} p_{i\nu})\Bigg\}\Bigg\{
\prod_{s=1\atop s\ne r,u}^n\ (p_i\cdot \tilde k_s)^{-1}\Bigg\}  \
\{p_i\cdot (\tilde k_r +\tilde k_u)\}^{-1} 
\MM(p_i; \tilde\ve_r, \tilde k_r, \tilde \ve_u, \tilde k_u) \, \eps_i^T \, \Gamma_{(i)}(p_i)\nonumber \\
&&\hskip -.2in =
\sum_{r,u\in A_i \atop r<u} \Bigg\{\prod_{s\in A_i\atop s\ne r, u}\
(\ve_s^{\mu\nu} p_{i\mu} p_{i\nu})\Bigg\} \Bigg\{\prod_{s\in A_i\atop s\ne r,u} (p_i\cdot k_s)^{-1} \Bigg\}
\{p_i\cdot (k_r + k_u)\}^{-1} 
\MM(p_i;  \ve_r,  k_r, \ve_u, k_u) \, \eps_i^T \, \Gamma_{(i)}(p_i)
\, , \nonumber \\
\een
where we have used the fact that the set $\{\tilde k_1,\cdots, \tilde k_n\}$ 
corresponds to the set  $\{k_a; a\in A_i\}$, and that a similar
relation exists also for the polarization tensors.

Summing over all insertions of all
other soft gravitons on other legs we now get the result
\ben\label{manip}
&& \sum_{A_1,\cdots A_N; \ A_i\subset \{1,\cdots , M\}
\atop A_i\cap A_j = \emptyset \, {\rm for} \, i\ne j; \
A_1\cup A_2\cup \cdots \cup A_N=\{1,\cdots M\}}
\sum_{i=1}^N  \Bigg[\prod_{j=1\atop j\ne i}^N \prod_{q\in A_j} \left\{ (p_j \cdot k_q)^{-1} 
 \ve_{q,\mu\nu} \, p_j^\mu  p_{j}^\nu\right\} \Bigg]\nonumber \\ && \hskip 1in
 \sum_{r,u\in A_i \atop r<u} \Bigg[ \{p_i\cdot (k_r+k_u)\}^{-1}\ 
\prod_{s\in A_i\atop s\ne r, u}  \left \{(p_i\cdot k_s)^{-1} \, \ve_{s,\mu\nu} 
 \, p_i^\mu  p_{i}^\nu\right\} \ 
\nonumber \\ && \hskip 1in \MM(p_i; \ve_r, k_r, \ve_u, k_u) \ 
\Gamma(\eps_1, p_1,\cdots , \eps_N, p_N) \Bigg]\, .
\een
After rearrangement of the sums and products, this reproduces the terms on the last line
of \refb{efullgen}. This completes our proof that amplitudes with multiple soft gravitons are
given by \refb{efullgen}.

Finally let us briefly discuss the gauge invariance of \refb{efullgen}. For this it will be useful to 
use the compact notation for the amplitude $A$ as 
given in eq.\refb{efullgenintro}. Let us suppose that we transform
$\ve_p$ by the gauge transformation $\delta_p$ defined in \refb{evardef}. Then the non-vanishing 
contribution to $\delta_p\, A$ is given by
\ben \label{efinvar1}
&& \Bigg\{ \prod_{i=1}^N \eps_{i, \alpha_i}\Bigg\}
\sum_{s=1\atop s\ne p}^M \Bigg\{ \prod_{r=1\atop r\ne s, p}^M S^{(0)}_r\Bigg\} \ \delta_p S^{(0)}_p \ 
[S^{(1)}_s \Gamma]^{\alpha_1\cdots\alpha_p} \nonumber \\
&& + \Bigg\{ \prod_{i=1}^N \eps_{i, \alpha_i}\Bigg\} \sum_{r,u=1\atop r<u}^M 
\Bigg\{ \prod_{s=1\atop s\ne r,u}^M \, S^{(0)}_s \Bigg\} 
\ \Bigg\{\sum_{j=1}^N \ \{p_j\cdot (k_r+k_u)\}^{-1}\ 
\ \delta_p\  \MM(p_j; \ve_r, k_r, \ve_u, k_u) \Bigg\}  \
\Gamma^{\alpha_1\cdots  \alpha_N} \, . \nonumber \\
\een
The first line of \refb{efinvar1} can be evaluated using \refb{evar1}, and yields the result
\be \label{efinvar2}
- 2\ \Bigg\{ \prod_{i=1}^N \eps_{i, \alpha_i}\Bigg\}
\sum_{s=1\atop s\ne p}^M \Bigg\{ \prod_{r=1\atop r\ne s, p}^M S^{(0)}_r\Bigg\} \  S^{(0)}_s \ 
k_s\cdot \xi_p  \ \Gamma^{\alpha_1\cdots\alpha_N} \, .
\ee
The second line of \refb{efinvar1} receives contribution from the choices $r=p$ or $u=p$.
Since $\MM(p_i; \ve_r, k_r, \ve_u, k_u)$ is symmetric under the exchange of $r$ and $u$, we
can take $u=p$ and replace the $r<u$ constraint in the sum by $r\ne p$.  Therefore 
the second line of \refb{efinvar1} takes the form
\be \label{efinvar3}
\Bigg\{ \prod_{i=1}^N \eps_{i, \alpha_i}\Bigg\} \sum_{r=1\atop r\ne p}^M 
\Bigg\{ \prod_{s=1\atop s\ne r,p}^M \, S^{(0)}_s \Bigg\} 
\ \Bigg\{\sum_{j=1}^N \ \{p_j\cdot (k_r+k_p)\}^{-1}\ 
\ \delta_p\  \MM(p_j; \ve_r, k_r, \ve_p, k_p) \Bigg\}  \
\Gamma^{\alpha_1\cdots  \alpha_N} \, .
\ee
Using \refb{evar2}
we can now express this as
\be \label{efinvar4}
2\ \Bigg\{ \prod_{i=1}^N \eps_{i, \alpha_i}\Bigg\} \sum_{r=1\atop r\ne p}^M 
\Bigg\{ \prod_{s=1\atop s\ne r,p}^M \, S^{(0)}_s \Bigg\} 
 \ S^{(0)}_r \  k_r\cdot \xi_p \ 
 \
\Gamma^{\alpha_1\cdots  \alpha_N}\, .
\ee
This precisely cancels \refb{efinvar2}, establishing gauge invariance of the amplitude.

\bigskip

\noindent{\bf Acknowledgments:}  
We would like to thank Anirban Basu, Dileep Jatkar, Alok Laddha,
Bhuwanesh Rao Patil and Arnab Priya Saha
for discussions. The work of A.S. was supported
in part by the JC Bose fellowship of the Department of Science and Technology,
India.
The work of MV was also supported by the SPM fellowship of CSIR. 
We also thank the people and Government of India for their continuous support for theoretical physics. 

\appendix

\sectiono{Summation identities} \label{esum}

In this appendix we list three summation identities that are used in 
the analysis in
section \ref{s4}.
\be \label{eap1}
\sum_{\hbox{all permutations of subscripts $1,\cdots, n$}} \ 
\prod_{m=1}^n (a_1+a_2+\cdots + a_m)^{-1}
=  \prod_{m=1}^n (a_m)^{-1} \ \, .
\ee
\ben \label{eap2}
&&
\sum_{\hbox{all permutations of subscripts $1,\cdots, n$}} 
\ \sum_{m=2}^n \sum_{r,u=1\atop r<u}^m
b_{ru} \, (a_1+\cdots + a_m)^{-1} \prod_{\ell=1}^n (a_1+\cdots +a_\ell)^{-1}
\nonumber \\ 
&& \hskip 1in = \prod_{m=1}^n (a_m)^{-1} \ \sum_{r,u=1\atop r<u}^n b_{ru} \ (a_r+a_u)^{-1}
\qquad \hbox{for $b_{rs}=b_{sr}$ for $1\le r<s\le n$}\, .
\een
\ben\label{eap3}
&&  \sum_{\hbox{all permutations of subscripts $1,\cdots, n$}} \ \sum_{r,u=1\atop r<u}^n
c_{ur} \, \prod_{\ell=1}^n (a_1+\cdots +a_\ell)^{-1}
\nonumber \\ 
&& \hskip 1in = \prod_{m=1}^n (a_m)^{-1} \ 
\sum_{r,u=1\atop r<u}^n  (a_r+a_u)^{-1} \ (a_u\ c_{ur}+a_r \ c_{ru})\, .
\een

The proof of these identities may be given as follows. Let us first consider \refb{eap1}.
The summand on the left hand side may be expressed as
\be \label{eap4}
\int_0^\infty ds_1 e^{-s_1 a_1} \int_0^\infty ds_2 e^{-s_2 (a_1+a_2)}
\cdots \int_0^\infty ds_n e^{-s_n (a_1+\cdots +a_n)}\, .
\ee
Defining new variables
\be\label{eap5}
t_1 = s_1+ s_2+\cdots s_n, \quad t_2=s_2+\cdots + s_{n}, \cdots , \quad
t_n = s_n\, ,
\ee
we may express \refb{eap4} as
\be\label{eap6}
\int_R \, dt_1\, dt_2\, \cdots dt_n \, e^{-t_1 a_1 - t_2 a_2-\cdots - t_n a_n}
\ee
where the integration range $R$ is
\be\label{eap7}
\infty > t_ 1\ge t_ 2\ge \cdots \ge t_ {n-1} \ge t_ n\ge 0\, .
\ee
Summing over all permutations of the subscripts $1,\cdots ,n$ can now be implemented by
summing over permutations of $t_1,\cdots t_n$. This has the effect of making the integration
range unrestricted, with each $t_i$ running from 0 to $\infty$. The corresponding
integral  generates the right hand side of \refb{eap1}.

The proof of \refb{eap3} follows from a simple variation of this. For this note that the
coefficient of the $c_{ur}$ term on the left hand side for $r<u$ 
is given by a sum over permutations
with the same summand as in \refb{eap1}, but with the restriction  that we sum over
those permutations in which $r$ comes before $u$. Translated to \refb{eap7} this
means that after  summing over permutations the restriction $t_r>t_u$ is still maintained.
Therefore the result is
\be \label{eap8}
\int_{t_r\ge t_u} \, dt_1\, dt_2\, \cdots dt_n \, e^{-t_1 a_1 - t_2 a_2-\cdots - t_n a_n}\, .
\ee
This integral can be easily evaluated to give
\be 
(a_1\cdots a_n)^{-1} \, a_u \, (a_r+a_u)^{-1}\, .
\ee
This is precisely the coefficient of $c_{ur}$ on the right hand side of \refb{eap3}. Similarly in the
computation of the coefficient of $c_{ru}$ for $r<u$
we only sum over those permutations for which $u$ 
comes before $r$. This has the effect the changing the constraint $t_r\ge t_u$ to $t_r\le t_u$ in
\refb{eap8} and reproduces correctly the coefficient on $c_{ru}$ on the right hand side
of \refb{eap3}.

Finally let us consider \refb{eap2}. We begin with a different sum
\be \label{eap9}
\sum_{\hbox{all permutations of subscripts $1,\cdots, n$}} \ 
\prod_{\ell=1}^n \bigg(a_1+a_2+\cdots + a_\ell - \sum_{r,u=1\atop r<u}^\ell b_{ru}\bigg)^{-1}\, ,
\ee
and note that the first subleading term 
in a Taylor series expansion of \refb{eap9} in powers of $b_{mn}$'s give the left
hand side of \refb{eap2}. We now manipulate this as before, arriving at the analog
of \refb{eap4}:
\be \label{eap10}
\int_0^\infty ds_1 e^{-s_1 a_1} \int_0^\infty ds_2 e^{-s_2 (a_1+a_2 -b_{12})}
\cdots \int_0^\infty ds_n e^{-s_n (a_1+\cdots +a_n -\sum_{r,u=1\atop r<u}^n b_{ru} 
)}\, .
\ee
The change of variables given in \refb{eap5} converts this to
\ben\label{eap6a}
&& \int_R \, dt_1\, dt_2\, \cdots dt_n \, e^{-t_1 a_1 - t_2 a_2-\cdots - t_n a_n}
\nonumber \\
&&
\, \exp\Bigg[ (t_2-t_3) b_{12} + (t_3-t_4) (b_{12}+b_{23}+b_{13})
+ \cdots + (t_{n-1} - t_{n})  \sum_{r,u=1\atop r<u}^{n-1} b_{ru} 
+ t_n \sum_{r,u=1\atop r<u}^{n} b_{ru} \Bigg]\, . \nonumber \\
&=& \int_R \, dt_1\, dt_2\, \cdots dt_n \, e^{-t_1 a_1 - t_2 a_2-\cdots - t_n a_n}
\exp\Bigg[\sum_{r,u=1\atop r<u}^n b_{ru} \, t_u
\Bigg]\, .
\een
We now expand the last factor of \refb{eap6a} in a Taylor series expansion and
pick the coefficient of the $b_{ru}$ term. This has the effect of multiplying the
integrand by $t_u$ and restrict the sum over permutations to those for which
$r$ remains to the left of $u$. However as $b_{ru}$ is symmetric in $r,u$, there is
also another term related to this one under the exchange of the subscripts $r$ and $u$.
Therefore the integral is given by
\be 
\int_{t_r>t_u} \, dt_1\, dt_2\, \cdots dt_n \, e^{-t_1 a_1 - t_2 a_2-\cdots - t_n a_n}\ t_u
 + (r\leftrightarrow u)\, 
\, .
\ee
Evaluation of this integral gives
\be
(a_1\cdots a_n)^{-1} \,  \left\{ {a_u\over (a_r+a_u)^2} +  {a_r\over (a_r+a_u)^2} \right\} =
(a_1\cdots a_n)^{-1} \, (a_r+a_u)^{-1}\, .
\ee
This is precisely the coefficient of $b_{ru}$ on the right hand side of \refb{eap2}.

We can also give recursive proof of all the identities without using the integral representations.
Let us begin with the identity \refb{eap1}. Let us suppose that it holds for
$(n-1)$ objects. We now organise the sum over permutations of all subscripts $1,\cdots, n$
in \refb{eap1} by first fixing the last element to be some integer $i$, and 
summing over all permutations of the subscripts other than $i$. This gives, using
\refb{eap1} for $(n-1)$ objects,
\be
(a_1\cdots a_{i-1} a_{i+1}\cdots a_n)^{-1} \, (a_1+\cdots a_n)^{-1}\, .
\ee
We now sum over all possible choices of $i$.
This gives
\be
\sum_{i=1}^{n} 
(a_1\cdots a_{i-1} a_{i+1}\cdots a_n)^{-1} \, (a_1+\cdots a_n)^{-1} \, .
\ee
This can be written as
\be
(a_1\cdots a_n)^{-1} (a_1+\cdots +a_n)^{-1} \sum_{i=1}^n a_i
= (a_1\cdots a_n)^{-1}\, ,
\ee
reproducing the right hand side of \refb{eap1}.

A recursive proof of \refb{eap3} can be given as follows.
Let us again assume that the
identity is valid for $(n-1)$ objects. Now for $u>r$, the coefficient of $c_{ur}$ on the
left hand side involves a sum over permutations of the subscripts $1,\cdots, n$, with the 
same summand as in identity \refb{eap1}, but with the restriction that $r$ always appears
to the left of $u$ in the permutation. We now organise the sum as follows. 
First we fix the last element and sum over permutations of the first $(n-1)$ elements.
If the last element is $i$ with $i\ne u$, 
then the result, using \refb{eap2}
for $(n-1)$ objects, is given by
\be
\Bigg\{\prod_{m=1\atop m\ne i}^n (a_m)^{-1} \Bigg\} \ a_u \ (a_u+a_r)^{-1} \
(a_1+\cdots a_n)^{-1} \, .
\ee
Note that $i$ cannot be $r$ since that will violate the rule that the $r$ always
appears to the left of $u$.
On the other hand if the last element is $u$ then the sum over permutations over the
first $(n-1)$ elements becomes unrestricted
and we can apply \refb{eap1} to get
\be 
\Bigg\{\prod_{m=1\atop m\ne u}^n (a_m)^{-1} \Bigg\} \ (a_1+\cdots +a_n)^{-1}\, .
\ee
Therefore the total answer, obtained by summing over all possible choices of the
last element (other than $r$), is
\be
\sum_{i\ne r, u} \Bigg\{\prod_{m=1\atop m\ne i}^n (a_m)^{-1} \Bigg\} 
\ a_u \ (a_u+a_r)^{-1} \
(a_1+\cdots a_n)^{-1} + \Bigg\{\prod_{m=1\atop m\ne u}^n (a_m)^{-1} \Bigg\} \ (a_1+\cdots +a_n)^{-1}\, .
\ee
Elementary algebra reduces this to 
\be
(a_1\cdots a_n)^{-1} \, a_u\, (a_r+a_u)^{-1}\, ,
\ee
which is the coefficient of $c_{ur}$ on the right hand side of \refb{eap3}. The analysis for
the case $r>u$ is identical, with the roles of $r$ and $u$ interchanged. 

Finally we turn to the proof of \refb{eap2}. By collecting the coefficients of $b_{ru}$ on both sides
and using the symmetry of $b_{ru}$, we can write this identity as
\ben \label{eap2a}
&& \sum_{\hbox{all permutations of subscripts $1,\cdots, n$}} 
\ \sum_{m=2\atop m\ge r, u}^n  \, (a_1+\cdots + a_m)^{-1} \prod_{\ell=1}^n (a_1+\cdots +a_\ell)^{-1}
\nonumber \\  &=& 
(a_r+a_u)^{-1} \ \prod_{m=1}^n (a_m)^{-1} \, .
\een
As before, we shall proceed by assuming this to be valid for $(n-1)$ objects and then prove this for $n$
objects.  Let us first consider the contribution from the $m=n$ term in the sum on the left hand side
of \refb{eap2a}. 
The contribution of
this term is given by
\be
(a_1+\cdots +a_n)^{-2}  \sum_{\hbox{all permutations of subscripts $1,\cdots, n$}} 
\prod_{\ell=1}^{n-1} (a_1+\cdots +a_\ell)^{-1} \, .
\ee
We now perform the sum over all permutations by fixing the last element to be some fixed number $i$,
sum over permutations of the rest for which we can use \refb{eap1}, and then sum over all choices
of $i$. This gives
\be \label{eff1}
(a_1+\cdots +a_n)^{-2}  \sum_{i=1}^n \Bigg\{ \prod_{m=1\atop m\ne i}^n (a_m)^{-1}\Bigg\}
=(a_1+\cdots +a_n)^{-1} \Bigg\{ \prod_{m=1}^n (a_m)^{-1}\Bigg\}\, .
\ee
Next we consider the contribution to the sum in the left hand side of \refb{eap2a} for $m\le (n-1)$. 
This is given by
\be 
(a_1+\cdots +a_n)^{-1}\,  \sum_{\hbox{all permutations of subscripts $1,\cdots, n$}} 
\ \sum_{m=2\atop m\ge r, u}^{n-1}  \, (a_1+\cdots + a_m)^{-1} \prod_{\ell=1}^{n-1} (a_1+\cdots +a_\ell)^{-1}\, .
\ee
We again perform the sum over permutations by fixing the last element to be some fixed number $i$, summing
over permutation of the rest of the objects, and then summing over $i$. Note however that now $i$ cannot be
either $r$ or $u$ since then we cannot satisfy the constraint $m\ge r, u$. The sum over permutations can now
be performed using \refb{eap2a} for $n-1$ objects and gives
\ben \label{eff2}
&& (a_1+\cdots +a_n)^{-1}  \sum_{i=1\atop i\ne r, s}^n \Bigg\{\prod_{m=1\atop m\ne i}^n (a_m)^{-1}\Bigg\}  \ 
(a_r+a_u)^{-1}\nonumber \\ 
&=& (a_1+\cdots +a_n)^{-1}  (a_r+a_u)^{-1} \Bigg\{\prod_{m=1}^n (a_m)^{-1}\Bigg\}  \ 
(a_1+\cdots +a_n-a_r-a_u)\, .
\een
Adding this to \refb{eff1} we get
\be 
(a_r+a_u)^{-1} \  \Bigg\{\prod_{m=1}^n (a_m)^{-1}\Bigg\}\, ,
\ee
which is precisely the right hand side of \refb{eap2a}.

\end{document}